\documentclass{aa}
\usepackage{graphicx}
\usepackage{txfonts}
\usepackage{subcaption}         
\usepackage{lscape}             
\usepackage{placeins}           
\usepackage{multirow}

\usepackage{amssymb}
\usepackage{epstopdf}

\begin{document}

   \title{XShooter DESI Lens Program: Sample characterization}


%

   \author{
   Eric Jullo\inst{1} \corrauth{eric.jullo@lam.fr} \and
   Christophe Boghossian\inst{1} \email{christophe.boghossian@etu.univ-amu.fr} \and
   Luderic Chapel\inst{1} \email{luderic.chapel@etu.univ-amu.fr} \and
   Felipe Urcelay\inst{2,3} \email{fjurcelay@uc.cl} \and
   Christopher Storfer\inst{4,6} \email{cstorfer@hawaii.edu} \and
   Xiaosheng Huang\inst{5,6} \email{xhuang@lbl.gov} \and
   Raphael Gavazzi\inst{1} \email{gavazzi@iap.fr} \and
   Jens-Kristian Krogager\inst{7,8} \email{jens-kristian.krogager@univ-lyon1.fr} \and
   Aleksandar Cikota\inst{9} \email{aleksandar.cikota@noirlab.edu}
        }

   \institute{
Aix Marseille Univ, CNRS, CNES, LAM, Marseille, France\and
Institute of Astrophysics, Pontificia Universidad Cat\'olica de Chile, Santiago, Chile\and
Astrophysics Sub-department, University of Oxford, Oxford, UK\and
Institute for Astronomy, University of Hawai’i, Honolulu, HI 96822, USA\and
Department of Physics and Astronomy, University of San Francisco, San Francisco, CA 94117, USA\and
Physics Division, Lawrence Berkeley National Laboratory, Berkeley, CA 94720, USA\and
Université Claude Bernard Lyon 1, Centre de Recherche Astrophysique de Lyon UMR5574, 9 Av. Charles André, 69230 Saint-Genis-Laval, France\and
French-Chilean Laboratory for Astronomy (FCLA), CNRS-IRL3386, U. de Chile, Camino el Observatorio 1515, Casilla 36-D, Santiago, Chile\and
Gemini Observatory, NSF’s NOIRLab, La Serena, Chile
   }

   \date{Received xx; accepted yy}

 
  \abstract
   {Large imaging surveys in cosmology are detecting orders of magnitude more lens systems than known so far.
   This unprecedented dataset will lead to robust constraints on cosmology and galaxy evolution models.  However, a preliminary
    careful characterization of the lens and source samples are mandatory.
   In this work, we report on a VLT/XShooter observation program of 67 lens systems to characterize their spectroscopic redshift distribution. These systems were previously detected on the Dark Energy Spectroscopic Instrument Legacy Imaging Surveys by Huang et al. 2021 and Storfer et al. 2022 with deep residual neural network.
   We manage to measure redshifts for 58 lenses and 57 sources. We also identify 2 sources with indication of outflow in the shape of the emission lines  and 7 sources with rotating disks in $[OII]$ or $H\alpha$.
   We find no particular bias associated to the redshift measurement operation, meaning that our measured source redshift distribution is likely representative of the true one and can be used to calibrate analyses in large imaging surveys.
}
   \keywords{Cosmology --
   	Strong gravitational lensing
	}

\maketitle
\nolinenumbers

\section{Introduction}

The large cosmological survey Euclid \citep{laureijs2011}  has already uncovered a large amount of strong lensing systems \citep[e.g.]{walmsley2025, rojas2025, lines2025a, lines2025b}. 
They can be used to constrain cosmological parameters \citep{li2024}, as well as investigate the galaxy evolution scenarios over a broad range of redshifts \citep[see e.g.][for a review]{shajib2024}. The number of rare events such as double source plane  lenses \citep[DSPL,][]{li2025b} or lensed supernovae \citep{sainzdemurieta2024} will naturally increase by several orders of magnitudes, hence improving further our constraints on cosmological parameters. 

Strong gravitational lensing is also particularly useful at shedding light on how galaxies merge in groups and clusters and how dark matter and baryons adjust their dynamical equilibrium and respective density profiles all along the infalling process \citep[e.g.][]{niemiec2022}. Investigating subhalos detected in two galaxy scale lenses, \cite{despali2025} revealed some tensions in simulations between required stellar content and observed luminosity function upper limits. In 11 galaxy-cluster scale lenses, \cite{meneghetti2020} uncovered an excess of galaxy-galaxy strong lensing events compared to CDM simulations, which could be partially resolved with subhalos with denser stellar cores \citep{meneghetti2023}. Finally, with a sample of 7 galaxy scale lenses, \cite{sahu2024} found no evolution of the slope of the matter density profile $\gamma$ and redshift, in spite of theoretical predictions. The large amount of forthcoming lensing systems will allow to explore further these discrepancies.

Recent wide-field imaging surveys from Stage III weak-lensing experiments, including the Dark Energy Survey (DES), the Kilo-Degree Survey (KiDS), and the Hyper Suprime-Cam Subaru Strategic Program (HSC), have enabled the construction of increasingly large and homogeneous samples of strong gravitational lens candidates. In the Kilo-Degree Survey (KiDS), convolutional neural network searches applied to luminous red galaxies have yielded of order a few thousand initial detections, refined to 268 high-quality candidates in the latest DR4 compilation (e.g. Petrillo et al. 2019; Li et al. 2020, 2021). The Hyper Suprime-Cam Subaru Strategic Program (HSC) has produced similarly large samples through multiple complementary searches: for instance, the HOLISMOKES analysis identified 206 galaxy-scale candidates with deep-learning pipelines (Cañameras et al. 2021) and the SuGOHI analysis identified 1522 candidates (Jaelani et al. 2020). In DES, automated searches based on convolutional neural networks have led to 511 strong-lens candidates \citep{jacobs2019}.

In parallel to Stage III surveys, the DESI Legacy Imaging Surveys (DECaLS, BASS, and MzLS) have recently emerged as a major source of strong-lensing candidates over very large sky areas. Early work by \cite{huang2020} applied convolutional neural networks to $\sim 9000$ deg$^2$ of DECaLS imaging, identifying 335 new candidates through a semi-automated search combining machine learning and visual inspection. This effort was subsequently extended in \cite{huang2021}, who analyzed a larger footprint from Data Release 8 and reported 1210 additional candidates, demonstrating the efficiency of deep residual networks on wide-field data . Continued analyses within the same framework have further expanded these samples: the DR9 study identified 1895 candidates (1512 new discoveries), bringing the cumulative number of candidates in the Legacy Surveys to $\le 3000$ systems \citep{storfer2024}. These results highlight the complementarity between Legacy Survey analyses and Stage III experiments, particularly in terms of sky coverage, and confirm the key role of deep-learning-based searches in building the next generation of large, homogeneous strong-lens catalogues.

The \cite{huang2021} lens catalog was obtained by running a ResNET deep learning neural network algorithm from \cite{lanusse2018} on the $grz$ images. The 5-$\sigma$ $z$-band median limiting magnitude is 22.5 mag for galaxies. The algorithm was trained on 632 known and candidate lens systems located in the DECaLS footprint, out of which an homogeneous set of image cutouts in the $grz$ bands was produced. In addition, a sample of 21,000 non-lens image cutouts was assembled, all with at least three passes in each of the $grz$ bands, a $z$-band mag $< 20.0$, and typed as DEV or COMP. The authors also included cutouts selected by eye that could confuse the neural net, such as spiral galaxies of different sizes and spiral arm configurations, elliptical galaxies, galaxy groups, images having objects with diﬀerent colors (typically a blue galaxy next to a red
one).

Several spectroscopic campaigns were carried out to measure the redshift of the \cite{huang2021} and \cite{storfer2024} lens systems. Systems observable from the Northern sky $Dec > -20^\circ$ were included in a secondary program of DESI \citep{adame2024}. The brightest image of each lensed source and $\sim 20$\% of putative lensing galaxies were scheduled for observations, resulting in a total sample of 3588 targets. The Early Data Release contains 73 spectra of lenses, and 36 spectra of lensed sources out of which 72 and 22 redshifts were measured respectively, hence allowing to confirm 20 lens systems \citep{huang2025}. \cite{shu2025} also matched spectra from DESI DR1 with HSC observations \citep{aihara2018} and measured redshifts for 27 systems. Using NIRES spectrograph on Keck, \cite{agarwal2026} determined 6 source redshifts, hence leading to a total of 127 redshifts.

In the Southern hemisphere $Dec < -20^\circ$, we initiated a campaign of observations with the MUSE \citep{bacon2010} and XShooter instruments \citep{vernet2011}. 75 and 96 lens systems were targeted with MUSE and XShooter respectively over several semesters, in filler programs. \cite{lin2025} report the MUSE measurements of 48 systems with redshifts for both the lens and the source, and 21 systems for which only the redshift of the lens was determined. In \cite{sheu2024}, they modeled the well known Carousel lens DESI-090.9854-35.9683 at redshift $z=0.49$, with seven observed lensed sources, out of which 5 were confirmed with MUSE observations. With these measurements, \cite{urcelay2026} estimated cosmological parameters. 

In this work, we present our XShooter follow-up observations and summary statistics of the DESI ($Dec < -20^\circ$) lens and source samples. In  section \ref{sec:data}, we present the selection of our targets, the observations and the data reduction processes. In section \ref{sec:results}, we report our redshift estimates and summary statistics, and we conclude in section \ref{sec:conclusion}.

\section{XShooter Observations}
\label{sec:data}

\begin{figure}
\centering
\includegraphics{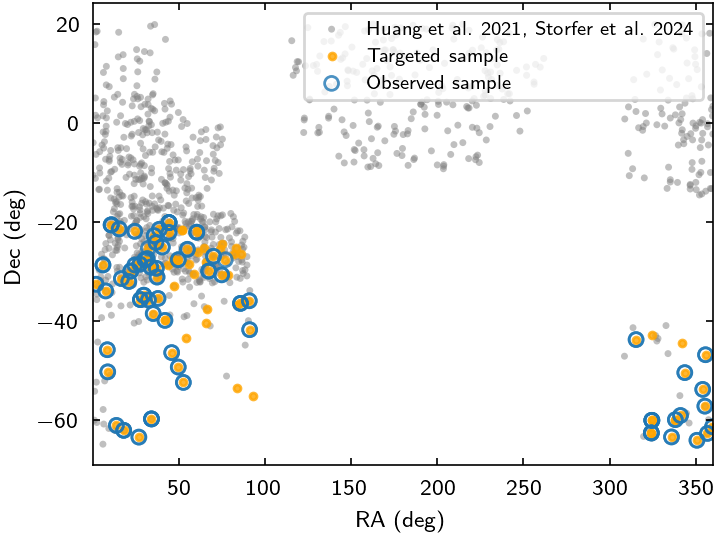}
\caption{Sky position of targeted and observed lens systems. In pink, the parent sample combining \cite{huang2021} and \cite{storfer2024} lens systems. In orange, the 96 grade A targeted system with declination ${\rm Dec} < -20^\circ$. Open blue circles represent the 67 observed systems.}
\label{fig:sky_positions}
\end{figure}

In \cite{huang2021}, the authors uncovered 1312 lens systems out of all central galaxies with $z$-band AB magnitude  $\sim 22.5$ in a region of 14,000 deg$^2$. In 2022, 1512 new candidate systems were found in the Data Release 9 of the Legacy Survey with central galaxies with $z_{\rm AB} < 20$ \citep{storfer2024}, benefiting from an extension of the coverage to 19,000 deg$^2$ \footnote{The list of candidates is available on NeuraLens at \url{https://sites.google.com/usfca.edu/neuralens}}.  Based on these detections, we proposed a XShooter filler program for 96 observations of the grade A systems located at ${\rm Dec} < -20^\circ$. This program ran over several semesters (see Table \ref{tab:observations}). Out of these observations, 67 systems were of sufficient quality to allow redshift estimates. The distribution of systems on the sky is shown in Fig~\ref{fig:sky_positions}.

For every system, we position the slit so that it covers both the lens and the source. This makes the data reduction more difficult because we cannot use the automatic data reduction routines, but spare some observing time. In some cases, the lens is the central galaxy of the lens system and is brighter than $z_{AB} < 20$, otherwise, it is a fainter galaxy close to the source. Still, the sources and the lenses appear as non-blended objects in the DECam ground-based images with seeing $\sim 1$ arcsec. Given that most of the expected emission lines ($[OII]$, $[OIII]$$H_\alpha$) for the sources are expected in the NIR, we requested only a 80\% fraction of lunar illumination (FLI). Since we are not interested in absolute flux calibration, we agreed to observe with thin clouds, an airmass up to 1.8 and a seeing $< 2''$.

The main goal being the detection of $[OII]$, $[OIII]$ and H$\alpha$ emissions, we took 2 dithers in SLIT, NODDING mode with 3 arcsec nodding along the slit, and orienting the slit to incorporate most of the source flux. We adopt a setup with a 11\arcsec x 1.6\arcsec slit size. In order to minimize overheads, and UVB and VIS arms not being the main focus, but having the largest readout time, we took exposures of 1400s for both the UVB and VIS arms, with a 2x1 slow read mode. For the NIR arm, we split this time into 5x300s sub-exposures. The total exposure time is 1h per Observation Block, including telescope pointing, offsets, and read out overheads. A summary of the observations is reported in Table~\ref{tab:observations}. Over the course of 4 semesters, we obtained 79 observations, out of which 12 have been re-observed, resulting in 67 observations fulfilling our requirements (night logs grade A).

\begin{table}
\caption{Summary of the XShooter spectroscopic observations}
\begin{center}
\begin{tabular}{|c|c|c|c|}
\hline
Program ID & \# OBs & Sky & Semester \\
 &  & Quality &  \\
\hline
113.26QD.001 & 28  &  15 grade A & Jun-Sep 2024 \\
112.260T.001 & 9  & 8 grade A & Oct 2023-Mar 2024 \\
111.250E.001 & 41  & 39 grade A & Jun-Sep 2023 \\
110.23U2.001 & 1  & 1 grade B  & Dec 2022 \\
\hline
\end{tabular}
\end{center}
\label{tab:observations}
\end{table}

Observations were reduced with the standard pipeline \textit{esoreflex} version 2.11.5, \textit{CPL} version 7.3.2 and XSchooter pipeline version 3.6.3. We adopted the SLIT, STARE data reduction mode for each of the two scientific raw frames, and wrote our own Python scripts to isolate the lens and source fluxes on 2D spectrograms and collapse them into 1D spectra. We used the two exposures to detect and confirm spectral features with more confidence, and used DS9 SAO \citep{ds9sao2000} to check emission lines in 2D spectrograms of the UVB, VIS and NIR arms successively. 
We used Marz web interface \citep{hinton2016} to display 1D spectra and estimate redshifts manually.
We graded robust redshift estimates with deep absorption lines and marked continuum with $Q_{z}=3$. Redshift estimates with emission lines are graded with $Q_{z} = 4$. In contrast, very uncertain continuum detection or detections with only one emission line and uncertain continuum are graded as $Q_{z} =1$ and $Q_{z}= 2$ respectively.

\section{Results}
\label{sec:results}

\begin{figure}
\begin{center}
\includegraphics{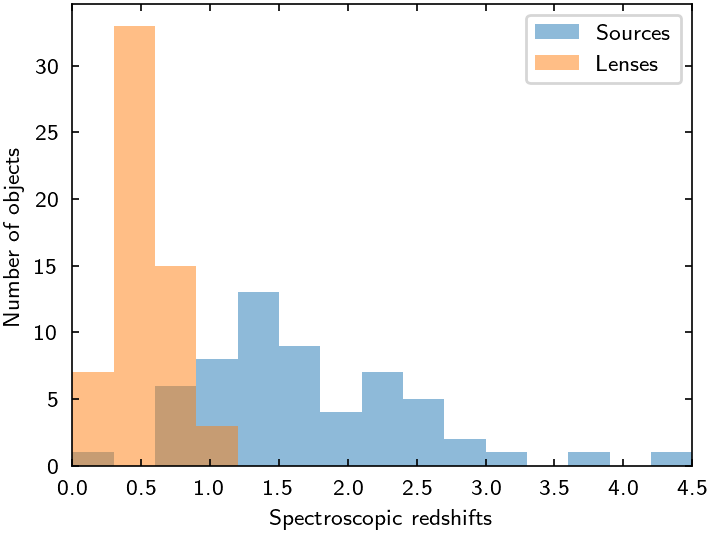}
\caption{Spectroscopic redshift distributions measured for the observed sources and lenses.}
\label{fig:redshifts}
\end{center}
\end{figure}

We managed to measure redshifts for 58 lenses and 57 sources, i.e. an efficiency of 82\% for the lenses and 85\% for the sources. 45 systems have both a lens and a source redshift. We report the redshift measurements in Table~\ref{tab:confirmed}. Fig~\ref{fig:redshifts} shows that most of the lenses fall in the redshift range $z < 1$. Their spectra usually present few emission lines, and redshifts can be derived from the Balmer break, calcium and sodium absorption lines. For some spectra with low signal to noise, data reduction imperfection confused the identification, in particular those related to the spectral orders tracing, and sky line subtraction residuals. The highest lens redshift is for system DESI-030.4360-27.6618 with $z = 1.02$ and quality flag $Q_{z}=2$. Most of the lens redshifts (35\%) are ranked with $Q_{z}=3$. Only 12\% of the spectra present emission lines and are graded as $Q_{z}=4$. 33\% of the lens redshifts are ranked with $Q_z < 2$. Sources cover a broad range in redshift $0.28 < z < 4.34$, and the median redshift is $z_{s} = 1.54$. Most of the spectra present emission lines, and as a result 70\% of the redshifts have a quality flag $Q_{z} = 4$. 

\begin{figure}
\begin{center}
\includegraphics{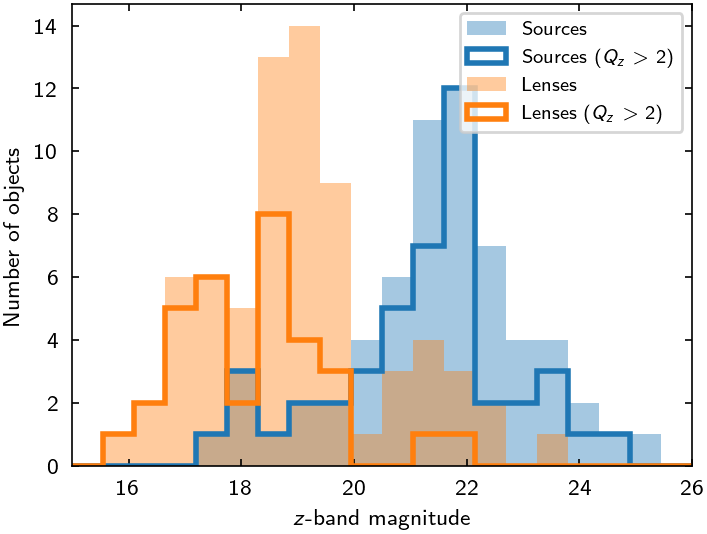}
\caption{Magnitude distribution of lenses and sources observed and for which a redshift could be measured with quality flag $Q_{z} > 2$.}
\label{fig:zband_mag}
\end{center}
\end{figure}

Fig~\ref{fig:zband_mag} shows that we only managed to robustly measure redshifts with quality flag $Q_{z} > 2$ for lenses with magnitude $z_\mathrm{AB} < 20$. The success rate for lenses with magnitudes $18 < z_\mathrm{AB} < 20$ is 40\%, whereas it is of 93\% for brighter lenses. Faint lenses with $z_\mathrm{AB} > 20$ are mostly satellite galaxies in small groups at redshift $z \sim 0.6$. In contrast to lenses, we manage to measure redshifts for sources of any magnitude, down to $z_\mathrm{AB} = 25.2$ with $Q_{z} > 2$. 

\begin{figure}
\begin{center}
\includegraphics{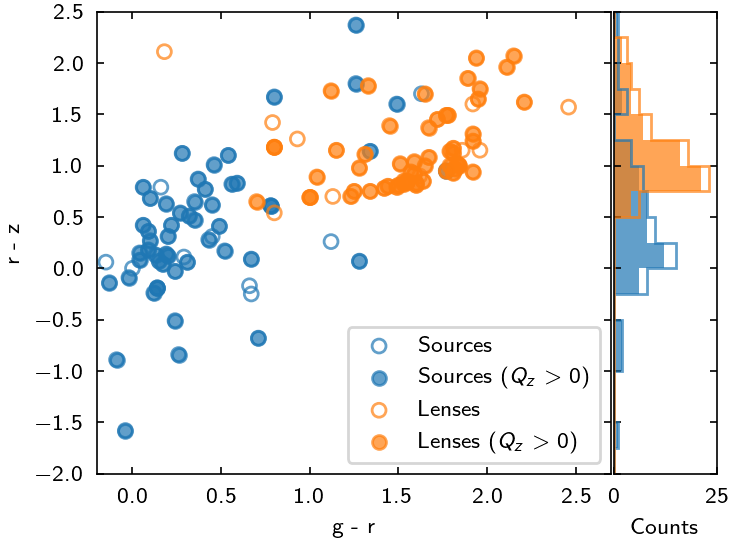}
\caption{Color color distributions and histograms of the lenses and sources observed and for which a redshift could be measured. We observe no particular bias between the two samples.}
\label{fig:cc_hist}
\end{center}
\end{figure}

In Fig.~\ref{fig:cc_hist}, we compare the $g-r$ and $r-z$ color distribution of lenses and sources. We do not identify any bias associated to the redshift measurement operation. Given that \cite{huang2021} did not perform any selection on the sources, this suggests that our redshift distribution of the sources is likely representative of the true one.

\section{Conclusion}
\label{sec:conclusion}

With this XShooter observation program, we observed 67 lens systems, out of which we managed to measure redshifts for 58 lenses and 57 sources. 45 systems have both a lens and a source redshift. Most of the lens redshifts were obtained from the continuum, and the source redshifts from emission lines.
We found no particular bias associated to the redshift measurement operation, neither for the lenses nor for the sources, meaning that our source redshift distribution is likely representative of the true redshift distribution of the sources. Therefore, it can be used in subsequent forward model analyses, or as an updated training set for other surveys such as Euclid or LSST.

\begin{acknowledgements}

The Legacy Surveys consist of three individual and complementary projects: the Dark Energy Camera Legacy Survey (DECaLS; Proposal ID \#2014B-0404; PIs: David Schlegel and Arjun Dey), the Beijing-Arizona Sky Survey (BASS; NOAO Prop. ID \#2015A-0801; PIs: Zhou Xu and Xiaohui Fan), and the Mayall z-band Legacy Survey (MzLS; Prop. ID \#2016A-0453; PI: Arjun Dey). DECaLS, BASS and MzLS together include data obtained, respectively, at the Blanco telescope, Cerro Tololo Inter-American Observatory, NSF’s NOIRLab; the Bok telescope, Steward Observatory, University of Arizona; and the Mayall telescope, Kitt Peak National Observatory, NOIRLab. Pipeline processing and analyses of the data were supported by NOIRLab and the Lawrence Berkeley National Laboratory (LBNL). The Legacy Surveys project is honored to be permitted to conduct astronomical research on Iolkam Du’ag (Kitt Peak), a mountain with particular significance to the Tohono O’odham Nation.

NOIRLab is operated by the Association of Universities for Research in Astronomy (AURA) under a cooperative agreement with the National Science Foundation. LBNL is managed by the Regents of the University of California under contract to the U.S. Department of Energy.

This project used data obtained with the Dark Energy Camera (DECam), which was constructed by the Dark Energy Survey (DES) collaboration. Funding for the DES Projects has been provided by the U.S. Department of Energy, the U.S. National Science Foundation, the Ministry of Science and Education of Spain, the Science and Technology Facilities Council of the United Kingdom, the Higher Education Funding Council for England, the National Center for Supercomputing Applications at the University of Illinois at Urbana-Champaign, the Kavli Institute of Cosmological Physics at the University of Chicago, Center for Cosmology and Astro-Particle Physics at the Ohio State University, the Mitchell Institute for Fundamental Physics and Astronomy at Texas A\&M University, Financiadora de Estudos e Projetos, Fundacao Carlos Chagas Filho de Amparo, Financiadora de Estudos e Projetos, Fundacao Carlos Chagas Filho de Amparo a Pesquisa do Estado do Rio de Janeiro, Conselho Nacional de Desenvolvimento Cientifico e Tecnologico and the Ministerio da Ciencia, Tecnologia e Inovacao, the Deutsche Forschungsgemeinschaft and the Collaborating Institutions in the Dark Energy Survey. The Collaborating Institutions are Argonne National Laboratory, the University of California at Santa Cruz, the University of Cambridge, Centro de Investigaciones Energeticas, Medioambientales y Tecnologicas-Madrid, the University of Chicago, University College London, the DES-Brazil Consortium, the University of Edinburgh, the Eidgenossische Technische Hochschule (ETH) Zurich, Fermi National Accelerator Laboratory, the University of Illinois at Urbana-Champaign, the Institut de Ciencies de l’Espai (IEEC/CSIC), the Institut de Fisica d’Altes Energies, Lawrence Berkeley National Laboratory, the Ludwig Maximilians Universitat Munchen and the associated Excellence Cluster Universe, the University of Michigan, NSF’s NOIRLab, the University of Nottingham, the Ohio State University, the University of Pennsylvania, the University of Portsmouth, SLAC National Accelerator Laboratory, Stanford University, the University of Sussex, and Texas A\&M University.

BASS is a key project of the Telescope Access Program (TAP), which has been funded by the National Astronomical Observatories of China, the Chinese Academy of Sciences (the Strategic Priority Research Program “The Emergence of Cosmological Structures” Grant \# XDB09000000), and the Special Fund for Astronomy from the Ministry of Finance. The BASS is also supported by the External Cooperation Program of Chinese Academy of Sciences (Grant \# 114A11KYSB20160057), and Chinese National Natural Science Foundation (Grant \# 12120101003, \# 11433005).

The Legacy Survey team makes use of data products from the Near-Earth Object Wide-field Infrared Survey Explorer (NEOWISE), which is a project of the Jet Propulsion Laboratory/California Institute of Technology. NEOWISE is funded by the National Aeronautics and Space Administration.

The Legacy Surveys imaging of the DESI footprint is supported by the Director, Office of Science, Office of High Energy Physics of the U.S. Department of Energy under Contract No. DE-AC02-05CH1123, by the National Energy Research Scientific Computing Center, a DOE Office of Science User Facility under the same contract; and by the U.S. National Science Foundation, Division of Astronomical Sciences under Contract No. AST-0950945 to NOAO.

\end{acknowledgements}

\bibliographystyle{aa}
\bibliography{all}

\begin{thebibliography}{54}
\expandafter\ifx\csname natexlab\endcsname\relax\def\natexlab#1{#1}\fi

\bibitem[{{Abell} {et~al.}(1989){Abell}, {Corwin}, \& {Olowin}}]{abell1989}
{Abell}, G.~O., {Corwin}, Jr., H.~G., \& {Olowin}, R.~P. 1989, \apjs, 70, 1

\bibitem[{{Adnan} {et~al.}(2025){Adnan}, {Hasan}, {Al-Imtiaz}, {Robin},
  {Shwadhin}, {Shajib}, {Nahid}, {Tanver}, {Akter}, {Jahan}, {Jafar}, {Rashid},
  {Biswas}, {Chowdhury}, {Feardous}, {Rahaman}, {Ridwan}, {Sharma},
  {Chowdhury}, \& {Hossain}}]{adnan2025}
{Adnan}, S.~M.~R., {Hasan}, M.~J., {Al-Imtiaz}, A., {et~al.} 2025, \aap, 699,
  A259

\bibitem[{{Agarwal} {et~al.}(2026){Agarwal}, {Huang}, {Sheu}, {Storfer},
  {Tamargo-Arizmendi}, {Tabares-Tarquinio}, {Schlegel}, {Aldering}, {Bolton},
  {Cikota}, {Dey}, {Filipp}, {Jullo}, {Kwon}, {Perlmutter}, {Shu}, {Sukay},
  {Suzuki}, {Aguilar}, {Ahlen}, {BenZvi}, {Brooks}, {Claybaugh}, {Doel},
  {Forero-Romero}, {Gazta{\~n}aga}, {Gontcho A Gontcho}, {Gutierrez},
  {Honscheid}, {Ishak}, {Juneau}, {Kehoe}, {Kisner}, {Koposov}, {Lambert},
  {Landriau}, {Le Guillou}, {de la Macorra}, {Meisner}, {Miquel}, {Moustakas},
  {Myers}, {Poppett}, {Prada}, {P{\'e}rez-R{\`a}fols}, {Rossi}, {Sanchez},
  {Schubnell}, {Sprayberry}, {Tarl{\'e}}, {Weaver}, \& {Zou}}]{agarwal2026}
{Agarwal}, S., {Huang}, X., {Sheu}, W., {et~al.} 2026, \apj, 1001, 162

\bibitem[{{Aihara} {et~al.}(2018){Aihara}, {Arimoto}, {Armstrong}, {Arnouts},
  {Bahcall}, {Bickerton}, {Bosch}, {Bundy}, {Capak}, {Chan}, {Chiba}, {Coupon},
  {Egami}, {Enoki}, {Finet}, {Fujimori}, {Fujimoto}, {Furusawa}, {Furusawa},
  {Goto}, {Goulding}, {Greco}, {Greene}, {Gunn}, {Hamana}, {Harikane},
  {Hashimoto}, {Hattori}, {Hayashi}, {Hayashi}, {He{\l}miniak}, {Higuchi},
  {Hikage}, {Ho}, {Hsieh}, {Huang}, {Huang}, {Ikeda}, {Imanishi}, {Inoue},
  {Iwasawa}, {Iwata}, {Jaelani}, {Jian}, {Kamata}, {Karoji}, {Kashikawa},
  {Katayama}, {Kawanomoto}, {Kayo}, {Koda}, {Koike}, {Kojima}, {Komiyama},
  {Konno}, {Koshida}, {Koyama}, {Kusakabe}, {Leauthaud}, {Lee}, {Lin}, {Lin},
  {Lupton}, {Mandelbaum}, {Matsuoka}, {Medezinski}, {Mineo}, {Miyama},
  {Miyatake}, {Miyazaki}, {Momose}, {More}, {More}, {Moritani}, {Moriya},
  {Morokuma}, {Mukae}, {Murata}, {Murayama}, {Nagao}, {Nakata}, {Niida},
  {Niikura}, {Nishizawa}, {Obuchi}, {Oguri}, {Oishi}, {Okabe}, {Okamoto},
  {Okura}, {Ono}, {Onodera}, {Onoue}, {Osato}, {Ouchi}, {Price}, {Pyo}, {Sako},
  {Sawicki}, {Shibuya}, {Shimasaku}, {Shimono}, {Shirasaki}, {Silverman},
  {Simet}, {Speagle}, {Spergel}, {Strauss}, {Sugahara}, {Sugiyama}, {Suto},
  {Suyu}, {Suzuki}, {Tait}, {Takada}, {Takata}, {Tamura}, {Tanaka}, {Tanaka},
  {Tanaka}, {Tanaka}, {Terai}, {Terashima}, {Toba}, {Tominaga}, {Toshikawa},
  {Turner}, {Uchida}, {Uchiyama}, {Umetsu}, {Uraguchi}, {Urata}, {Usuda},
  {Utsumi}, {Wang}, {Wang}, {Wong}, {Yabe}, {Yamada}, {Yamanoi}, {Yasuda},
  {Yeh}, {Yonehara}, \& {Yuma}}]{aihara2018}
{Aihara}, H., {Arimoto}, N., {Armstrong}, R., {et~al.} 2018, \pasj, 70, S4

\bibitem[{{Bacon} {et~al.}(2010){Bacon}, {Accardo}, {Adjali}, {Anwand},
  {Bauer}, {Biswas}, {Blaizot}, {Boudon}, {Brau-Nogue}, {Brinchmann},
  {Caillier}, {Capoani}, {Carollo}, {Contini}, {Couderc}, {Daguis{\'e}},
  {Deiries}, {Delabre}, {Dreizler}, {Dubois}, {Dupieux}, {Dupuy}, {Emsellem},
  {Fechner}, {Fleischmann}, {Fran{\c{c}}ois}, {Gallou}, {Gharsa}, {Glindemann},
  {Gojak}, {Guiderdoni}, {Hansali}, {Hahn}, {Jarno}, {Kelz}, {Koehler},
  {Kosmalski}, {Laurent}, {Le Floch}, {Lilly}, {Lizon}, {Loupias}, {Manescau},
  {Monstein}, {Nicklas}, {Olaya}, {Pares}, {Pasquini}, {P{\'e}contal-Rousset},
  {Pell{\'o}}, {Petit}, {Popow}, {Reiss}, {Remillieux}, {Renault}, {Roth},
  {Rupprecht}, {Serre}, {Schaye}, {Soucail}, {Steinmetz}, {Streicher}, {Stuik},
  {Valentin}, {Vernet}, {Weilbacher}, {Wisotzki}, \& {Yerle}}]{bacon2010}
{Bacon}, R., {Accardo}, M., {Adjali}, L., {et~al.} 2010, in Society of
  Photo-Optical Instrumentation Engineers (SPIE) Conference Series, Vol. 7735,
  Ground-based and Airborne Instrumentation for Astronomy III, ed. I.~S.
  {McLean}, S.~K. {Ramsay}, \& H.~{Takami}, 773508

\bibitem[{{Bleem} {et~al.}(2020){Bleem}, {Bocquet}, {Stalder}, {Gladders},
  {Ade}, {Allen}, {Anderson}, {Annis}, {Ashby}, {Austermann}, {Avila}, {Avva},
  {Bayliss}, {Beall}, {Bechtol}, {Bender}, {Benson}, {Bertin}, {Bianchini},
  {Blake}, {Brodwin}, {Brooks}, {Buckley-Geer}, {Burke}, {Carlstrom}, {Rosell},
  {Carrasco Kind}, {Carretero}, {Chang}, {Chiang}, {Citron}, {Moran},
  {Costanzi}, {Crawford}, {Crites}, {da Costa}, {de Haan}, {De Vicente},
  {Desai}, {Diehl}, {Dietrich}, {Dobbs}, {Eifler}, {Everett}, {Flaugher},
  {Floyd}, {Frieman}, {Gallicchio}, {Garc{\'\i}a-Bellido}, {George}, {Gerdes},
  {Gilbert}, {Gruen}, {Gruendl}, {Gschwend}, {Gupta}, {Gutierrez}, {Halverson},
  {Harrington}, {Henning}, {Heymans}, {Holder}, {Hollowood}, {Holzapfel},
  {Honscheid}, {Hrubes}, {Huang}, {Hubmayr}, {Irwin}, {James}, {Jeltema},
  {Joudaki}, {Khullar}, {Klein}, {Knox}, {Kuropatkin}, {Lee}, {Li}, {Lidman},
  {Lowitz}, {MacCrann}, {Mahler}, {Maia}, {Marshall}, {McDonald}, {McMahon},
  {Melchior}, {Menanteau}, {Meyer}, {Miquel}, {Mocanu}, {Mohr}, {Montgomery},
  {Nadolski}, {Natoli}, {Nibarger}, {Noble}, {Novosad}, {Padin}, {Palmese},
  {Parkinson}, {Patil}, {Paz-Chinch{\'o}n}, {Plazas}, {Pryke}, {Ramachandra},
  {Reichardt}, {Remolina Gonz{\'a}lez}, {Romer}, {Roodman}, {Ruhl}, {Rykoff},
  {Saliwanchik}, {Sanchez}, {Saro}, {Sayre}, {Schaffer}, {Schrabback},
  {Serrano}, {Sharon}, {Sievers}, {Smecher}, {Smith}, {Soares-Santos}, {Stark},
  {Story}, {Suchyta}, {Tarle}, {Tucker}, {Vanderlinde}, {Veach}, {Vieira},
  {Wang}, {Weller}, {Whitehorn}, {Wu}, {Yefremenko}, \& {Zhang}}]{bleem2020}
{Bleem}, L.~E., {Bocquet}, S., {Stalder}, B., {et~al.} 2020, \apjs, 247, 25

\bibitem[{{Bleem} {et~al.}(2015{\natexlab{a}}){Bleem}, {Stalder}, {Brodwin},
  {Busha}, {Gladders}, {High}, {Rest}, \& {Wechsler}}]{bleem2015b}
{Bleem}, L.~E., {Stalder}, B., {Brodwin}, M., {et~al.} 2015{\natexlab{a}},
  \apjs, 216, 20

\bibitem[{{Bleem} {et~al.}(2015{\natexlab{b}}){Bleem}, {Stalder}, {de Haan},
  {Aird}, {Allen}, {Applegate}, {Ashby}, {Bautz}, {Bayliss}, {Benson},
  {Bocquet}, {Brodwin}, {Carlstrom}, {Chang}, {Chiu}, {Cho}, {Clocchiatti},
  {Crawford}, {Crites}, {Desai}, {Dietrich}, {Dobbs}, {Foley}, {Forman},
  {George}, {Gladders}, {Gonzalez}, {Halverson}, {Hennig}, {Hoekstra},
  {Holder}, {Holzapfel}, {Hrubes}, {Jones}, {Keisler}, {Knox}, {Lee}, {Leitch},
  {Liu}, {Lueker}, {Luong-Van}, {Mantz}, {Marrone}, {McDonald}, {McMahon},
  {Meyer}, {Mocanu}, {Mohr}, {Murray}, {Padin}, {Pryke}, {Reichardt}, {Rest},
  {Ruel}, {Ruhl}, {Saliwanchik}, {Saro}, {Sayre}, {Schaffer}, {Schrabback},
  {Shirokoff}, {Song}, {Spieler}, {Stanford}, {Staniszewski}, {Stark}, {Story},
  {Stubbs}, {Vanderlinde}, {Vieira}, {Vikhlinin}, {Williamson}, {Zahn}, \&
  {Zenteno}}]{bleem2015}
{Bleem}, L.~E., {Stalder}, B., {de Haan}, T., {et~al.} 2015{\natexlab{b}},
  \apjs, 216, 27

\bibitem[{{CHEX-MATE Collaboration} {et~al.}(2021){CHEX-MATE Collaboration},
  {Arnaud}, {Ettori}, {Pratt}, {Rossetti}, {Eckert}, {Gastaldello}, {Gavazzi},
  {Kay}, {Lovisari}, {Maughan}, {Pointecouteau}, {Sereno}, {Bartalucci},
  {Bonafede}, {Bourdin}, {Cassano}, {Duffy}, {Iqbal}, {Maurogordato}, {Rasia},
  {Sayers}, {Andrade-Santos}, {Aussel}, {Barnes}, {Barrena}, {Borgani},
  {Burkutean}, {Clerc}, {Corasaniti}, {Cuillandre}, {De Grandi}, {De Petris},
  {Dolag}, {Donahue}, {Ferragamo}, {Gaspari}, {Ghizzardi}, {Gitti}, {Haines},
  {Jauzac}, {Johnston-Hollitt}, {Jones}, {K{\'e}ruzor{\'e}}, {Le Brun},
  {Mayet}, {Mazzotta}, {Melin}, {Molendi}, {Nonino}, {Okabe}, {Paltani},
  {Perotto}, {Pires}, {Radovich}, {Rubino-Martin}, {Salvati}, {Saro},
  {Sartoris}, {Schellenberger}, {Streblyanska}, {Tarr{\'\i}o}, {Tozzi},
  {Umetsu}, {van der Burg}, {Vazza}, {Venturi}, {Yepes}, \&
  {Zarattini}}]{chexmate2021}
{CHEX-MATE Collaboration}, {Arnaud}, M., {Ettori}, S., {et~al.} 2021, \aap,
  650, A104

\bibitem[{{Coe} {et~al.}(2019){Coe}, {Salmon}, {Brada{\v{c}}}, {Bradley},
  {Sharon}, {Zitrin}, {Acebron}, {Cerny}, {Cibirka}, {Strait},
  {Paterno-Mahler}, {Mahler}, {Avila}, {Ogaz}, {Huang}, {Pelliccia}, {Stark},
  {Mainali}, {Oesch}, {Trenti}, {Carrasco}, {Dawson}, {Rodney}, {Strolger},
  {Riess}, {Jones}, {Frye}, {Czakon}, {Umetsu}, {Vulcani}, {Graur}, {Jha},
  {Graham}, {Molino}, {Nonino}, {Hjorth}, {Selsing}, {Christensen},
  {Kikuchihara}, {Ouchi}, {Oguri}, {Welch}, {Lemaux}, {Andrade-Santos}, {Hoag},
  {Johnson}, {Peterson}, {Past}, {Fox}, {Agulli}, {Livermore}, {Ryan}, {Lam},
  {Sendra-Server}, {Toft}, {Lovisari}, \& {Su}}]{coe2019}
{Coe}, D., {Salmon}, B., {Brada{\v{c}}}, M., {et~al.} 2019, \apj, 884, 85

\bibitem[{Collaboration {et~al.}(2024)Collaboration, Adame, Aguilar, Ahlen,
  Alam, Aldering, Alexander, Alfarsy, Allende~Prieto, Alvarez, Alves, Anand,
  Andrade-Oliveira, Armengaud, Asorey, Avila, Aviles, Bailey,
  Balaguera-Antol{\'\i}nez, Ballester, Baltay, Bault, Bautista, Behera,
  Beltran, BenZvi, Beraldo~e Silva, Bermejo-Climent, Berti, Besuner, Beutler,
  Bianchi, Blake, Blum, Bolton, Brieden, Brodzeller, Brooks, Brown,
  Buckley-Geer, Burtin, Cabayol-Garcia, Cai, Canning, Cardiel-Sas,
  Carnero~Rosell, Castander, Cervantes-Cota, Chabanier, Chaussidon,
  Chaves-Montero, Chen, Chen, Chuang, Claybaugh, Cole, Cooper, Cuceu, Davis,
  Dawson, de~Belsunce, de~la Cruz, de~la Macorra, Della~Costa, de~Mattia,
  Demina, Demirbozan, DeRose, Dey, Dey, Dhungana, Ding, Ding, Doel, Doshi,
  Douglass, Edge, Eftekharzadeh, Eisenstein, Elliott, Ereza, Escoffier,
  Fagrelius, Fan, Fanning, Fawcett, Ferraro, Flaugher, Font-Ribera,
  Forero-Romero, Forero-S{\'a}nchez, Frenk, G{\"a}nsicke, Garc{\'\i}a,
  Garc{\'\i}a-Bellido, Garcia-Quintero, Garrison, Gil-Mar{\'\i}n, Golden-Marx,
  Gontcho A~Gontcho, Gonzalez-Morales, Gonzalez-Perez, Gordon, Graur, Green,
  Gruen, Guy, Hadzhiyska, Hahn, Han, Hanif, Herrera-Alcantar, Honscheid, Hou,
  Howlett, Huterer, Ir{\v s}i{\v c}, Ishak, Jacques, Jana, Jiang, Jimenez,
  Jing, Joudaki, Joyce, Jullo, Juneau, Kara{\c c}aylı, Karim, Kehoe, Kent,
  Khederlarian, Kim, Kirkby, Kisner, Kitaura, Kizhuprakkat, Kneib, Koposov,
  Kov{\'a}cs, Kremin, Krolewski, L'Huillier, Lahav, Lambert, Lamman, Lan,
  Landriau, Lang, Lange, Lasker, Leauthaud, Le~Guillou, Levi, Li, Linder,
  Lyons, Magneville, Manera, Manser, Margala, Martini, McDonald, Medina,
  Medina-Varela, Meisner, Mena-Fern{\'a}ndez, Meneses-Rizo, Mezcua, Miquel,
  Montero-Camacho, Moon, Moore, Moustakas, Mueller, Mundet,
  Mu{\~n}oz-Guti{\'e}rrez, Myers, Nadathur, Napolitano, Neveux, Newman, Nie,
  Nikutta, Niz, Norberg, Noriega, Paillas, Palanque-Delabrouille, Palmese, Pan,
  Parkinson, Penmetsa, Percival, P{\'e}rez-Fern{\'a}ndez, P{\'e}rez-R{\`a}fols,
  Pieri, Poppett, Porredon, Pothier, Prada, Pucha, Raichoor,
  Ram{\'\i}rez-P{\'e}rez, Ramirez-Solano, Rashkovetskyi, Ravoux, Rocher,
  Rockosi, Ross, Rossi, Ruggeri, Ruhlmann-Kleider, Sabiu, Said, Saintonge,
  Samushia, Sanchez, Saulder, Schaan, Schlafly, Schlegel, Scholte, Schubnell,
  Seo, Shafieloo, Sharples, Sheu, Silber, Sinigaglia, Siudek, Slepian, Smith,
  Soumagnac, Sprayberry, Stephey, Su{\'a}rez-P{\'e}rez, Sun, Tan, Tarl{\'e},
  Tojeiro, Ure{\~n}a-L{\'o}pez, Vaisakh, Valcin, Valdes, Valluri,
  Vargas-Maga{\~n}a, Variu, Verde, Walther, Wang, Wang, Weaver, Weaverdyck,
  Wechsler, White, Xie, Yang, Y{\`e}che, Yu, Yuan, Zhang, Zhang, Zhao, Zheng,
  Zhou, Zhou, Zou, Zou, \& Zu}]{adame2024}
Collaboration, D., Adame, A.~G., Aguilar, J., {et~al.} 2024, The Astronomical
  Journal, 168, 58

\bibitem[{{Despali} {et~al.}(2025){Despali}, {Heinze}, {Fassnacht}, {Vegetti},
  {Spingola}, {Klessen}, \& {Tajalli}}]{despali2025}
{Despali}, G., {Heinze}, F.~M., {Fassnacht}, C.~D., {et~al.} 2025, \aap, 699,
  A222

\bibitem[{{D{\'\i}az-S{\'a}nchez} {et~al.}(2021){D{\'\i}az-S{\'a}nchez},
  {Dannerbauer}, {Sulzenauer}, {Iglesias-Groth}, \& {Rebolo}}]{diaz2021}
{D{\'\i}az-S{\'a}nchez}, A., {Dannerbauer}, H., {Sulzenauer}, N.,
  {Iglesias-Groth}, S., \& {Rebolo}, R. 2021, \apj, 919, 48

\bibitem[{{Diehl} {et~al.}(2017){Diehl}, {Buckley-Geer}, {Lindgren}, {Nord},
  {Gaitsch}, {Gaitsch}, {Lin}, {Allam}, {Collett}, {Furlanetto}, {Gill},
  {More}, {Nightingale}, {Odden}, {Pellico}, {Tucker}, {da Costa}, {Fausti
  Neto}, {Kuropatkin}, {Soares-Santos}, {Welch}, {Zhang}, {Frieman}, {Abdalla},
  {Annis}, {Benoit-L{\'e}vy}, {Bertin}, {Brooks}, {Burke}, {Carnero Rosell},
  {Carrasco Kind}, {Carretero}, {Cunha}, {D'Andrea}, {Desai}, {Dietrich},
  {Drlica-Wagner}, {Evrard}, {Finley}, {Flaugher}, {Garc{\'\i}a-Bellido},
  {Gerdes}, {Goldstein}, {Gruen}, {Gruendl}, {Gschwend}, {Gutierrez}, {James},
  {Kuehn}, {Kuhlmann}, {Lahav}, {Li}, {Lima}, {Maia}, {Marshall}, {Menanteau},
  {Miquel}, {Nichol}, {Nugent}, {Ogando}, {Plazas}, {Reil}, {Romer}, {Sako},
  {Sanchez}, {Santiago}, {Scarpine}, {Schindler}, {Schubnell},
  {Sevilla-Noarbe}, {Sheldon}, {Smith}, {Sobreira}, {Suchyta}, {Swanson},
  {Tarle}, {Thomas}, {Walker}, \& {DES Collaboration}}]{diehl2017}
{Diehl}, H.~T., {Buckley-Geer}, E.~J., {Lindgren}, K.~A., {et~al.} 2017, \apjs,
  232, 15

\bibitem[{{Ebeling} {et~al.}(2001){Ebeling}, {Edge}, \& {Henry}}]{ebeling2001}
{Ebeling}, H., {Edge}, A.~C., \& {Henry}, J.~P. 2001, \apj, 553, 668

\bibitem[{{Euclid Collaboration} {et~al.}(2025{\natexlab{a}}){Euclid
  Collaboration}, {Lines}, {Collett}, {Holloway}, {Rojas}, {Schuldt},
  {Metcalf}, {Li}, {Verma}, {Despali}, {Courbin}, {Gavazzi}, {Tortora},
  {Cl{\'e}ment}, {Aghanim}, {Altieri}, {Amendola}, {Andreon}, {Auricchio},
  {Baccigalupi}, {Baldi}, {Balestra}, {Bardelli}, {Battaglia}, {Biviano},
  {Branchini}, {Brescia}, {Camera}, {Ca{\~n}as-Herrera}, {Capobianco},
  {Carbone}, {Carretero}, {Castellano}, {Castignani}, {Cavuoti}, {Cimatti},
  {Colodro-Conde}, {Congedo}, {Conselice}, {Conversi}, {Copin}, {Courtois},
  {Cropper}, {Degaudenzi}, {De Lucia}, {Dole}, {Dubath}, {Dupac}, {Dusini},
  {Ealet}, {Escoffier}, {Farina}, {Farinelli}, {Faustini}, {Ferriol},
  {Finelli}, {Frailis}, {Franceschi}, {Fumana}, {Galeotta}, {George}, {Gillis},
  {Giocoli}, {G{\'o}mez-Alvarez}, {Gracia-Carpio}, {Grazian}, {Grupp},
  {Haugan}, {Holmes}, {Hook}, {Hormuth}, {Hornstrup}, {Jahnke}, {Jhabvala},
  {Joachimi}, {Keih{\"a}nen}, {Kermiche}, {Kiessling}, {Kubik}, {K{\"u}mmel},
  {Kunz}, {Kurki-Suonio}, {Le Brun}, {Ligori}, {Lilje}, {Lindholm}, {Lloro},
  {Mainetti}, {Maino}, {Maiorano}, {Mansutti}, {Marcin}, {Marggraf},
  {Martinelli}, {Martinet}, {Marulli}, {Massey}, {Medinaceli}, {Mei},
  {Melchior}, {Mellier}, {Meneghetti}, {Merlin}, {Meylan}, {Mora}, {Moresco},
  {Moscardini}, {Nakajima}, {Neissner}, {Niemi}, {Nightingale}, {Padilla},
  {Paltani}, {Pasian}, {Pedersen}, {Percival}, {Pettorino}, {Pires}, {Polenta},
  {Poncet}, {Popa}, {Pozzetti}, {Raison}, {Renzi}, {Rhodes}, {Riccio},
  {Romelli}, {Roncarelli}, {Rosset}, {Saglia}, {Sakr}, {S{\'a}nchez}, {Sapone},
  {Sartoris}, {Schewtschenko}, {Schneider}, {Schrabback}, {Secroun}, {Seidel},
  {Serrano}, {Sirignano}, {Sirri}, {Stanco}, {Steinwagner},
  {Tallada-Cresp{\'\i}}, {Taylor}, {Tereno}, {Tessore}, {Toft}, {Toledo-Moreo},
  {Torradeflot}, {Tutusaus}, {Valiviita}, {Vassallo}, {Veropalumbo}, {Wang},
  {Weller}, {Zacchei}, {Zamorani}, {Zerbi}, {Zucca}, {Ballardini},
  {Bolzonella}, {Bozzo}, {Burigana}, {Cabanac}, {Calabrese}, {Cappi}, {Castro},
  {Escartin Vigo}, {Gabarra}, {Garc{\'\i}a-Bellido}, {Gautard}, {Hemmati},
  {Huertas-Company}, {Macias-Perez}, {Maoli}, {Mart{\'\i}n-Fleitas}, {Maturi},
  {Mauri}, {Monaco}, {P{\"o}ntinen}, {Porciani}, {Risso}, {Scottez}, {Sereno},
  {Tenti}, {Tucci}, {Viel}, {Wiesmann}, {Akrami}, {Andika}, {Angora},
  {Anselmi}, {Archidiacono}, {Atrio-Barandela}, {Aubourg}, {Bazzanini},
  {Bertacca}, \& {Bethermin}}]{lines2025b}
{Euclid Collaboration}, {Lines}, N.~E.~P., {Collett}, T.~E., {et~al.}
  2025{\natexlab{a}}, arXiv e-prints, arXiv:2512.05899

\bibitem[{{Euclid Collaboration} {et~al.}(2025{\natexlab{b}}){Euclid
  Collaboration}, {Lines}, {Collett}, {Walmsley}, {Rojas}, {Li}, {Leuzzi},
  {Manj{\'o}n-Garc{\'\i}a}, {Vincken}, {Wilde}, {Holloway}, {Verma}, {Metcalf},
  {Andika}, {Melo}, {Melchior}, {Dom{\'\i}nguez S{\'a}nchez},
  {D{\'\i}az-S{\'a}nchez}, {Acevedo Barroso}, {Cl{\'e}ment}, {Krawczyk},
  {Pearce-Casey}, {Serjeant}, {Courbin}, {Despali}, {Gavazzi}, {Schuldt},
  {Degaudenzi}, {Ecker}, {Enzi}, {Finner}, {Galan}, {Giocoli}, {Hogg},
  {Jahnke}, {Kruk}, {Mahler}, {More}, {Nagam}, {Pearson}, {Sainz de Murieta},
  {Scarlata}, {Sluse}, {Sonnenfeld}, {Spiniello}, {Thai}, {Tortora}, {Ulivi},
  {Weisenbach}, {Zumalacarregui}, {Aghanim}, {Altieri}, {Amara}, {Andreon},
  {Auricchio}, {Aussel}, {Baccigalupi}, {Baldi}, {Balestra}, {Bardelli},
  {Battaglia}, {Bender}, {Bernardeau}, {Biviano}, {Bonchi}, {Bonino},
  {Branchini}, {Brescia}, {Brinchmann}, {Camera}, {Ca{\~n}as-Herrera},
  {Capobianco}, {Carbone}, {Cardone}, {Carretero}, {Casas}, {Castellano},
  {Castignani}, {Cavuoti}, {Chambers}, {Cimatti}, {Colodro-Conde}, {Congedo},
  {Conselice}, {Conversi}, {Copin}, {Costille}, {Courtois}, {Cropper}, {Da
  Silva}, {De Lucia}, {Di Giorgio}, {Dolding}, {Dole}, {Dubath}, {Duncan},
  {Dupac}, {Escoffier}, {Fabricius}, {Farina}, {Farinelli}, {Faustini},
  {Ferriol}, {Finelli}, {Fotopoulou}, {Frailis}, {Franceschi}, {Fumana},
  {Galeotta}, {George}, {Gillard}, {Gillis}, {G{\'o}mez-Alvarez},
  {Gracia-Carpio}, {Granett}, {Grazian}, {Grupp}, {Guzzo}, {Gwyn}, {Haugan},
  {Holmes}, {Hook}, {Hormuth}, {Hornstrup}, {Hudelot}, {Jhabvala},
  {Keih{\"a}nen}, {Kermiche}, {Kiessling}, {Kubik}, {K{\"u}mmel}, {Kunz},
  {Kurki-Suonio}, {Le Boulc'h}, {Le Brun}, {Le Mignant}, {Ligori}, {Lilje},
  {Lindholm}, {Lloro}, {Mainetti}, {Maino}, {Maiorano}, {Mansutti}, {Marcin},
  {Marggraf}, {Martinelli}, {Martinet}, {Marulli}, {Massey}, {Maurogordato},
  {Medinaceli}, {Mei}, {Mellier}, {Meneghetti}, {Merlin}, {Meylan}, {Mora},
  {Moresco}, {Moscardini}, {Nakajima}, {Neissner}, {Nichol}, {Niemi},
  {Nightingale}, {Padilla}, {Paltani}, {Pasian}, {Pedersen}, {Percival},
  {Pettorino}, {Pires}, {Polenta}, {Poncet}, {Popa}, {Pozzetti}, {Raison},
  {Rebolo}, {Renzi}, {Rhodes}, {Riccio}, {Romelli}, {Roncarelli}, {Saglia},
  {Sakr}, {S{\'a}nchez}, {Sapone}, {Sartoris}, {Schewtschenko}, {Schirmer},
  {Schneider}, {Schrabback}, {Secroun}, {Seidel}, {Seiffert}, {Serrano},
  {Simon}, {Sirignano}, {Sirri}, \& {Spurio Mancini}}]{lines2025a}
{Euclid Collaboration}, {Lines}, N.~E.~P., {Collett}, T.~E., {et~al.}
  2025{\natexlab{b}}, arXiv e-prints, arXiv:2503.15326

\bibitem[{{Euclid Collaboration} {et~al.}(2025{\natexlab{c}}){Euclid
  Collaboration}, {Rojas}, {Collett}, {Acevedo Barroso}, {Nightingale},
  {Stern}, {Moustakas}, {Schuldt}, {Despali}, {Melo}, {Walmsley}, {Ballard},
  {Enzi}, {Li}, {Sainz de Murieta}, {Andika}, {Cl{\'e}ment}, {Courbin},
  {Ecker}, {Gavazzi}, {Jackson}, {Kov{\'a}cs}, {Matavulj}, {Meneghetti},
  {Serjeant}, {Sluse}, {Tortora}, {Verma}, {Marchetti}, {O'Riordan},
  {McCarthy}, {Suyu}, {Metcalf}, {Aghanim}, {Altieri}, {Amara}, {Andreon},
  {Auricchio}, {Aussel}, {Baccigalupi}, {Baldi}, {Balestra}, {Bardelli},
  {Battaglia}, {Bender}, {Biviano}, {Bonchi}, {Branchini}, {Brescia},
  {Brinchmann}, {Camera}, {Ca{\~n}as-Herrera}, {Capobianco}, {Carbone},
  {Cardone}, {Carretero}, {Casas}, {Castellano}, {Castignani}, {Cavuoti},
  {Chambers}, {Cimatti}, {Colodro-Conde}, {Congedo}, {Conselice}, {Conversi},
  {Copin}, {Courtois}, {Cropper}, {Da Silva}, {Degaudenzi}, {De Lucia}, {Di
  Giorgio}, {Dolding}, {Dole}, {Dubath}, {Dupac}, {Escoffier}, {Fabricius},
  {Farina}, {Farinelli}, {Faustini}, {Ferriol}, {Finelli}, {Fotopoulou},
  {Frailis}, {Franceschi}, {Galeotta}, {George}, {Gillard}, {Gillis},
  {Giocoli}, {G{\'o}mez-Alvarez}, {Gracia-Carpio}, {Granett}, {Grazian},
  {Grupp}, {Guzzo}, {Gwyn}, {Haugan}, {Holmes}, {Hook}, {Hormuth}, {Hornstrup},
  {Hudelot}, {Jahnke}, {Jhabvala}, {Keih{\"a}nen}, {Kermiche}, {Kiessling},
  {Kubik}, {Kuijken}, {K{\"u}mmel}, {Kunz}, {Kurki-Suonio}, {Le Boulc'h}, {Le
  Brun}, {Le Mignant}, {Liebing}, {Ligori}, {Lilje}, {Lindholm}, {Lloro},
  {Mainetti}, {Maino}, {Maiorano}, {Mansutti}, {Marcin}, {Marggraf},
  {Martinelli}, {Martinet}, {Marulli}, {Massey}, {Maurogordato}, {McCracken},
  {Medinaceli}, {Mei}, {Melchior}, {Mellier}, {Merlin}, {Meylan}, {Mora},
  {Moresco}, {Moscardini}, {Nakajima}, {Neissner}, {Nichol}, {Niemi},
  {Padilla}, {Paltani}, {Pasian}, {Pedersen}, {Percival}, {Pettorino}, {Pires},
  {Polenta}, {Poncet}, {Popa}, {Pozzetti}, {Raison}, {Rebolo}, {Renzi},
  {Rhodes}, {Riccio}, {Romelli}, {Roncarelli}, {Saglia}, {Sakr}, {S{\'a}nchez},
  {Sapone}, {Sartoris}, {Schewtschenko}, {Schirmer}, {Schneider}, {Schrabback},
  {Secroun}, {Seidel}, {Seiffert}, {Serrano}, {Simon}, {Sirignano}, {Sirri},
  {Stanco}, {Steinwagner}, {Tallada-Cresp{\'\i}}, {Taylor}, {Tereno}, {Toft},
  {Toledo-Moreo}, {Torradeflot}, {Tutusaus}, {Valenziano}, {Valiviita},
  {Vassallo}, {Verdoes Kleijn}, {Veropalumbo}, {Wang}, {Weller}, {Zacchei}, \&
  {Zamorani}}]{rojas2025}
{Euclid Collaboration}, {Rojas}, K., {Collett}, T.~E., {et~al.}
  2025{\natexlab{c}}, arXiv e-prints, arXiv:2503.15325

\bibitem[{{Euclid Collaboration} {et~al.}(2025{\natexlab{d}}){Euclid
  Collaboration}, {Walmsley}, {Holloway}, {Lines}, {Rojas}, {Collett}, {Verma},
  {Li}, {Nightingale}, {Despali}, {Schuldt}, {Gavazzi}, {Melo}, {Metcalf},
  {Andika}, {Leuzzi}, {Manj{\'o}n-Garc{\'\i}a}, {Pearce-Casey}, {Vincken},
  {Wilde}, {Busillo}, {Tortora}, {Acevedo Barroso}, {Dole}, {Ecker}, {Pearson},
  {Marshall}, {More}, {Saifollahi}, {Gracia-Carpio}, {Baeten}, {Cornen},
  {Johnson}, {Macmillan}, {Kruk}, {Remmelgas}, {Cl{\'e}ment}, {Degaudenzi},
  {Courbin}, {Bovy}, {Casas}, {Dannerbauer}, {Diego}, {Finner}, {Galan},
  {Giocoli}, {Hogg}, {Jahnke}, {Katona}, {Kov{\'a}cs}, {De Leo}, {Mahler},
  {Millon}, {Nagam}, {Nugent}, {Sainz de Murieta}, {O'Riordan}, {Sluse},
  {Sonnenfeld}, {Spiniello}, {Serjeant}, {Thai}, {Ulivi}, {Walth},
  {Weisenbach}, {Zumalacarregui}, {Aghanim}, {Altieri}, {Amara}, {Andreon},
  {Auricchio}, {Aussel}, {Baccigalupi}, {Baldi}, {Balestra}, {Bardelli},
  {Battaglia}, {Bernardeau}, {Biviano}, {Bonchi}, {Bonino}, {Branchini},
  {Brescia}, {Brinchmann}, {Camera}, {Ca{\~n}as-Herrera}, {Capobianco},
  {Carbone}, {Cardone}, {Carretero}, {Castander}, {Castellano}, {Castignani},
  {Cavuoti}, {Chambers}, {Cimatti}, {Colodro-Conde}, {Congedo}, {Conselice},
  {Conversi}, {Copin}, {Corcione}, {Courtois}, {Cropper}, {Da Silva}, {De
  Lucia}, {Di Giorgio}, {Dolding}, {Dubath}, {Duncan}, {Dupac}, {Ealet},
  {Escoffier}, {Fabricius}, {Farina}, {Farinelli}, {Faustini}, {Finelli},
  {Fotopoulou}, {Frailis}, {Franceschi}, {Fumana}, {Galeotta}, {George},
  {Gillard}, {Gillis}, {G{\'o}mez-Alvarez}, {Granett}, {Grazian}, {Grupp},
  {Guzzo}, {Gwyn}, {Haugan}, {Hoekstra}, {Holmes}, {Hook}, {Hormuth},
  {Hornstrup}, {Hudelot}, {Jhabvala}, {Joachimi}, {Keih{\"a}nen}, {Kermiche},
  {Kiessling}, {Kubik}, {K{\"u}mmel}, {Kunz}, {Kurki-Suonio}, {Lahav}, {Le
  Boulc'h}, {Le Brun}, {Le Mignant}, {Ligori}, {Lilje}, {Lindholm}, {Lloro},
  {Mainetti}, {Maino}, {Maiorano}, {Mansutti}, {Marcin}, {Marggraf},
  {Martinelli}, {Martinet}, {Marulli}, {Massey}, {Maurogordato}, {McCracken},
  {Medinaceli}, {Mei}, {Mellier}, {Meneghetti}, {Merlin}, {Meylan}, {Mora},
  {Moresco}, {Moscardini}, {Nakajima}, {Neissner}, {Nichol}, {Niemi},
  {Padilla}, {Paltani}, {Pasian}, {Pedersen}, {Percival}, {Pettorino}, {Pires},
  {Polenta}, {Poncet}, {Popa}, {Pozzetti}, {Raison}, {Rebolo}, {Renzi},
  {Rhodes}, {Riccio}, {Romelli}, {Roncarelli}, \& {Saglia}}]{walmsley2025}
{Euclid Collaboration}, {Walmsley}, M., {Holloway}, P., {et~al.}
  2025{\natexlab{d}}, arXiv e-prints, arXiv:2503.15324

\bibitem[{{Hilton} {et~al.}(2021){Hilton}, {Sif{\'o}n}, {Naess},
  {Madhavacheril}, {Oguri}, {Rozo}, {Rykoff}, {Abbott}, {Adhikari}, {Aguena},
  {Aiola}, {Allam}, {Amodeo}, {Amon}, {Annis}, {Ansarinejad}, {Aros-Bunster},
  {Austermann}, {Avila}, {Bacon}, {Battaglia}, {Beall}, {Becker}, {Bernstein},
  {Bertin}, {Bhandarkar}, {Bhargava}, {Bond}, {Brooks}, {Burke}, {Calabrese},
  {Carrasco Kind}, {Carretero}, {Choi}, {Choi}, {Conselice}, {da Costa},
  {Costanzi}, {Crichton}, {Crowley}, {D{\"u}nner}, {Denison}, {Devlin},
  {Dicker}, {Diehl}, {Dietrich}, {Doel}, {Duff}, {Duivenvoorden}, {Dunkley},
  {Everett}, {Ferraro}, {Ferrero}, {Fert{\'e}}, {Flaugher}, {Frieman},
  {Gallardo}, {Garc{\'\i}a-Bellido}, {Gaztanaga}, {Gerdes}, {Giles}, {Golec},
  {Gralla}, {Grandis}, {Gruen}, {Gruendl}, {Gschwend}, {Gutierrez}, {Han},
  {Hartley}, {Hasselfield}, {Hill}, {Hilton}, {Hincks}, {Hinton}, {Ho},
  {Honscheid}, {Hoyle}, {Hubmayr}, {Huffenberger}, {Hughes}, {Jaelani}, {Jain},
  {James}, {Jeltema}, {Kent}, {Knowles}, {Koopman}, {Kuehn}, {Lahav}, {Lima},
  {Lin}, {Lokken}, {Loubser}, {MacCrann}, {Maia}, {Marriage}, {Martin},
  {McMahon}, {Melchior}, {Menanteau}, {Miquel}, {Miyatake}, {Moodley},
  {Morgan}, {Mroczkowski}, {Nati}, {Newburgh}, {Niemack}, {Nishizawa},
  {Ogando}, {Orlowski-Scherer}, {Page}, {Palmese}, {Partridge},
  {Paz-Chinch{\'o}n}, {Phakathi}, {Plazas}, {Robertson}, {Romer}, {Carnero
  Rosell}, {Salatino}, {Sanchez}, {Schaan}, {Schillaci}, {Sehgal}, {Serrano},
  {Shin}, {Simon}, {Smith}, {Soares-Santos}, {Spergel}, {Staggs}, {Storer},
  {Suchyta}, {Swanson}, {Tarle}, {Thomas}, {To}, {Trac}, {Ullom}, {Vale}, {Van
  Lanen}, {Vavagiakis}, {De Vicente}, {Wilkinson}, {Wollack}, {Xu}, \&
  {Zhang}}]{hilton2021}
{Hilton}, M., {Sif{\'o}n}, C., {Naess}, S., {et~al.} 2021, \apjs, 253, 3

\bibitem[{{Hinton} {et~al.}(2016){Hinton}, {Davis}, {Lidman}, {Glazebrook}, \&
  {Lewis}}]{hinton2016}
{Hinton}, S.~R., {Davis}, T.~M., {Lidman}, C., {Glazebrook}, K., \& {Lewis},
  G.~F. 2016, Astronomy and Computing, 15, 61

\bibitem[{{Huang} {et~al.}(2025){Huang}, {Inchausti}, {Storfer},
  {Tabares-Tarquinio}, {Moustakas}, {Sheu}, {Agarwal}, {Tamargo-Arizmendi},
  {Schlegel}, {Aguilar}, {Ahlen}, {Aldering}, {Bailey}, {Banka}, {BenZvi},
  {Bianchi}, {Bolton}, {Brooks}, {Cikota}, {Claybaugh}, {Dawson}, {de la
  Macorra}, {Dey}, {Doel}, {Edelstein}, {Forero-Romero}, {Gaztanaga},
  {Gontcho}, {Gonzalez-Morales}, {Gu}, {Honscheid}, {Ishak}, {Juneau}, {Kehoe},
  {Kisner}, {Koposov}, {Kwon}, {Lambert}, {Landriau}, {Lang}, {Le Guillou},
  {Levi}, {Liu}, {Meisner}, {Miquel}, {Myers}, {Perlmutter},
  {Palanque-Delabrouille}, {Perez-Rafols}, {Poppett}, {Prada}, {Rossi},
  {Rubin}, {Sanchez}, {Schubnell}, {Shu}, {Silver}, {Sprayberry}, {Suzuki},
  {Tarle}, {Weaver}, \& {Zou}}]{huang2025}
{Huang}, X., {Inchausti}, J.~C., {Storfer}, C.~J., {et~al.} 2025, arXiv
  e-prints, arXiv:2509.18089

\bibitem[{{Huang} {et~al.}(2021){Huang}, {Storfer}, {Gu}, {Ravi}, {Pilon},
  {Sheu}, {Venguswamy}, {Banka}, {Dey}, {Landriau}, {Lang}, {Meisner},
  {Moustakas}, {Myers}, {Sajith}, {Schlafly}, \& {Schlegel}}]{huang2021}
{Huang}, X., {Storfer}, C., {Gu}, A., {et~al.} 2021, \apj, 909, 27

\bibitem[{{Huang} {et~al.}(2020){Huang}, {Storfer}, {Ravi}, {Pilon}, {Domingo},
  {Schlegel}, {Bailey}, {Dey}, {Gupta}, {Herrera}, {Juneau}, {Landriau},
  {Lang}, {Meisner}, {Moustakas}, {Myers}, {Schlafly}, {Valdes}, {Weaver},
  {Yang}, \& {Y{\`e}che}}]{huang2020}
{Huang}, X., {Storfer}, C., {Ravi}, V., {et~al.} 2020, \apj, 894, 78

\bibitem[{{Jacobs} {et~al.}(2019){Jacobs}, {Collett}, {Glazebrook},
  {Buckley-Geer}, {Diehl}, {Lin}, {McCarthy}, {Qin}, {Odden}, {Caso Escudero},
  {Dial}, {Yung}, {Gaitsch}, {Pellico}, {Lindgren}, {Abbott}, {Annis}, {Avila},
  {Brooks}, {Burke}, {Carnero Rosell}, {Carrasco Kind}, {Carretero}, {da
  Costa}, {De Vicente}, {Fosalba}, {Frieman}, {Garc{\'\i}a-Bellido},
  {Gaztanaga}, {Goldstein}, {Gruen}, {Gruendl}, {Gschwend}, {Hollowood},
  {Honscheid}, {Hoyle}, {James}, {Krause}, {Kuropatkin}, {Lahav}, {Lima},
  {Maia}, {Marshall}, {Miquel}, {Plazas}, {Roodman}, {Sanchez}, {Scarpine},
  {Serrano}, {Sevilla-Noarbe}, {Smith}, {Sobreira}, {Suchyta}, {Swanson},
  {Tarle}, {Vikram}, {Walker}, {Zhang}, \& {DES Collaboration}}]{jacobs2019}
{Jacobs}, C., {Collett}, T., {Glazebrook}, K., {et~al.} 2019, \apjs, 243, 17

\bibitem[{{Klein} {et~al.}(2019){Klein}, {Grandis}, {Mohr}, {Paulus}, {Abbott},
  {Annis}, {Avila}, {Bertin}, {Brooks}, {Buckley-Geer}, {Rosell}, {Kind},
  {Carretero}, {Castander}, {Cunha}, {D'Andrea}, {da Costa}, {De Vicente},
  {Desai}, {Diehl}, {Dietrich}, {Doel}, {Evrard}, {Flaugher}, {Fosalba},
  {Frieman}, {Garc{\'\i}a-Bellido}, {Gaztanaga}, {Giles}, {Gruen}, {Gruendl},
  {Gschwend}, {Gutierrez}, {Hartley}, {Hollowood}, {Honscheid}, {Hoyle},
  {James}, {Jeltema}, {Kuehn}, {Kuropatkin}, {Lima}, {Maia}, {March},
  {Marshall}, {Menanteau}, {Miquel}, {Ogando}, {Plazas}, {Romer}, {Roodman},
  {Sanchez}, {Scarpine}, {Schindler}, {Serrano}, {Sevilla-Noarbe}, {Smith},
  {Smith}, {Soares-Santos}, {Sobreira}, {Suchyta}, {Swanson}, {Tarle},
  {Thomas}, {Vikram}, \& {DES Collaboration}}]{klein2019}
{Klein}, M., {Grandis}, S., {Mohr}, J.~J., {et~al.} 2019, \mnras, 488, 739

\bibitem[{{Knowles} {et~al.}(2022){Knowles}, {Cotton}, {Rudnick}, {Camilo},
  {Goedhart}, {Deane}, {Ramatsoku}, {Bietenholz}, {Br{\"u}ggen}, {Button},
  {Chen}, {Chibueze}, {Clarke}, {de Gasperin}, {Ianjamasimanana}, {J{\'o}zsa},
  {Hilton}, {Kesebonye}, {Kolokythas}, {Kraan-Korteweg}, {Lawrie}, {Lochner},
  {Loubser}, {Marchegiani}, {Mhlahlo}, {Moodley}, {Murphy}, {Namumba},
  {Oozeer}, {Parekh}, {Pillay}, {Passmoor}, {Ramaila}, {Ranchod},
  {Retana-Montenegro}, {Sebokolodi}, {Sikhosana}, {Smirnov}, {Thorat},
  {Venturi}, {Abbott}, {Adam}, {Adams}, {Aldera}, {Bauermeister}, {Bennett},
  {Bode}, {Botha}, {Botha}, {Brederode}, {Buchner}, {Burger}, {Cheetham}, {de
  Villiers}, {Dikgale-Mahlakoana}, {du Toit}, {Esterhuyse}, {Fadana},
  {Fanaroff}, {Fataar}, {Foley}, {Fourie}, {Frank}, {Gamatham}, {Gatsi},
  {Geyer}, {Gouws}, {Gumede}, {Heywood}, {Hlakola}, {Hokwana}, {Hoosen},
  {Horn}, {Horrell}, {Hugo}, {Isaacson}, {Jonas}, {Jordaan}, {Joubert},
  {Julie}, {Kapp}, {Kasper}, {Kenyon}, {Kotz{\'e}}, {Kotze}, {Kriek}, {Kriel},
  {Krishnan}, {Kusel}, {Legodi}, {Lehmensiek}, {Liebenberg}, {Lord}, {Lunsky},
  {Madisa}, {Magnus}, {Main}, {Makhaba}, {Makhathini}, {Malan}, {Manley},
  {Marais}, {Maree}, {Martens}, {Mauch}, {McAlpine}, {Merry}, {Millenaar},
  {Mokone}, {Monama}, {Mphego}, {New}, {Ngcebetsha}, {Ngoasheng}, {Ockards},
  {Otto}, {Patel}, {Peens-Hough}, {Perkins}, {Ramanujam}, {Ramudzuli},
  {Ratcliffe}, {Renil}, {Robyntjies}, {Rust}, {Salie}, {Sambu}, {Schollar},
  {Schwardt}, {Schwartz}, {Serylak}, {Siebrits}, {Sirothia}, {Slabber},
  {Sofeya}, {Taljaard}, {Tasse}, {Tiplady}, {Toruvanda}, {Twum}, {van Balla},
  {van der Byl}, {van der Merwe}, {van Dyk}, {Van Tonder}, {Van Wyk}, {Venter},
  {Venter}, {Welz}, {Williams}, \& {Xaia}}]{knowles2022}
{Knowles}, K., {Cotton}, W.~D., {Rudnick}, L., {et~al.} 2022, \aap, 657, A56

\bibitem[{{Koulouridis} {et~al.}(2021){Koulouridis}, {Clerc}, {Sadibekova},
  {Chira}, {Drigga}, {Faccioli}, {Le F{\`e}vre}, {Garrel}, {Gaynullina},
  {Gkini}, {Kosiba}, {Pacaud}, {Pierre}, {Ridl}, {Tazhenova}, {Adami},
  {Altieri}, {Baguley}, {Cabanac}, {Cucchetti}, {Khalikova}, {Lieu}, {Melin},
  {Molham}, {Ramos-Ceja}, {Soucail}, {Takey}, \&
  {Valtchanov}}]{koulouridis2021}
{Koulouridis}, E., {Clerc}, N., {Sadibekova}, T., {et~al.} 2021, \aap, 652, A12

\bibitem[{{Lanusse} {et~al.}(2018){Lanusse}, {Ma}, {Li}, {Collett}, {Li},
  {Ravanbakhsh}, {Mandelbaum}, \& {P{\'o}czos}}]{lanusse2018}
{Lanusse}, F., {Ma}, Q., {Li}, N., {et~al.} 2018, \mnras, 473, 3895

\bibitem[{{Laureijs} {et~al.}(2011){Laureijs}, {Amiaux}, {Arduini},
  {Augu{\`e}res}, {Brinchmann}, {Cole}, {Cropper}, {Dabin}, {Duvet}, {Ealet},
  \& et~al.}]{laureijs2011}
{Laureijs}, R., {Amiaux}, J., {Arduini}, S., {et~al.} 2011, ArXiv e-prints
  [\eprint[arXiv]{1110.3193}]

\bibitem[{{Li} {et~al.}(2025){Li}, {Collett}, {Acebron}, {Ecker}, {Enzi},
  {Galan}, {Gavazzi}, {Granata}, {Grillo}, {Lines}, {Metcalf}, {Moustakas},
  {Nightingale}, {O'Riordan}, {Rojas}, {Schuldt}, \& {Sonnenfeld}}]{li2025b}
{Li}, T., {Collett}, T.~E., {Acebron}, A., {et~al.} 2025, {Breaking the mass
  sheet degeneracy in strong lensing time delay cosmology with a new sample of
  double source plane lenses.}, HST Proposal. Cycle 33, ID. \#18097

\bibitem[{{Li} {et~al.}(2024){Li}, {Collett}, {Krawczyk}, \& {Enzi}}]{li2024}
{Li}, T., {Collett}, T.~E., {Krawczyk}, C.~M., \& {Enzi}, W. 2024, \mnras, 527,
  5311

\bibitem[{{Lin} {et~al.}(2025){Lin}, {Toro Bertolla}, {Cikota}, {Huang},
  {Storfer}, {Tamargo-Arizmendi}, {Schlegel}, {Sheu}, \& {Suzuki}}]{lin2025}
{Lin}, E., {Toro Bertolla}, I., {Cikota}, A., {et~al.} 2025, arXiv e-prints,
  arXiv:2509.18078

\bibitem[{{May}(2013)}]{may2013}
{May}, P.~E. 2013, PhD thesis, University of Birmingham, UK

\bibitem[{{Mehrtens} {et~al.}(2012){Mehrtens}, {Romer}, {Hilton},
  {Lloyd-Davies}, {Miller}, {Stanford}, {Hosmer}, {Hoyle}, {Collins}, {Liddle},
  {Viana}, {Nichol}, {Stott}, {Dubois}, {Kay}, {Sahl{\'e}n}, {Young}, {Short},
  {Christodoulou}, {Watson}, {Davidson}, {Harrison}, {Baruah}, {Smith},
  {Burke}, {Mayers}, {Deadman}, {Rooney}, {Edmondson}, {West}, {Campbell},
  {Edge}, {Mann}, {Sabirli}, {Wake}, {Benoist}, {da Costa}, {Maia}, \&
  {Ogando}}]{mehrtens2012}
{Mehrtens}, N., {Romer}, A.~K., {Hilton}, M., {et~al.} 2012, \mnras, 423, 1024

\bibitem[{{Meneghetti} {et~al.}(2023){Meneghetti}, {Cui}, {Rasia}, {Yepes},
  {Acebron}, {Angora}, {Bergamini}, {Borgani}, {Calura}, {Despali}, {Giocoli},
  {Granata}, {Grillo}, {Knebe}, {Macci{\`o}}, {Mercurio}, {Moscardini},
  {Natarajan}, {Ragagnin}, {Rosati}, \& {Vanzella}}]{meneghetti2023}
{Meneghetti}, M., {Cui}, W., {Rasia}, E., {et~al.} 2023, \aap, 678, L2

\bibitem[{{Meneghetti} {et~al.}(2020){Meneghetti}, {Davoli}, {Bergamini},
  {Rosati}, {Natarajan}, {Giocoli}, {Caminha}, {Metcalf}, {Rasia}, {Borgani},
  {Calura}, {Grillo}, {Mercurio}, \& {Vanzella}}]{meneghetti2020}
{Meneghetti}, M., {Davoli}, G., {Bergamini}, P., {et~al.} 2020, Science, 369,
  1347

\bibitem[{{Newman} {et~al.}(2018){Newman}, {Belli}, {Ellis}, \&
  {Patel}}]{newman2018}
{Newman}, A.~B., {Belli}, S., {Ellis}, R.~S., \& {Patel}, S.~G. 2018, \apj,
  862, 126

\bibitem[{{Niemiec} {et~al.}(2022){Niemiec}, {Giocoli}, {Cohen}, {Jauzac},
  {Jullo}, \& {Limousin}}]{niemiec2022}
{Niemiec}, A., {Giocoli}, C., {Cohen}, E., {et~al.} 2022, \mnras, 512, 6021

\bibitem[{{Nord} {et~al.}(2016){Nord}, {Buckley-Geer}, {Lin}, {Diehl},
  {Helsby}, {Kuropatkin}, {Amara}, {Collett}, {Allam}, {Caminha}, {De Bom},
  {Desai}, {D{\'u}met-Montoya}, {Pereira}, {Finley}, {Flaugher}, {Furlanetto},
  {Gaitsch}, {Gill}, {Merritt}, {More}, {Tucker}, {Saro}, {Rykoff}, {Rozo},
  {Birrer}, {Abdalla}, {Agnello}, {Auger}, {Brunner}, {Carrasco Kind},
  {Castander}, {Cunha}, {da Costa}, {Foley}, {Gerdes}, {Glazebrook},
  {Gschwend}, {Hartley}, {Kessler}, {Lagattuta}, {Lewis}, {Maia}, {Makler},
  {Menanteau}, {Niernberg}, {Scolnic}, {Vieira}, {Gramillano}, {Abbott},
  {Banerji}, {Benoit-L{\'e}vy}, {Brooks}, {Burke}, {Capozzi}, {Carnero Rosell},
  {Carretero}, {D'Andrea}, {Dietrich}, {Doel}, {Evrard}, {Frieman},
  {Gaztanaga}, {Gruen}, {Honscheid}, {James}, {Kuehn}, {Li}, {Lima},
  {Marshall}, {Martini}, {Melchior}, {Miquel}, {Neilsen}, {Nichol}, {Ogando},
  {Plazas}, {Romer}, {Sako}, {Sanchez}, {Scarpine}, {Schubnell},
  {Sevilla-Noarbe}, {Smith}, {Soares-Santos}, {Sobreira}, {Suchyta}, {Swanson},
  {Tarle}, {Thaler}, {Walker}, {Wester}, {Zhang}, \& {DES
  Collaboration}}]{nord2016}
{Nord}, B., {Buckley-Geer}, E., {Lin}, H., {et~al.} 2016, \apj, 827, 51

\bibitem[{{Richard} {et~al.}(2021){Richard}, {Claeyssens}, {Lagattuta},
  {Guaita}, {Bauer}, {Pello}, {Carton}, {Bacon}, {Soucail}, {Lyon}, {Kneib},
  {Mahler}, {Cl{\'e}ment}, {Mercier}, {Variu}, {Tamone}, {Ebeling}, {Schmidt},
  {Nanayakkara}, {Maseda}, {Weilbacher}, {Bouch{\'e}}, {Bouwens}, {Wisotzki},
  {de la Vieuville}, {Martinez}, \& {Patr{\'\i}cio}}]{richard2021}
{Richard}, J., {Claeyssens}, A., {Lagattuta}, D., {et~al.} 2021, \aap, 646, A83

\bibitem[{{Rodney} {et~al.}(2021){Rodney}, {Brammer}, {Pierel}, {Richard},
  {Toft}, {O'Connor}, {Akhshik}, \& {Whitaker}}]{rodney2021}
{Rodney}, S.~A., {Brammer}, G.~B., {Pierel}, J. D.~R., {et~al.} 2021, Nature
  Astronomy, 5, 1118

\bibitem[{{Ruel} {et~al.}(2014){Ruel}, {Bazin}, {Bayliss}, {Brodwin}, {Foley},
  {Stalder}, {Aird}, {Armstrong}, {Ashby}, {Bautz}, {Benson}, {Bleem},
  {Bocquet}, {Carlstrom}, {Chang}, {Chapman}, {Cho}, {Clocchiatti}, {Crawford},
  {Crites}, {de Haan}, {Desai}, {Dobbs}, {Dudley}, {Forman}, {George},
  {Gladders}, {Gonzalez}, {Halverson}, {Harrington}, {High}, {Holder},
  {Holzapfel}, {Hrubes}, {Jones}, {Joy}, {Keisler}, {Knox}, {Lee}, {Leitch},
  {Liu}, {Lueker}, {Luong-Van}, {Mantz}, {Marrone}, {McDonald}, {McMahon},
  {Mehl}, {Meyer}, {Mocanu}, {Mohr}, {Montroy}, {Murray}, {Natoli},
  {Nurgaliev}, {Padin}, {Plagge}, {Pryke}, {Reichardt}, {Rest}, {Ruhl},
  {Saliwanchik}, {Saro}, {Sayre}, {Schaffer}, {Shaw}, {Shirokoff}, {Song},
  {{\v{S}}uhada}, {Spieler}, {Stanford}, {Staniszewski}, {Starsk}, {Story},
  {Stubbs}, {van Engelen}, {Vanderlinde}, {Vieira}, {Vikhlinin}, {Williamson},
  {Zahn}, \& {Zenteno}}]{ruel2014}
{Ruel}, J., {Bazin}, G., {Bayliss}, M., {et~al.} 2014, \apj, 792, 45

\bibitem[{{Sahu} {et~al.}(2024){Sahu}, {Tran}, {Suyu}, {Shajib}, {Ertl},
  {Kacprzak}, {Glazebrook}, {Jones}, {G.~C.}, {Barone}, {Baker}, {Skobe},
  {Derkenne}, {Lewis}, {Sweet}, \& {Lopez}}]{sahu2024}
{Sahu}, N., {Tran}, K.-V., {Suyu}, S.~H., {et~al.} 2024, \apj, 970, 86

\bibitem[{{Sainz de Murieta} {et~al.}(2024){Sainz de Murieta}, {Collett},
  {Magee}, {Pierel}, {Enzi}, {Lokken}, {Gagliano}, \&
  {Ryczanowski}}]{sainzdemurieta2024}
{Sainz de Murieta}, A., {Collett}, T.~E., {Magee}, M.~R., {et~al.} 2024,
  \mnras, 535, 2523

\bibitem[{{Shajib} {et~al.}(2024){Shajib}, {Vernardos}, {Collett}, {Motta},
  {Sluse}, {Williams}, {Saha}, {Birrer}, {Spiniello}, \& {Treu}}]{shajib2024}
{Shajib}, A.~J., {Vernardos}, G., {Collett}, T.~E., {et~al.} 2024, \ssr, 220,
  87

\bibitem[{{Sheu} {et~al.}(2024){Sheu}, {Cikota}, {Huang}, {Glazebrook},
  {Storfer}, {Agarwal}, {Schlegel}, {Suzuki}, {Barone}, {Bian}, {Jeltema},
  {Jones}, {Kacprzak}, {O'Donnell}, \& {G.~C.}}]{sheu2024}
{Sheu}, W., {Cikota}, A., {Huang}, X., {et~al.} 2024, \apj, 973, 3

\bibitem[{{Shu} \& {Li}(2025)}]{shu2025}
{Shu}, Y. \& {Li}, S. 2025, Science China Physics, Mechanics, and Astronomy,
  68, 129511

\bibitem[{{Sif{\'o}n} {et~al.}(2025){Sif{\'o}n}, {Finoguenov}, {Haines},
  {Jaff{\'e}}, {Amrutha}, {Demarco}, {Lima}, {Lima-Dias},
  {M{\'e}ndez-Hern{\'a}ndez}, {Merluzzi}, {Monachesi}, {Teixeira}, {Tejos},
  {Almeida-Fernandes}, {Araya-Araya}, {Argudo-Fern{\'a}ndez}, {Baier-Soto},
  {Bilton}, {Bom}, {Calder{\'o}n}, {Cassar{\`a}}, {Comparat}, {Courtois},
  {D'Ago}, {Dupuy}, {Fritz}, {Haack}, {Herpich}, {Ibar}, {Kuchner}, {Lacerna},
  {Lopes}, {Lopez}, {L{\"o}sch}, {McGee}, {Mendes de Oliveira}, {Morelli},
  {Moretti}, {Pallero}, {Piraino-Cerda}, {Pompei}, {Rescigno}, {Smith
  Castelli}, {Smith}, {Sodr{\'e}}, \& {Tempel}}]{sifon2025}
{Sif{\'o}n}, C., {Finoguenov}, A., {Haines}, C.~P., {et~al.} 2025, \aap, 697,
  A92

\bibitem[{{Smithsonian Astrophysical Observatory}(2000)}]{ds9sao2000}
{Smithsonian Astrophysical Observatory}. 2000, {SAOImage DS9: A utility for
  displaying astronomical images in the X11 window environment}, Astrophysics
  Source Code Library, record ascl:0003.002

\bibitem[{{Storfer} {et~al.}(2024){Storfer}, {Huang}, {Gu}, {Sheu}, {Banka},
  {Dey}, {Inchausti Reyes}, {Jain}, {Kwon}, {Lang}, {Lee}, {Meisner},
  {Moustakas}, {Myers}, {Tabares-Tarquinio}, {Schlafly}, \&
  {Schlegel}}]{storfer2024}
{Storfer}, C., {Huang}, X., {Gu}, A., {et~al.} 2024, \apjs, 274, 16

\bibitem[{{Urcelay} {et~al.}(2026){Urcelay}, {Huang}, {Sheu}, {O'Donnell},
  {Jeltema}, {Williams}, {Xu}, {Agarwal}, {Aldering},
  {{\'A}lvarez-Garc{\'\i}a}, {Ambardekar}, {Barone}, {Bian}, {Bolton},
  {Cikota}, {Farren}, {Glazebrook}, {Hoyt}, {Jain}, {Jones}, {Kacprzak}, {Lin},
  {Perlmutter}, {Rubin}, {Schlegel}, {Silver}, {Storfer}, {Suzuki}, {Truong},
  {{\'U}beda}, \& {C}}]{urcelay2026}
{Urcelay}, F., {Huang}, X., {Sheu}, W., {et~al.} 2026, arXiv e-prints,
  arXiv:2602.16077

\bibitem[{{Urcelay} {et~al.}(2025){Urcelay}, {Jullo}, {Barrientos}, {Huang}, \&
  {Hernandez}}]{urcelay2025}
{Urcelay}, F., {Jullo}, E., {Barrientos}, L.~F., {Huang}, X., \& {Hernandez},
  J. 2025, \aap, 694, A35

\bibitem[{{Vernet} {et~al.}(2011){Vernet}, {Dekker}, {D'Odorico}, {Kaper},
  {Kjaergaard}, {Hammer}, {Randich}, {Zerbi}, {Groot}, {Hjorth}, {Guinouard},
  {Navarro}, {Adolfse}, {Albers}, {Amans}, {Andersen}, {Andersen}, {Binetruy},
  {Bristow}, {Castillo}, {Chemla}, {Christensen}, {Conconi}, {Conzelmann},
  {Dam}, {de Caprio}, {de Ugarte Postigo}, {Delabre}, {di Marcantonio},
  {Downing}, {Elswijk}, {Finger}, {Fischer}, {Flores}, {Fran{\c{c}}ois},
  {Goldoni}, {Guglielmi}, {Haigron}, {Hanenburg}, {Hendriks}, {Horrobin},
  {Horville}, {Jessen}, {Kerber}, {Kern}, {Kiekebusch}, {Kleszcz}, {Klougart},
  {Kragt}, {Larsen}, {Lizon}, {Lucuix}, {Mainieri}, {Manuputy}, {Martayan},
  {Mason}, {Mazzoleni}, {Michaelsen}, {Modigliani}, {Moehler}, {M{\o}ller},
  {Norup S{\o}rensen}, {N{\o}rregaard}, {P{\'e}roux}, {Patat}, {Pena}, {Pragt},
  {Reinero}, {Rigal}, {Riva}, {Roelfsema}, {Royer}, {Sacco}, {Santin},
  {Schoenmaker}, {Spano}, {Sweers}, {Ter Horst}, {Tintori}, {Tromp}, {van
  Dael}, {van der Vliet}, {Venema}, {Vidali}, {Vinther}, {Vola}, {Winters},
  {Wistisen}, {Wulterkens}, \& {Zacchei}}]{vernet2011}
{Vernet}, J., {Dekker}, H., {D'Odorico}, S., {et~al.} 2011, \aap, 536, A105

\end{thebibliography}

\onecolumn
\begin{longtable}{llllllllll}
\caption{Lens system spectroscopic redshift measurements}
\label{tab:confirmed} \\
\hline\hline
Name & Object & RA & Dec & z & Q$_z$ & DESI $g$ & DESI $r$ & DESI $i$ & DESI $z$\\
&& deg & deg &&& mag & mag & mag & mag\\
\hline
\endfirsthead
\caption{continued.}\\
\hline\hline
Name & Object & RA & Dec & z & Q$_z$ & DESI $g$ & DESI $r$ & DESI $i$ & DESI $z$\\
&& deg & deg &&& mag & mag & mag & mag\\
\hline
\endhead
\hline
\multicolumn{10}{l}{
\begin{minipage}{.9\linewidth}
Note -- Redshift quality (Q$_{z}$); (1) Very uncertain continuum detection, (2) One emission line and uncertain continuum detection, (3) Good continuum detection, (4) Several emission lines or very good continuum detection
\end{minipage}}
\endfoot
DESI-002.0668-32.6211 & Lens & 2.0671 & -32.6207 &  0.295 & 3 & 18.57 & 17.15 & - & 16.37 \\
 & Source & 2.0667 & -32.6211 & 1.21525 & 4 & 20.07 & 19.98 & - & 19.62 \\
DESI-006.0673-28.7195 & Lens 1 & 6.0673 & -28.7200 & 0.51658 & 2 & 20.38 & 18.46 & 17.60 & 17.15 \\
 & Lens 2 & 6.0673 & -28.7200 & 0.602 & 1 & 24.94 & 23.35 & 22.67 & 22.31 \\
 & Source & 6.0691 & -28.7196 & 2.441 & 4 & 23.20 & 23.37 & 24.11 & - \\
DESI-007.6741-33.9765 & Lens & 7.6741 & -33.9765 & 0.71424 & 3 & 21.69 & 20.89 & 20.05 & 19.71 \\
 & Source & 7.6740 & -33.9770 & 1.583 & 4 & 20.75 & 20.55 & 20.26 & 20.24 \\
DESI-008.6235-45.8583 & Lens & 8.6210 & -45.8601 & 0.5330 & 3 & 20.48 & 18.56 & 17.72 & 17.32 \\
 & Source & 8.6235 & -45.8583 & 1.9260 & 4 & 21.31 & 21.11 & 21.00 & 20.99 \\
DESI-008.8615-50.3335 & Lens & 8.8610 & -50.3323 & 0.5470 & 3 & 21.93 & 20.14 & - & 19.00 \\
 & Source & 8.8630 & -50.3330 & 1.5992 & 2 & 22.32 & 22.05 & - & 21.51 \\
DESI-010.8534-20.6214 & Source & 10.8556 & -20.6210 & 0.8216 & 3 & 21.27 & 19.78 & 18.68 & 18.18 \\
DESI-010.8534-20.6214\_B & Source & 10.8515 & -20.6236 & 0.8211 & 4 & 22.11 & 20.77 & 19.97 & 19.63 \\
 & Lens 1 & 10.8509 & -20.6222 & 0.2860 & 3 & 20.48 & 19.04 & 18.55 & 18.24 \\
 & Lens 2 & 10.8523 & -20.6249 & 0.2875 & 3 & 19.38 & 17.85 & 17.33 & 17.00 \\
DESI-013.8728-61.1576 & Lens & 13.8723 & -61.1577 & 0.5880 & 4 & 21.53 & 20.38 & 19.60 & 19.23 \\
 & Source & 13.8712 & -61.1579 & 2.44873 & 4 & 23.93 & 23.67 & 24.11 & 24.51 \\
DESI-015.4984-21.4473 & Lens & 15.4982 & -21.4487 & 0.39750 & 3 & 20.16 & 18.33 & 17.69 & 17.31 \\
 & Source & 15.4994 & -21.4486 & 2.2011 & 4 & 23.71 & 24.05 & 25.45 & - \\
DESI-016.7705-31.4780 & Lens & 16.7705 & -31.4780 & 0.7724 & 3 & 22.26 & 20.37 & 19.10 & 18.52 \\
 & Source & 16.7716 & -31.4767 & 2.70329 & 4 & 21.88 & 21.36 & 21.24 & 21.19 \\
DESI-018.1307-62.1502 & Lens & 18.1308 & -62.1502 & 0.38525 & 3 & 20.08 & 18.27 & 17.67 & 17.34 \\
 & Source 1 & 18.1317 & -62.1511 & 1.3101 & 4 & 22.05 & 21.68 & 21.47 & 20.81 \\
 & Source 2 & 18.1313 & -62.1518 & 1.0027 & 3 & 22.46 & 22.14 & 21.99 & 21.63 \\
DESI-021.0887-32.0706 & Lens & 21.0887 & -32.0706 & 0.82 & 2 & 22.51 & 20.86 & 19.68 & 19.16 \\
 & Source & 21.0894 & -32.0710 & 3.6395 & 2 & 23.63 & 22.35 & 22.24 & 22.28 \\
DESI-022.2123-29.9602 & Lens 1 & 22.2123 & -29.9602 & 0.6580 & 2 & 21.47 & 19.69 & 18.64 & 18.20 \\
 & Lens 2 & 22.2133 & -29.9591 & 0.6576 & 3 & 23.85 & 22.57 & 21.83 & 21.59 \\
 & Source & 22.2127 & -29.9593 & 1.4685 & 4 & 22.40 & 22.27 & 22.33 & 22.14 \\
DESI-024.5133-21.9299 & Lens & 24.5149 & -21.9277 &  0.3315 & 3 & 22.34 & 20.74 & 20.20 & 19.93 \\
 & Source & 24.5132 & -21.9299 & 1.9469 & 4 & 22.30 & 21.04 & 20.33 & 19.24 \\
DESI-024.5974-28.7358 & Lens & 24.5957 & -28.7356 & 0.41854 & 3 & 20.59 & 18.82 & 18.22 & 17.87 \\
 & Source & 24.5957 & -28.7356 & 2.28515 & 4 & 20.59 & 18.82 & 18.22 & 17.87 \\
DESI-026.9149-63.5238 & Lens & 26.9170 & -63.5248 & 0.908 & 3 & 23.60 & 21.45 & 20.19 & 19.38 \\
 & Source & 26.9148 & -63.5238 & 1.86284 & 4 & 21.74 & 21.52 & 21.28 & 21.10 \\
DESI-027.4831-28.5753 & Lens & 27.4830 & -28.5753 & 0.561 & 2 & - & 19.33 & 18.46 & 18.05 \\
 & Source & 27.4825 & -28.5769 & 1.73369 & 1 & 21.81 & 21.62 & 21.63 & 21.48 \\
DESI-027.8201-35.7384 & Source & 27.8200 & -35.7385 & 1.4912 & 4 & 22.06 & 21.92 & 21.86 & 22.11 \\
DESI-027.8201-35.7384\_B & Lens & 27.8197 & -35.7400 & 0.49 & 1 & 23.97 & 22.16 & 21.44 & 21.04 \\
 & Source & 27.8200 & -35.7385 & 2.27579 & 4 & 22.06 & 21.92 & 21.86 & 22.11 \\
DESI-029.7682-34.8347 & Lens & 29.7678 & -34.8361 & 0.766 & 2 & 22.94 & 20.98 & 19.76 & 19.23 \\
 & Source 1 & 29.7667 & -34.8351 & 2.3405 & 4 & 22.46 & 21.68 & 21.35 & 21.07 \\
 & Source 2 & 29.7667 & -34.8351 & 2.28 & 4 & 22.46 & 21.68 & 21.35 & 21.07 \\
DESI-030.4360-27.6618 & Lens & 30.4360 & -27.6618 & 1.02 & 2 & 23.15 & 22.03 & 21.16 & 20.30 \\
 & Source & 30.4364 & -27.6619 & 2.52232 & 4 & 22.02 & 21.78 & 21.74 & 21.81 \\
DESI-031.7778-27.4457 & Lens & 31.7784 & -27.4475 & 0.35039 & 2 & 23.99 & 22.07 & 21.47 & 21.13 \\
 & Source & 31.7762 & -27.4456 & 1.67886 & 4 & 18.59 & 18.16 & 17.99 & 17.88 \\
DESI-032.4765-35.7990 & Lens & 32.4765 & -35.7990 & 0.3130 & 4 & 19.58 & 17.98 & 17.46 & 17.15 \\
 & Source & 32.4761 & -35.8003 & 1.1263 & 4 & 21.87 & 21.78 & 21.61 & 21.60 \\
DESI-033.8095-29.1570 & Source 1 & 33.8100 & -29.1572 & 1.85914 & 4 & 23.86 & 23.55 & 23.23 & 23.49 \\
DESI-034.2366-59.8485 & Lens 1 & 34.2367 & -59.8484 & 0.479 & 2 & 21.16 & 19.49 & 18.79 & 18.41 \\
 & Lens 2 & 34.2359 & -59.8478 & 0.47319 & 4 & 21.01 & 20.01 & 19.61 & 19.32 \\
DESI-035.2405-38.5511 & Lens & 35.2368 & -38.5517 & 0.357 & 3 & 20.81 & 19.22 & 18.62 & 18.30 \\
 & Source & 35.2382 & -38.5530 & 1.4802 & 4 & 22.04 & 21.76 & 21.20 & 20.64 \\
DESI-035.7946-22.8049 & Lens & 35.7946 & -22.8049 & 0.78 & 1 & 22.07 & 20.74 & 19.69 & 18.96 \\
 & Source & 35.7951 & -22.8038 & 2.5126 & 4 & 22.46 & 22.48 & 22.63 & 22.57 \\
DESI-036.8326-24.1921 & Lens & 36.8299 & -24.1941 & 0.39117 & 3 & 22.39 & 20.57 & 19.95 & 19.60 \\
 & Source & 36.8302 & -24.1953 & 2.798 & 4 & 23.46 & 23.59 & 23.91 & 23.73 \\
DESI-037.0621-29.3949 & Lens & 37.0621 & -29.3949 & 0.30451 & 4 & 19.60 & 17.96 & 17.43 & 17.11 \\
DESI-037.4837-31.1730 & Source & 37.4856 & -31.1734 & 1.30027 & 4 & 22.22 & 22.18 & 22.11 & 22.10 \\
DESI-038.0264-35.4899 & Lens & 38.0264 & -35.4899 & 0.48892 & 2 & 21.22 & 19.91 & 19.18 & 18.80 \\
 & Source & 38.0259 & -35.4897 & 2.13564 & 4 & 23.55 & 23.57 & 24.14 & - \\
DESI-038.8136-21.5095 & Lens & 38.8137 & -21.5094 & 0.32018 & 3 & 19.70 & 18.15 & 17.63 & 17.31 \\
 & Source & 38.8146 & -21.5099 & 1.4696 & 4 & 23.35 & 23.29 & 23.32 & 22.87 \\
DESI-040.5664-25.1054 & Lens & 40.5663 & -25.1054 & 0.28726 & 3 & 19.09 & 17.59 & 17.09 & 16.78 \\
 & Source & 40.5656 & -25.1062 & 1.59588 & 4 & 24.40 & 23.60 & 22.63 & 21.93 \\
 DESI-042.0412-39.928 & Lens & 42.0401 & -39.9301 & 0.685 & 2 & 20.85 & 19.33 & 18.29 & 17.82 \\
 & Source & 42.0420 & -39.9309 & 1.2722 & 4 & 22.92 & 22.98 & 23.07 & 22.34 \\
DESI-044.4121-22.1575 & Lens & 44.4121 & -22.1575 & 0.33026 & 3 & 21.11 & 19.54 & 18.99 & 18.60 \\
 & Source & 44.4122 & -22.1569 & 1.1694 & 3 & 24.61 & 23.35 & 22.21 & 20.98 \\
DESI-044.4344-20.1488 & Lens & 44.4322 & -20.1478 & 0.69114 & 4 & 23.27 & 22.57 & 22.02 & 21.92 \\
 & Source & 44.4323 & -20.1468 & 0.94483 & 4 & 22.93 & 22.39 & 21.60 & 21.29 \\
DESI-044.4344-20.1488\_B & Source & 44.4286 & -20.1474 & 1.68251 & 4 & 22.69 & 22.73 & 23.16 & 24.31 \\
DESI-045.9507-46.4407 & Lens & 45.9517 & -46.4407 & 0.59 & 2 & 21.96 & 20.29 & 19.34 & 18.92 \\
 & Source & 45.9507 & -46.4407 & 1.43636 & 4 & 21.36 & 21.26 & 21.27 & 20.99 \\
DESI-049.7686-49.3612 & Lens & 49.7685 & -49.3621 & 0.9375 & 3 & 22.51 & 20.57 & 19.36 & 18.52 \\
 & Source & 49.7685 & -49.3612 & 1.1178 & 3 & 22.55 & 22.20 & - & 21.73 \\
DESI-049.7786-27.6615 & Lens & 49.7786 & -27.6615 & 0.7 & 2 & 21.96 & 20.51 & 19.57 & 19.12 \\
 & Source & 49.7785 & -27.6612 & 3.18426 & 4 & 22.92 & 22.21 & - & 22.89 \\
DESI-052.7344-52.4708 & Lens & 52.7373 & -52.4703 & 0.4404 & 3 & 19.42 & 17.58 & 16.92 & 16.57 \\
 & Source & 52.7344 & -52.4708 & 1.45470 & 4 & 21.74 & 21.56 & - & 2.27 \\
DESI-055.0891-25.5584 & Lens & 55.0891 & -25.5584 & 0.657 & 2 & 21.06 & 19.34 & 18.30 & 17.89 \\
 & Source & 55.0892 & -25.5573 & 1.603 & 2 & 22.86 & 22.82 & 22.90 & 22.67 \\
DESI-060.5238-22.0990 & Lens 1 & 60.5233 & -22.0995 & 0.46 & 2 & 21.35 & 19.70 & 19.04 & 18.70 \\
 & Lens 2 & 60.5217 & -22.1009 & 0.46 & 1 & 21.25 & 19.46 & 18.81 & 18.46 \\
 & Source & 60.5226 & -22.0998 & 0.8209 & 4 & 22.67 & 22.48 & 22.24 & 21.85 \\
DESI-065.6447-28.0652 & Lens & 34.2359 & -59.8478 & 0.49 & 1 & 21.01 & 20.01 & 19.61 & 19.32 \\
 & Source & 34.2347 & -59.8485 & 0.75 & 1 & 23.19 & 22.95 & 23.11 & 23.46 \\
DESI-067.4729-29.9597 & Lens 1 & 67.4732 & -29.9610 & 0.61 & 2 & 24.64 & 22.69 & 21.52 & 21.04 \\
 & Lens 2 & 67.4735 & -29.9624 & 0.61 & 2 & 25.28 & 23.07 & 21.95 & 21.45 \\
DESI-074.9643-30.7236 & Lens & 74.9642 & -30.7236 & 0.4410 & 3 & 20.32 & 18.48 & 17.84 & 17.48 \\
 & Source & 74.9647 & -30.7227 & 1.4485 & 4 & 22.06 & 21.71 & - & 21.06 \\
DESI-077.0456-27.6752 & Source & 77.0491 & -27.6754 & 2.30996 & 4 & 24.35 & 23.68 & 23.75 & 23.59 \\
DESI-091.1807-41.8102 & Source & 91.1801 & -41.8101 & 2.5894 & 4 & 24.58 & 24.78 & 27.90 & - \\
DESI-315.3648-43.8104 & Lens & 315.3648 & -43.8107 & 0.452 & 1 & 21.50 & 20.46 & 19.90 & 19.57 \\
 & Source & 315.3648 & -43.8103 & 1.1985 & 4 & 21.71 & 21.80 & 22.30 & 22.69 \\
DESI-324.2124-62.6803 & Lens & 324.2066 & -62.6827 & 0.23 & 1 & 21.37 & 20.14 & - & 19.44 \\
DESI-324.2176-62.6801 & Lens & 324.2148 & -62.6776 & 0.2275 & 3 & 20.87 & 19.53 & 19.07 & 18.78 \\
DESI-324.5127-60.1291 & Source 1 & 324.5127 & -60.1291 & 0.84156 & 4 & 21.62 & 21.06 & - & 20.24 \\
 & Source 2 & 324.5152 & -60.1304 & 1.2581 & 3 & 21.71 & 21.54 & - & 21.50 \\
DESI-324.5146-60.1293 & Lens & 324.5073 & -60.1290 & 0.317 & 2 & 22.18 & 20.65 & 20.13 & 19.83 \\
 & Source 1 & 324.5127 & -60.1291 & 0.84156 & 4 & 21.59 & 21.00 & 20.38 & 20.17 \\
 & Source 2 & 324.5145 & -60.1292 & 1.2599 & 4 & 21.68 & 21.53 & 21.53 & 21.46 \\
DESI-335.9189-63.4914 & Lens & 335.9213 & -63.4922 & 0.683 & 1 & 22.43 & 20.66 & - & 19.17 \\
 & Source & 335.9187 & -63.4906 & 4.33815 & 2 & 25.09 & 24.97 & - & 25.21 \\
DESI-338.1418-60.002 & Lens & 338.1424 & -60.0002 & 0.553 & 1 & 20.51 & 19.71 & 18.91 & 18.53 \\
 & Source & 338.1448 & -59.9994 & 0.9442 & 4 & 22.46 & 22.36 & - & 21.68 \\
DESI-341.1432-59.1531 & Lens & 341.1450 & -59.1528 & 0.8882 & 2 & 23.66 & 21.55 & 20.33 & 19.59 \\
DESI-343.6329-50.4884 & Source & 343.6329 & -50.4884 & 0.2857 & 4 & 18.52 & 18.03 & - & 17.62 \\
DESI-350.6766-64.1644 & Source & 350.6765 & -64.1648 & 1.540 & 4 & 19.79 & 19.34 & 19.07 & 18.72 \\
DESI-354.0277-53.8763 & Lens & 354.0298 & -53.8766 & 0.516 & 3 & 19.94 & 18.13 & 17.35 & 16.96 \\
 & Source & 354.0277 & -53.8762 &  1.15103 & 4 & 22.12 & 21.66 & - & 20.65 \\
DESI-355.2727-57.2679 & Lens & 355.2727 & -57.2662 & 0.4410 & 3 & 19.95 & 18.44 & 17.82 & 17.42 \\
 & Source & 355.2707 & -57.2664 & 1.24752 & 4 & 20.33 & 19.92 & 19.66 & 19.15 \\
DESI-355.6468-46.8755 & Lens & 355.6501 & -46.8744 & 0.2805 & 4 & 18.17 & 16.68 & 16.18 & 15.89 \\
 & Source & 355.6468 & -46.8755 & 1.6076 & 4 & 22.47 & 22.41 & 22.08 & 21.62 \\
DESI-356.7894-62.7765 & Lens & 356.7868 & -62.7787 & 0.35569 & 4 & 20.57 & 19.32 & - & 18.57 \\
DESI-359.7003-61.433 & Lens & 359.6974 & -61.4323 & 0.37243 & 4 & 21.11 & 19.56 & 19.01 & 18.71 \\
\end{longtable}
\FloatBarrier
\twocolumn

\begin{appendix}




\onecolumn
\begin{figure}
\centering
\begin{tabular}{c}
\includegraphics{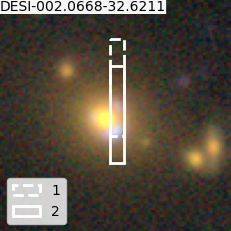} \\
\includegraphics[width=\textwidth]{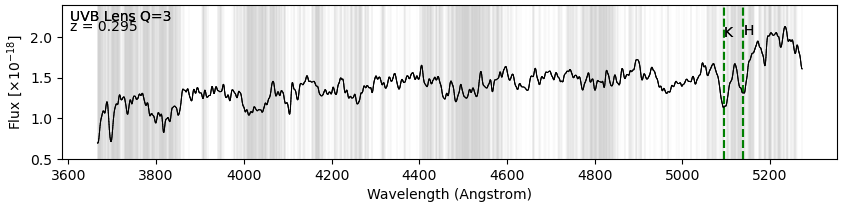} \\
\includegraphics[width=\textwidth]{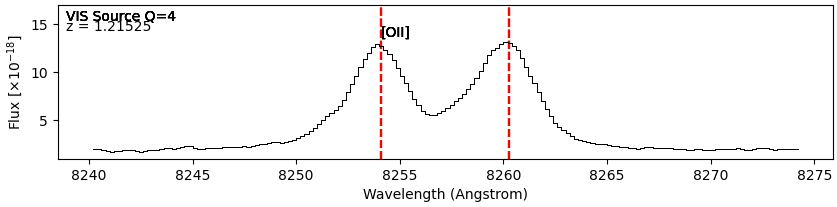} \\
\end{tabular}
\caption{ {\it Top: } RGB image of gravitational lens system DESI-002.0668-32.6211 taken from DESI Legacy Imaging Surveys. The solid and dashed white boxes represent the slit position for the two observations of the system. The central lens galaxy and the two images of the source in blue are observed in both observations {\it Bottom: } Spectra of the lens and the source. The shape of the continuum and the $H$ and $K$ absorption lines give us confidence in the estimate of the lens redshift $z=0.295$, in spite of the sky lines and imaging artifacts masked in grey. The source redshift is identified without any ambiguity thanks to the presence of the $[OII]$ doublet in the VIS arm spectrum. }
\end{figure}
\clearpage
\begin{figure}
\centering
\begin{tabular}{c}
\includegraphics{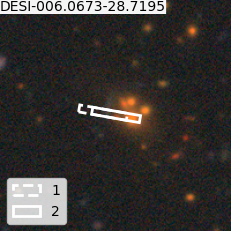} \\
\includegraphics[width=\textwidth]{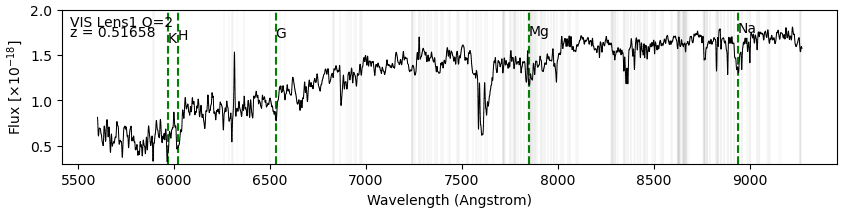} \\
\includegraphics[width=\textwidth]{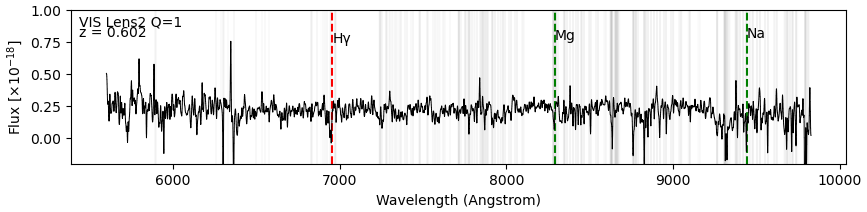} \\
\includegraphics[width=\textwidth]{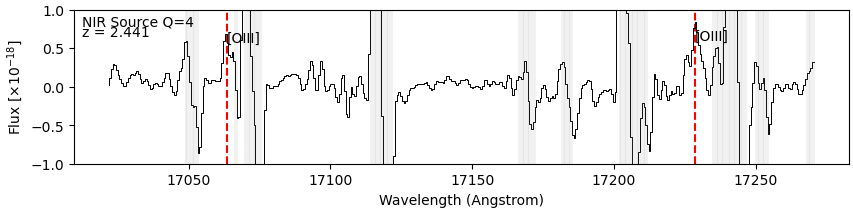} \\
\end{tabular}
\caption{ {\it Top: } RGB image of gravitational lens system DESI-006.0673-28.7195 taken from DESI Legacy Imaging Surveys. The two slit positions (solid and dashed white boxes) cover the central (Lens1) and a background (Lens2) galaxies {\it Bottom: } Spectra of the two lenses and the source. Strong $H$ and $K$ lines in the spectrum of the second VIS observation of Lens1, allow us to give a redshift $z = 0.51658$. This redshift is slightly smaller than the redshift of the background lens (Lens2) $z=0.602$ that we infer from $H\gamma$, $Mg$ and $Na$ absorption lines. There is also a very faint $H\alpha$ line in the two NIR arm spectra confirming this redshift. We infer the redshift of the source $z=2.441$ from the $[OIII]$ doublet in the NIR arm spectra of the two observations.}
\end{figure}
\clearpage
\begin{figure}
\begin{tabular}{c}
\includegraphics{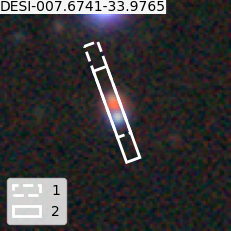} \\
\includegraphics[width=\textwidth]{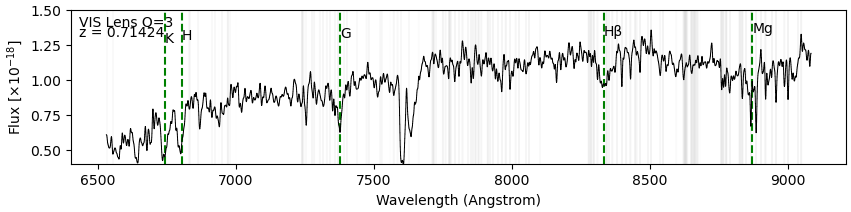} \\
\includegraphics[width=\textwidth]{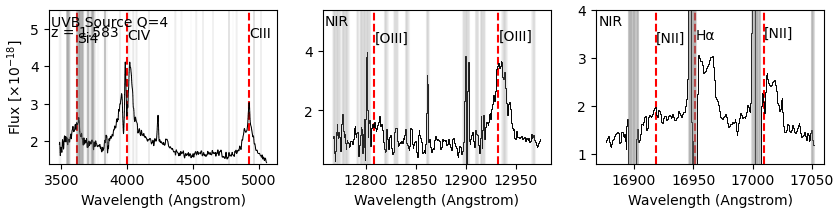} \\
\end{tabular}
\caption{ {\it Top: } RGB image of gravitational lens system DESI-007.6741-33.9765 taken from DESI Legacy Imaging Surveys. The two slit positions (solid and dashed white boxes) cover the lens in red, and the two images of the sources, one bright to the south, and one faint to the north. {\it Bottom: } The lens spectrum presents bright $H$, $K$ and $G$ absorption lines. The strong $H\beta$ and weak $Mg$ lines suggest the presence of a young stellar population (1-3 Gyr) in a low metallicity environment mixed with the old population traced by the $H$ and $K$ lines. The source spectrum is typical of a QSO with broad $SiIV$, $CIV$ and $CIII$ emission in the UV, and a strong $[OIII]$ doublet in the NIR. The $H\alpha$ line is present as well, but covered by a sky line. Combined with the $NII$ doublet (more pronounced in the 2D spectrogram), it seems that the $H\alpha$, $[NII]$ complex is slightly shifted in velocity ($v \sim +230$ km/s) characteristic of an outflow.}
\end{figure}
\clearpage
\begin{figure}
\begin{tabular}{c}
\includegraphics{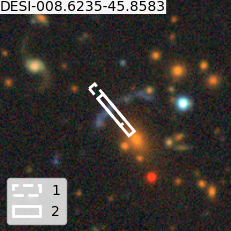} \\
\includegraphics[width=\textwidth]{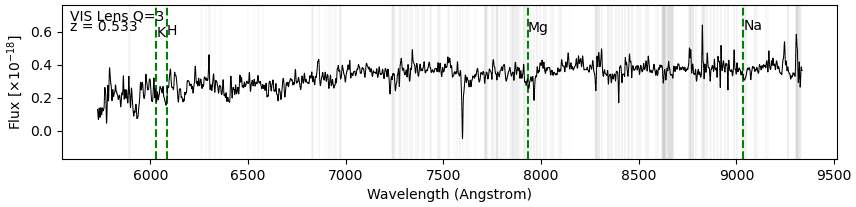} \\
\includegraphics[width=\textwidth]{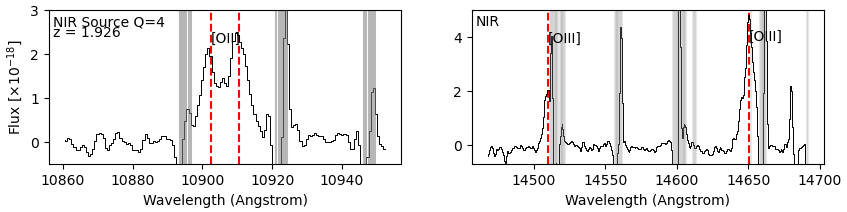} \\
\end{tabular}
\caption{ {\it Top: } RGB image of gravitational lens system DESI-008.6235-45.8583 taken from DESI Legacy Imaging Surveys. The two slit positions (solid and dashed white boxes) cross the $\sim 15\arcsec$ giant arc of the source. The main lens galaxy is observed only in the second exposure. {\it Bottom: } We infer the redshift $z=0.533$ of the lens from obvious $H$, $K$ and $Mg$ lines. At this redshift, there is maybe a $[NII]$ line at $\lambda=10324\AA$ and $H\alpha$ in absorption in the NIR spectrum, but this could be image artifacts. The redshift of the source $z=1.926$ is determined from the $[OII]$ and $[OIII]$ lines.}
\end{figure}
\clearpage
\begin{figure}
\begin{tabular}{c}
\includegraphics{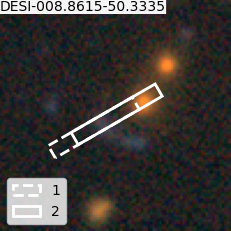} \\
\includegraphics[width=\textwidth]{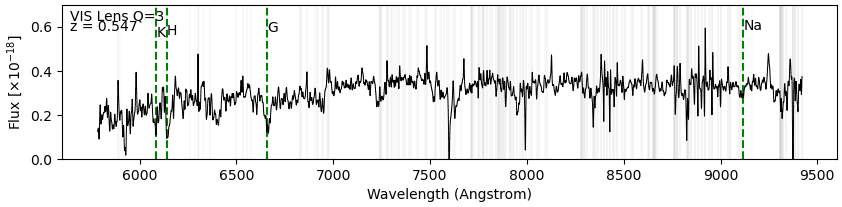} \\
\includegraphics[width=\textwidth]{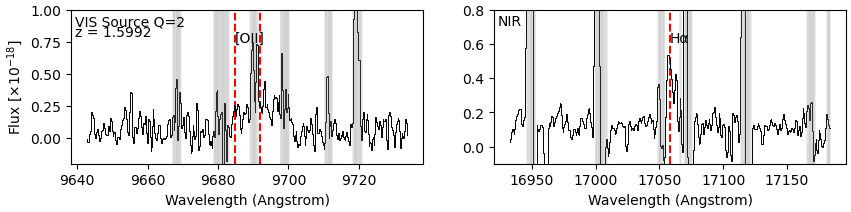} \\
\end{tabular}
\caption{ {\it Top: } RGB image of gravitational lens system DESI-008.8615-50.3335 taken from DESI Legacy Imaging Surveys. The two slit positions (solid and dashed white boxes) cross the $\sim 6\arcsec$ giant arc of the source. The main lens galaxy is observed only in the second exposure.{\it Bottom: } We determine the lens redshift $z=0.547$ thanks to the presence of prominent $H$, $K$ and $G$ lines, as well as a small $Na$ line. A strong $H\alpha$ emission line in the source NIR arm spectrum gives us confidence in the redshift $z=1.5992$, confirmed by a faint $[OII]$ emission in the VIS arm spectrum. }
\end{figure}
\clearpage
\begin{figure}
\begin{tabular}{c}
\includegraphics{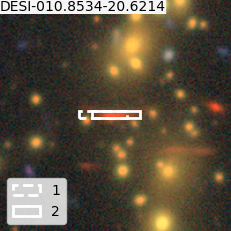} \\
\includegraphics[width=\textwidth]{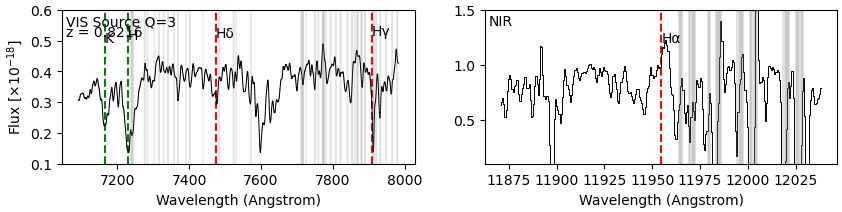} \\
\end{tabular}
\caption{ {\it Top: } RGB image of gravitational lens system DESI-010.8534-20.6214 taken from DESI Legacy Imaging Surveys. This galaxy cluster at redshift $z=0.2924$ is also known as Abell 2813, SPT-CLJ0043-2037 or PSZ2 G106.87-83.83. It is part of several cluster surveys (e.g. RELICS, \cite{coe2019}; \cite{chexmate2021}; MeerKAT Galaxy Cluster Legacy Survey, \cite{knowles2022}; CHANCES, \cite{sifon2025}). The two slit positions (solid and dashed white boxes) are aligned with the $\sim 8\arcsec$ red giant arc. {\it Bottom: } We detect an extended $H\alpha$ emission line at the location of the giant arc in the NIR arm 2D spectrogram, leading to a redshift $z=0.8216$. This redshift is confirmed by strong $H$, $K$ and faint $H\delta$, $H\gamma$ absorption lines in the VIS arm 1D spectrum.}
\label{fig:DESI-10.8534}
\end{figure}
\clearpage
\begin{figure}
\begin{tabular}{c}
\includegraphics{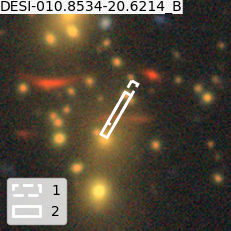} \\
\includegraphics[width=\textwidth]{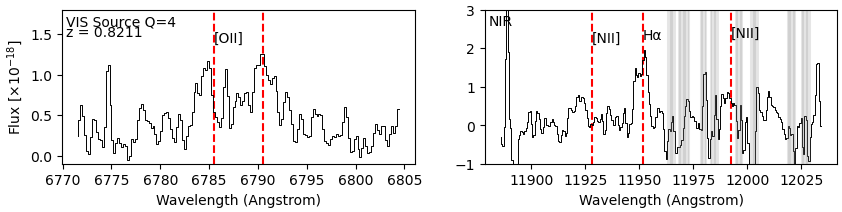} \\
\includegraphics[width=\textwidth]{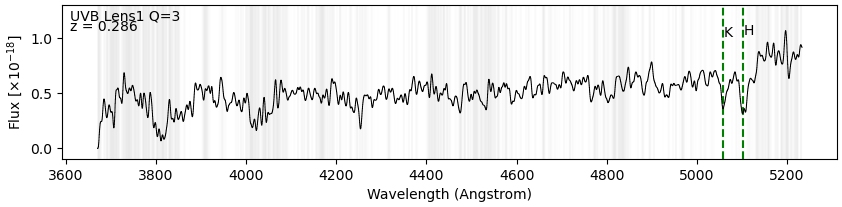} \\
\includegraphics[width=\textwidth]{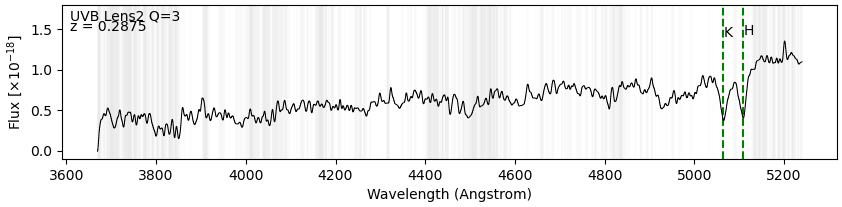} \\
\end{tabular}
\caption{ {\it Top: } RGB image of gravitational lens system DESI-010.8534-20.6214\_B taken from DESI Legacy Imaging Surveys. This observation focuses on a $\sim 10\arcsec$ giant arc southward of the one in Fig.~\ref{fig:DESI-10.8534}. The two slit positions(solid and dashed white boxes) are crossing it. Slit position 1 also contains a small cluster galaxy northward of the arc (Lens1), and slit position 2 contains a bright one southward (Lens2). {\it Bottom: } We detect $H\alpha$ and $[OII]$ in emission in the NIR and the VIS arm 2D spectrograms respectively. The S/N is better in the slit 2 exposure. These detections correspond to a redshift $z=0.8211$, consistent with previous measurements only based on the $H$ and $K$ lines \citep{may2013}. This is slightly smaller ($\Delta v \sim 180\ \mathrm{km/s}$) than the redshift of the giant arc northward suggesting a group of galaxies behind this cluster.  The redshifts $z=0.286$ for the galaxy northward of the arc (Lens1) and $z=0.2875$ for the one southward (Lens2) are clearly determined from the $H$ and $K$ absorption lines and the shape of the continuum in the VIS arm 1D spectra.}
\end{figure}
\clearpage
\begin{figure}
\begin{tabular}{c}
\includegraphics{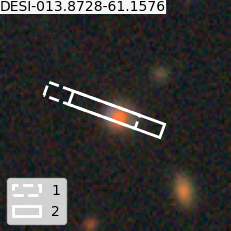} \\
\includegraphics[width=\textwidth]{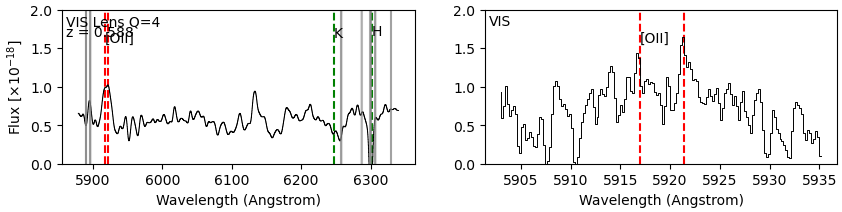} \\
\includegraphics[width=\textwidth]{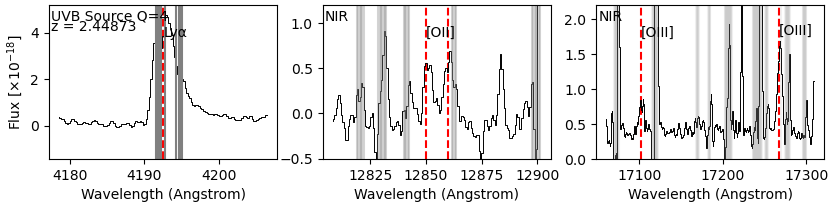} \\
\end{tabular}
\caption{ {\it Top: } RGB image of gravitational lens system DESI-013.8728-61.1576 taken from DESI Legacy Imaging Surveys. The two slit positions (solid and dashed white boxes) cover the two images of the blue source on both sides of the lens in red {\it Bottom: } For the lens, the $[OII]$ emission in the VIS arm spectrum, confirmed by the $H$ and $K$ absorption lines allow us to infer a redshift $z=0.5880$. For the source, the $[OII]$ and $[OIII]$ doublets in the NIR, confirmed by the $Ly\alpha$ line in the UV allow us to infer a redshift of $z=2.44873$.}
\end{figure}
\clearpage
\begin{figure}
\begin{tabular}{c}
\includegraphics{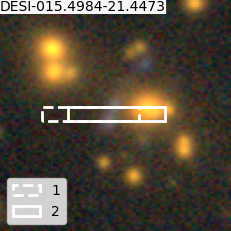} \\
\includegraphics[width=\textwidth]{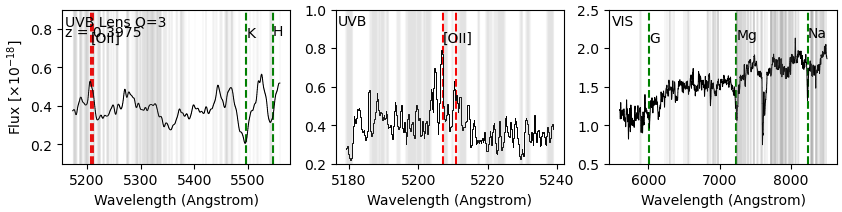} \\
\includegraphics[width=\textwidth]{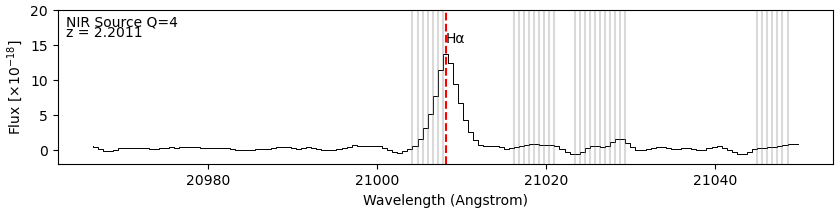} \\
\end{tabular}
\caption{ {\it Top: } RGB image of gravitational lens system DESI-015.4984-21.4473 taken from DESI Legacy Imaging Surveys. In this group scale lens, the 2 slit positions cover both the central galaxy and the $\sim 10\arcsec$ long arc. {\it Bottom: } The deep $H$ and $K$ and the small $[OII]$ doublet in the UV arm spectrum of the lens allow us to infer a redshift $z=0.3975$. For the source, a strong $H\alpha$ line in the NIR arm spectrum, and faint $Ly\alpha$ and $NV$ lines in the UV allow us to infer a redshift $z=2.2011$. }
\end{figure}
\clearpage
\begin{figure}
\begin{tabular}{c}
\includegraphics{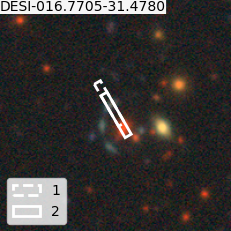} \\
\includegraphics[width=\textwidth]{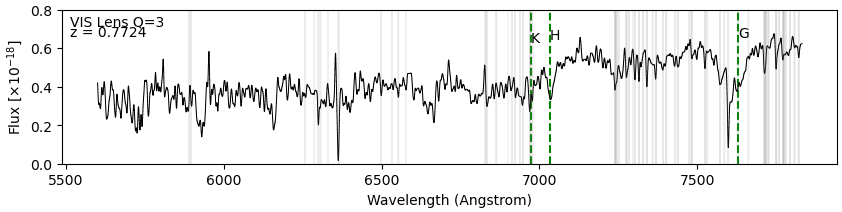} \\
\includegraphics[width=\textwidth]{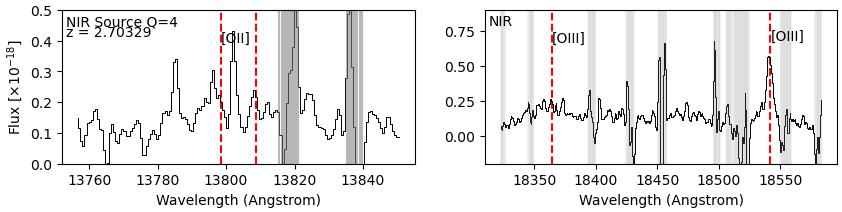} \\
\end{tabular}
\caption{ {\it Top: } RGB image of gravitational lens system DESI-016.7705-31.4780 taken from DESI Legacy Imaging Surveys. In this group scale lens, the 2 slit positions cover both the central red galaxy and the third image of the multiply imaged source in blue. {\it Bottom: } The $H$ and $K$, as well as the faint $G$ absorption lines in the VIS arm spectrum of the lens allow us to infer a redshift $z=0.7724$. For the source, strong $[OIII]$ emission lines and a faint $[OII]$ doublet in the NIR arc spectrum allow us to infer a redshift $z=2.70329$. }
\end{figure}
\clearpage
\begin{figure}
\begin{tabular}{c}
\includegraphics{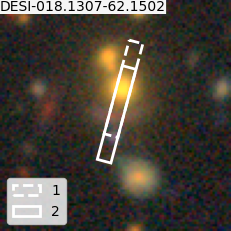} \\
\includegraphics[width=\textwidth]{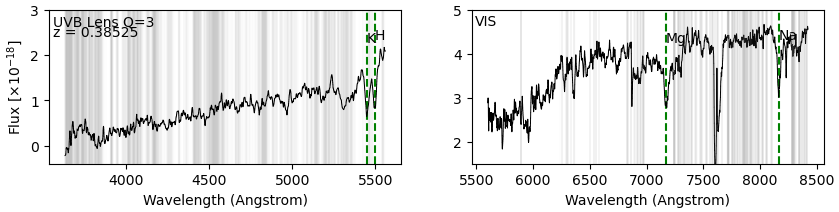} \\
\includegraphics[width=\textwidth]{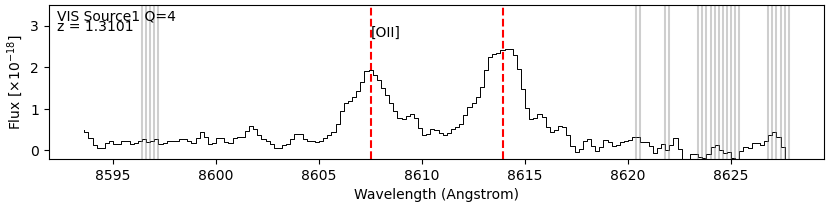} \\
\includegraphics[width=\textwidth]{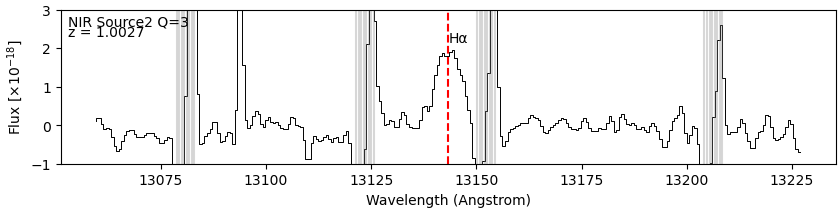} \\
\end{tabular}
\caption{ {\it Top: } RGB image of gravitational lens system DESI-018.1307-62.1502 taken from DESI Legacy Imaging Surveys. The 2 slit positions cover both the central galaxy and two sources southward, a giant arc (Source 1) and a galaxy not necessarily lensed (Source 2). {\it Bottom: } The $H$ and $K$ lines in the UVB arm spectrum and the $Mg$ and $Na$ absorption lines in the $VIS$ arm spectrum of the lens allow us to infer a redshift $z=0.37525$. The strong $[OII]$ doublet in the VIS arm spectrum of Source 1 allow us to infer a redshift $z=1.3101$. For Source 2, a strong $H\alpha$ line in the NIR arm spectrum corresponds to a redshift $z=1.0027$. }
\end{figure}
\clearpage
\begin{figure}
\begin{tabular}{c}
\begin{tabular}{cc}
\includegraphics{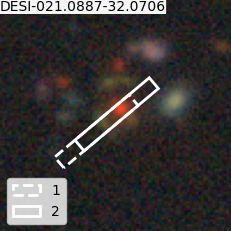} &
\includegraphics{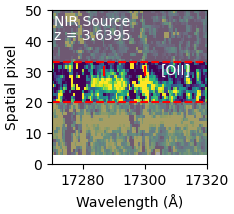} \\
\end{tabular} \\
\includegraphics[width=\textwidth]{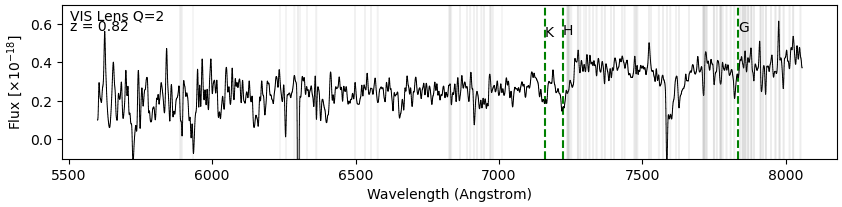} \\
\includegraphics[width=\textwidth]{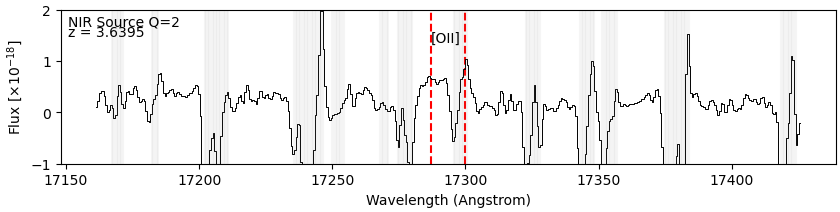} \\
\end{tabular}
\caption{ {\it Top: } RGB image of gravitational lens system DESI-021.0887-32.0706 taken from DESI Legacy Imaging Surveys. The 2 slit positions cover both the central red galaxy and one image of the threefold multiply imaged source in blue. On the right, the window extraction of the $[OII]$ doublet of the source with marks at $\lambda\lambda = 3726.032\AA, 3728.815\AA$. {\it Bottom: } For the lens, the $H$ and $K$ lines as well as the faint $G$ absorption line allow to infer a redshift $z=0.82$. For the source, the clear $[OII]$ doublet in the 2D spectrum is less evident in the collapsed NIR arc spectrum, but still allow us to infer a redshift $z=3.6395$. }
\end{figure}
\clearpage
\begin{figure}
\begin{tabular}{c}
\includegraphics{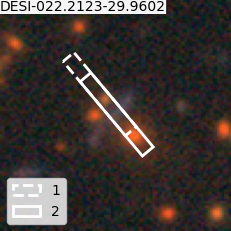} \\
\includegraphics[width=\textwidth]{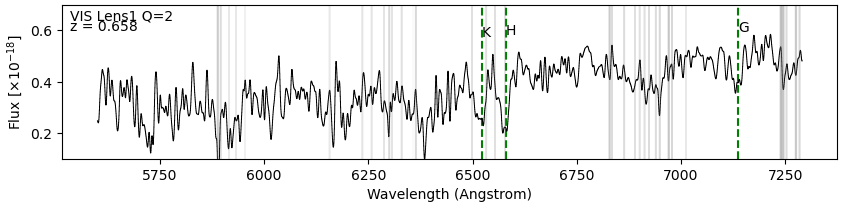} \\
\includegraphics[width=\textwidth]{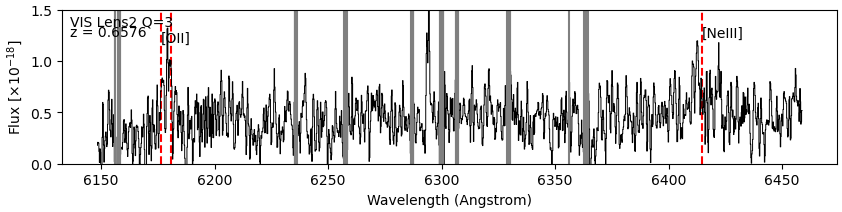} \\
\includegraphics[width=\textwidth]{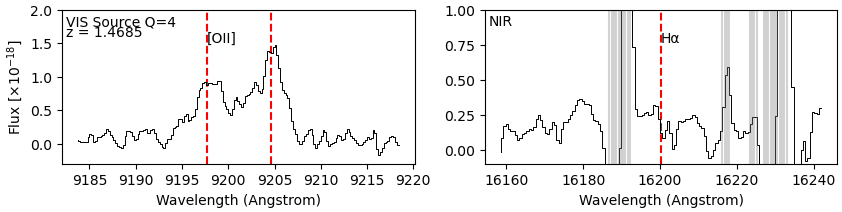} \\
\end{tabular}
\caption{ {\it Top: } RGB image of gravitational lens system DESI-022.2123-29.9602 taken from DESI Legacy Imaging Surveys. The 2 slit positions cover the central galaxy (Lens1), a satellite galaxy (Lens2), and in between a $\sim 6\arcsec$ long arc. {\it Bottom: } For lens1, the $H$ and $K$ lines as well as the faint $G$ absorption line allow to infer a redshift $z=0.658$. For lens2, an $[OII]$ doublet and a $[NeIII]$ emission suggest a redshift $z=0.6576$. For the source, a strong $[OII]$ doublet and a $H\alpha$ emission allow us to infer a redshift $z=1.4685$. }
\end{figure}
\clearpage
\begin{figure}
\begin{tabular}{c}
\includegraphics{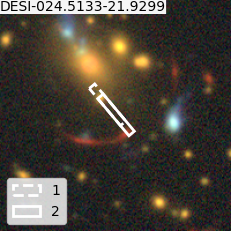} \\
\includegraphics[width=\textwidth]{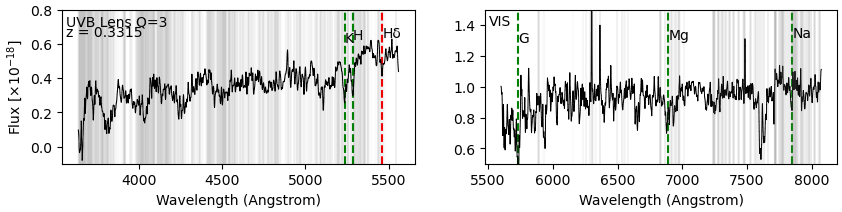} \\
\includegraphics[width=\textwidth]{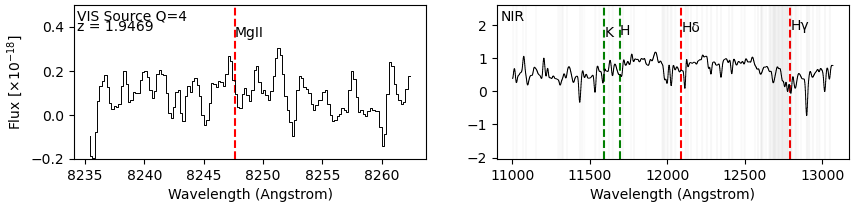} \\
\end{tabular}
\caption{ {\it Top: } RGB image of gravitational lens system DESI-024.5133-21.9299 taken from DESI Legacy Imaging Surveys. This lens system is also known MACS J0138.0-2155 \citep{ebeling2001} and the redshift of the red giant arc $z = 1.95$ was published in \cite{newman2018}. This arc is famous for hosting the Requiem supernovae \citep{rodney2021}. The slit in position 1 covers a satellite galaxy, and in position 2 it crosses the red giant arc. {\it Bottom: } The clear $H$ and $K$ lines in the UVB arm of the spectrum as well as the deep $G$, $Mg$ and $Na$ absorption lines in the VIS arm spectrum allow us to infer a redshift of $z=0.3315$ for the lens. This is slightly smaller than the central galaxy redshift $z=0.338$, which places it in the cluster foreground. The redshift of the source is more uncertain. We clearly detect the continuum in the NIR arm spectrum, but we do not detect any prominent emission line. The closest features $H$, $K$ and $H\delta$ lines in the NIR and $MgII$ in the VIS arm spectrum allow us to infer a redshift $z=1.9469$. }
\end{figure}
\clearpage
\begin{figure}
\begin{tabular}{c}
\includegraphics{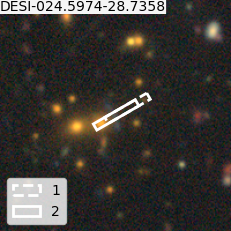} \\
\includegraphics[width=\textwidth]{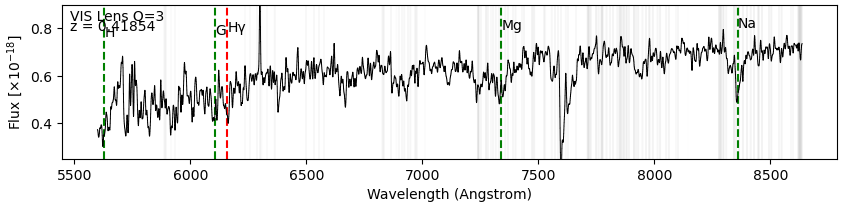} \\
\includegraphics[width=\textwidth]{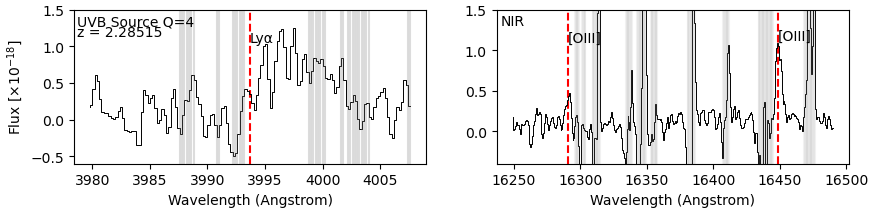} \\
\end{tabular}
\caption{ {\it Top: } RGB image of gravitational lens system DESI-024.5974-28.7358 taken from DESI Legacy Imaging Surveys. The 2 slit positions cover a bright satellite galaxy of the group and cross a $\sim 3\arcsec$ long arc. {\it Bottom: } For the lens, we detect $H$, $G$, $Mg$ and $Na$ absorption lines as well as $H\gamma$ very faint. $H\alpha$ also falls in the VIS arm spectrum but in absorption, and contaminated by sky lines. The $K$ line falls in-between the VIS and UVB arm spectra. This allows us to infer a redshift for the lens $z=0.41854$. For the source, strong $[OIII]$ doublet in the NIR arm spectrum allows us to infer a redshift $z=2.28515$. There is also a detection of $Ly\alpha$ in the UVB arm but contaminated by sky lines and shifted with respect to the systemic redshift of the galaxy.}
\end{figure}
\clearpage
\begin{figure}
\begin{tabular}{c}
\includegraphics{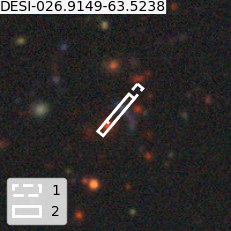} \\
\includegraphics[width=\textwidth]{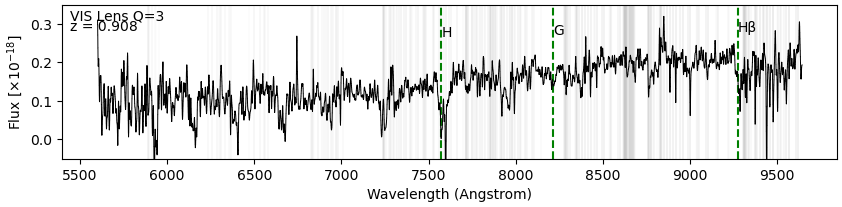} \\
\includegraphics[width=\textwidth]{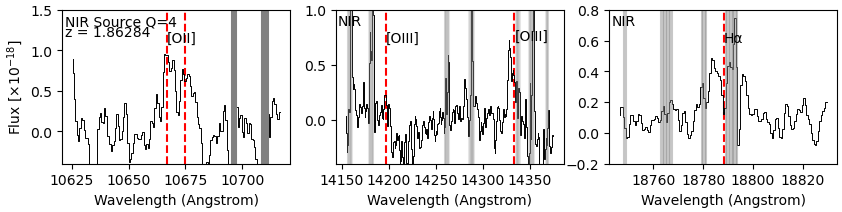} \\
\end{tabular}
\caption{ {\it Top: } RGB image of gravitational lens system DESI-026.9149-63.5238 taken from DESI Legacy Imaging Surveys. The 2 slit positions cross a $\sim 3\arcsec$ long blue arc, and the slit in position 2 covers the central red galaxy of the group. {\it Bottom: } For the lens, the shape of the continuum as well as the $H$ and $G$ absorption lines allow us to infer redshift $z=0.908$. A faint $H\beta$ emission line in a broader absorption system is consistent with this redshift. For the source, clear $[OII]$ and $[OIII]$ doublets, as well as a $H\alpha$ emission, although contaminated by sky lines allow us to infer a redshift $z=1.86284$.} 
\end{figure}
\clearpage
\begin{figure}
\begin{tabular}{c}
\includegraphics{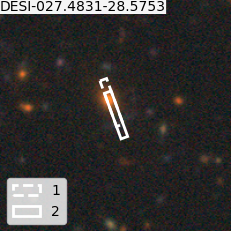} \\
\includegraphics[width=\textwidth]{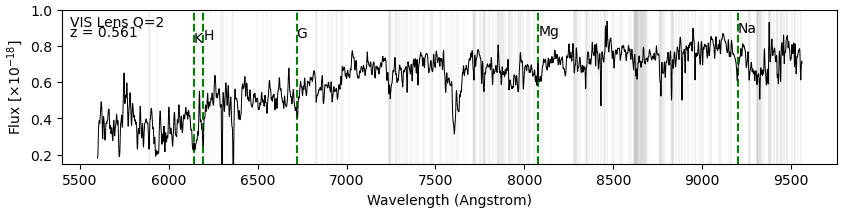} \\
\includegraphics[width=\textwidth]{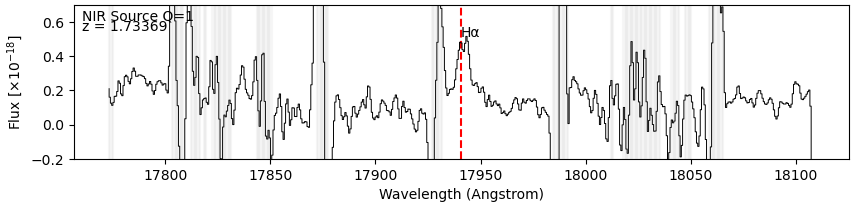} \\
\end{tabular}
\caption{ {\it Top: } RGB image of gravitational lens system DESI-027.4831-28.5753 taken from DESI Legacy Imaging Surveys. The 2 slits cover the bright yellow lens, and a faint $\sim 2\arcsec$ long arc southward. {\it Bottom: } For the lens, $H$, $K$, $G$, $Mg$ and $Na$ absorption lines allow us to infer a redshift $z=0.561$. For the source, we detect a single bright line in the 2 exposures of the 2D NIR arm spectrum that we attribute to $H\alpha$, allowing us to infer a redshift $z=1.73369$ with low confidence. }
\end{figure}
\clearpage
\begin{figure}
\begin{tabular}{c}
\begin{tabular}{cc}
\includegraphics{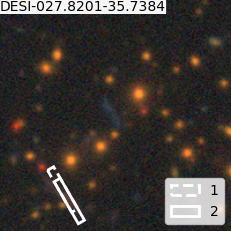} &
\includegraphics{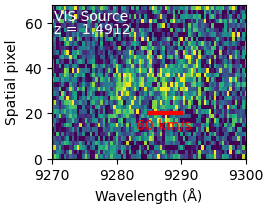} 
\end{tabular} \\
\includegraphics[width=\textwidth]{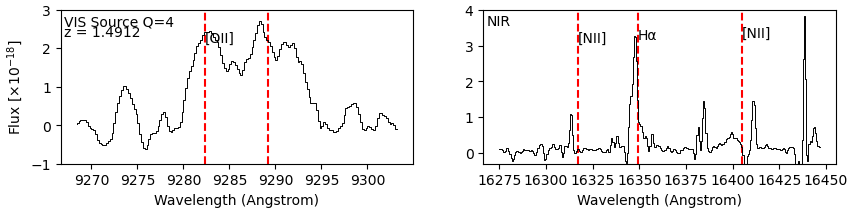} \\
\end{tabular}
\caption{ {\it Top: } RGB image of gravitational lens system DESI-027.8201-35.7384 taken from DESI Legacy Imaging Surveys, also known as SPT-CL J0151-354 \citep{bleem2020} or ACT-CL J0151.2-3544 \citep{hilton2021} at photometric redshift $z=0.53$. The 2 slit positions cover a red $\sim 2\arcsec$ long arc, in the southern side of this galaxy cluster {\it Bottom: } We detect the $[OII]$, $H\alpha$ and $[NII]$ emission lines in the VIS and NIR spectra, allowing us to infer a redshift $z=1.4912$. In addition, the $[OII]$ and $H\alpha$ lines present the same velocity gradient suggesting that this galaxy is rotating at $\sim 30$ km/s.}
\end{figure}
\clearpage
\begin{figure}
\begin{tabular}{c}
\includegraphics{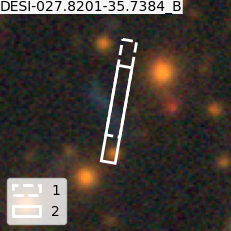} \\
\includegraphics[width=\textwidth]{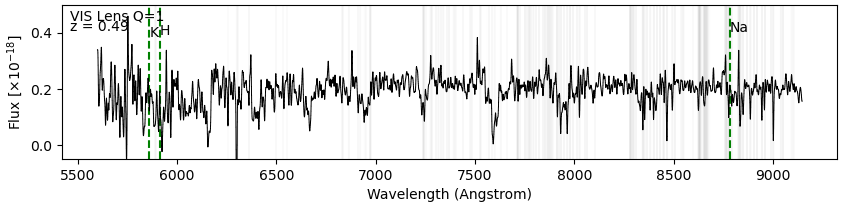} \\
\includegraphics[width=\textwidth]{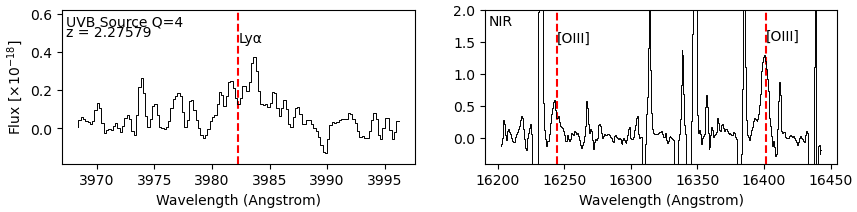} \\
\end{tabular}
\caption{ {\it Top: } RGB image of gravitational lens system DESI-027.8201-35.7384\_B taken from DESI Legacy Imaging Surveys, also known as SPT-CL J0151-354 \citep{bleem2020} or ACT-CL J0151.2-3544 \citep{hilton2021} at photometric redshift $z=0.53$. The 2 slit positions cross a blue arc, and slit 2 covers a cluster member galaxy in yellow. {\it Bottom: } The spectrum of the lens present almost no feature, apart from a faint continuum and uncertain $K$, $H$ and $Na$ lines, hence allowing us to infer a redshift $z=0.49$. Note that this observation was performed in degraded conditions with a seeing of 2\arcsec. For the source, the clear $[OIII]$ and $Ly\alpha$ lines in the NIR and UVB spectra allow us to infer a redshift $z=2.27579$. }
\end{figure}
\clearpage
\begin{figure}
\begin{tabular}{c}
\begin{tabular}{cc}
\includegraphics{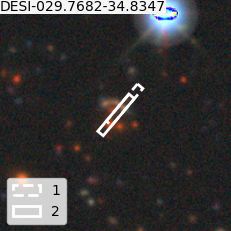} &
\includegraphics{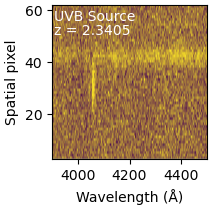} 
\end{tabular} \\
\includegraphics[width=\textwidth]{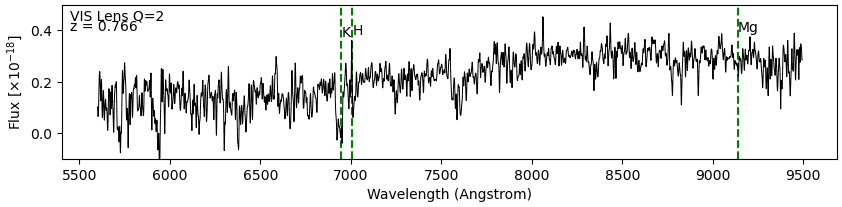} \\
\includegraphics[width=\textwidth]{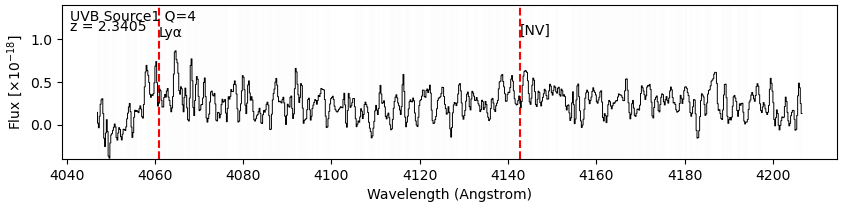} \\
\includegraphics[width=\textwidth]{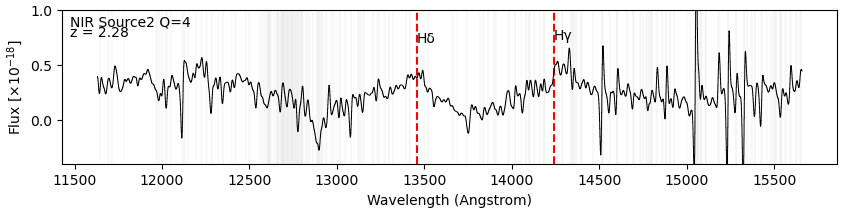} \\
\end{tabular}
\caption{ {\it Top: } RGB image of gravitational lens system DESI-029.7682-34.8347 taken from DESI Legacy Imaging Surveys. The 2 slit positions cover one of the red central galaxy of the group and cross a $\sim 4\arcsec$ long arc. {\it Bottom: } For the lens, the continuum as well as $H$, $K$ and $Mg$ lines allow us to infer a redshift $z=0766$. For the source, a $\sim 3\arcsec$ extended $Ly\alpha$ emission along the slit suggests the presence of a large $Ly\alpha$ blob at redshift $z=2.3405$. In addition, coincident with the red part of the arc, $H\delta$ and $H\gamma$ lines suggests a QSO emission at redshift $z=2.28$. }
\end{figure}
\clearpage
\begin{figure}
\begin{tabular}{c}
\includegraphics{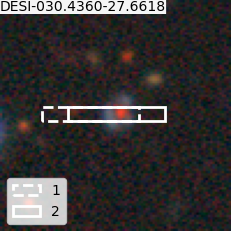} \\
\includegraphics[width=\textwidth]{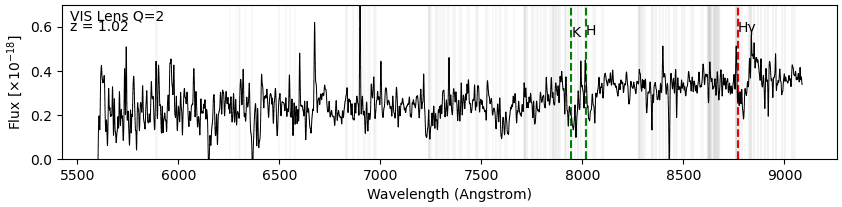} \\
\includegraphics[width=\textwidth]{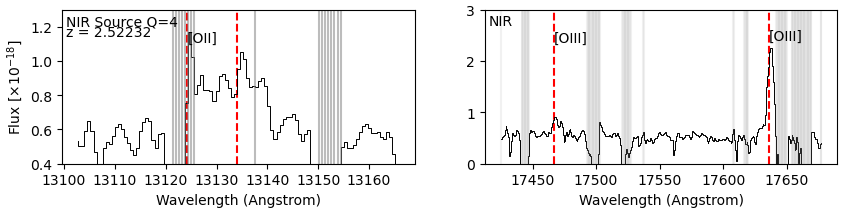} \\
\end{tabular}
\caption{ {\it Top: } RGB image of gravitational lens system DESI-030.4360-27.6618 taken from DESI Legacy Imaging Surveys. This system is also part of HST program 15867 (PI: X. Huang). The 2 slit positions cover both the red central galaxy and the 2 blue multiple images on both sides of it. {\it Bottom: } For the lens, $H$ and $H$ absorption lines as well as a faint $H\gamma$ emission allow us to infer a redshift $z=1.02$. For the source, a strong $[OIII]$ and a faint $[OII]$ emission in the NIR allow us to infer a redshift $z=2.52232$.}
\end{figure}
\clearpage
\begin{figure}
\begin{tabular}{c}
\includegraphics{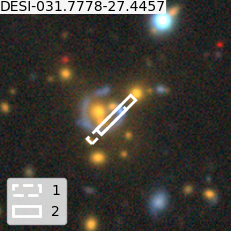} \\
\includegraphics[width=\textwidth]{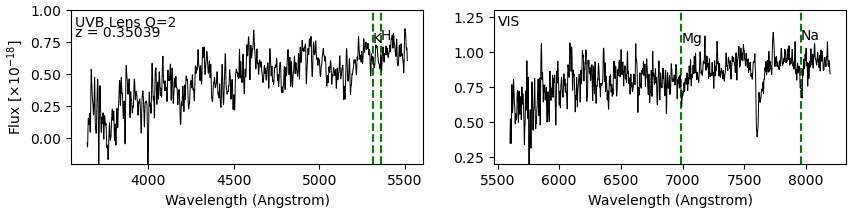} \\
\includegraphics[width=\textwidth]{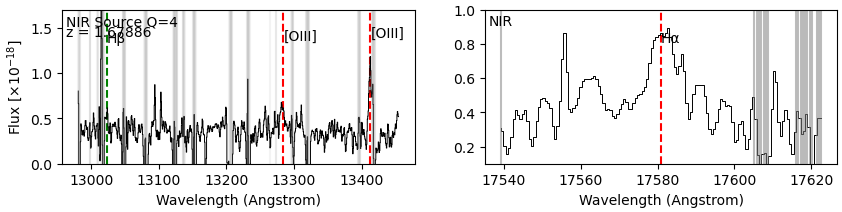} \\
\end{tabular}
\caption{ {\it Top: } RGB image of gravitational lens system DESI-031.7778-27.4457 taken from DESI Legacy Imaging Surveys. This system is also part of \cite{lin2025} sample, but without redshift for the source. The 2 slit positions both cover a bright blue counter image of the giant arc, and a satellite galaxy of the group.  {\it Bottom: } For the lens, the $H$ and $K$ in the UVB and $Mg$ and $Na$ absorption lines in the VIS allow us to infer a redshift $z=0.3509$, slightly lower ($\sim -600\ \mathrm{km/s}$) than the group redshift $z=0.354$ reported in \cite{lin2025}. For the source, bright $[OIII]$, $H\alpha$ and $H\beta$ emission lines in the NIR allow us to infer a redshift $z=1.67886$. }
\end{figure}
\clearpage
\begin{figure}
\begin{tabular}{c}
\includegraphics{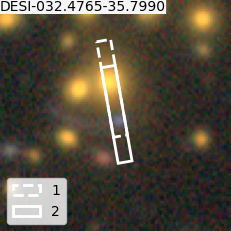} \\
\includegraphics[width=\textwidth]{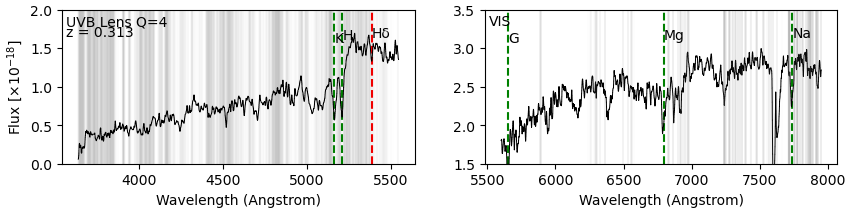} \\
\includegraphics[width=\textwidth]{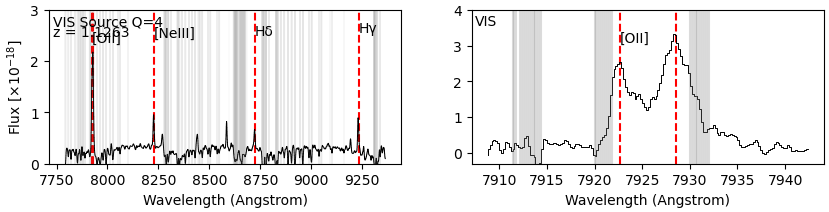} \\
\end{tabular}
\caption{ {\it Top: } RGB image of gravitational lens system DESI-032.4765-35.7990 taken from DESI Legacy Imaging Surveys. This galaxy cluster is also known as ACT-CL J0209.9-3548 \citep{hilton2021}. The 2 slit positions both cover the central galaxy and the bright side of a giant arc southward. {\it Bottom: } For the lens, the deep $H$, $K$, $G$, $Mg$ and $Na$ absorption lines allow us to infer a redshift $z=0.313$. For the source, the $[OII]$, $[NeIII]$, $H\delta$ and $H\gamma$ emission lines allow us to infer a redshift $z=1.1263$. We also note a $MgII$ emission in VIS spectrum, but no $H\alpha$ emission in the NIR. We estimate $[OII] / [NeIII] \sim 10$. This suggests a galaxy with a peculiar state of evolution, either a post-starburst or a galaxy with  escaping ionising photons. Unfortunately, the UVB spectral coverage does not allow to observe $Ly\alpha$ at this redshift to conclude. }
\end{figure}
\clearpage
\begin{figure}
\begin{tabular}{c}
\includegraphics{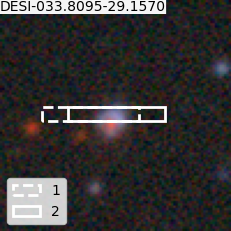} \\
\includegraphics[width=\textwidth]{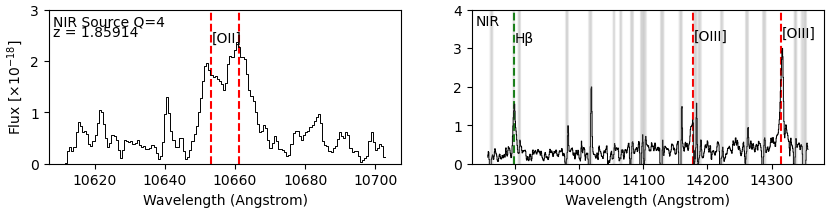} \\
\end{tabular}
\caption{ {\it Top: } RGB image of gravitational lens system DESI-033.8095-29.1570 taken from DESI Legacy Imaging Surveys. This system looks like an Einstein ring, and is part of the \cite{jacobs2019} DES lens sample. The 2 slit positions cover the upper and bright part of the Einstein ring, and a fraction of the lens galaxy. {\it Bottom: } For the source, the $[OII]$, $[OIII]$ and $H\beta$ emissions in the NIR allow us to infer a redshift $z=1.85914$. The $H\alpha$ line is buried into sky lines. For the lens, the photometric redshift of the lens is $z_\mathrm{phot} = 0.94$ \citep{adnan2025}. In the NIR spectrum, we identify an emission line at $\lambda = 16802\AA$, that could be attributed to $H\alpha$ at $z=1.56$, or a Paschen line at $z\sim0.82$, if we were willing to match it with the photometric redshift. }
\end{figure}
\clearpage
\begin{figure}
\begin{tabular}{c}
\begin{tabular}{cc}
\includegraphics{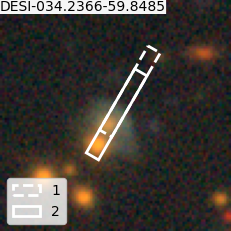} &
\includegraphics{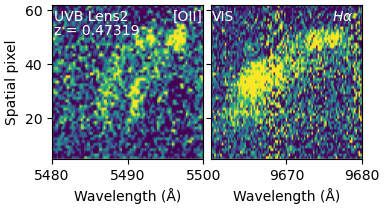} 
\end{tabular} \\
\includegraphics[width=\textwidth]{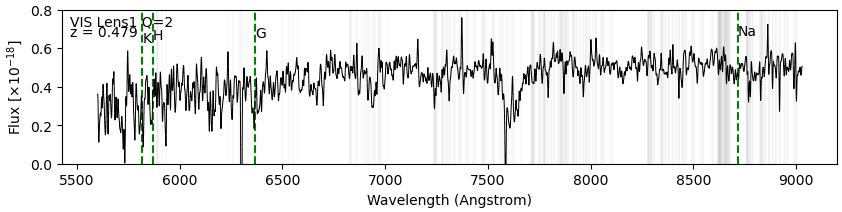} \\
\includegraphics[width=\textwidth]{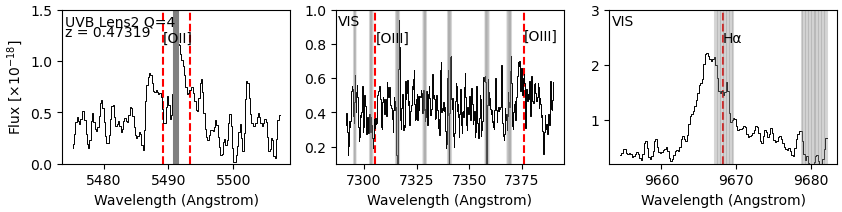} \\
\end{tabular}
\caption{ {\it Top: } RGB image of gravitational lens system DESI-034.2366-59.8485 taken from DESI Legacy Imaging Surveys. The slit position 2 covers the central galaxy of the group (Lens 1), its satellite (Lens 2) and the arc. The slit position 1 only covers the satellite galaxy and the arc. The strong velocity field in the 2D spectrogram of the satellite galaxy suggests a rotating disk. Therefore, the blue region would not be an arc but the spiral arms of the satellite galaxy. {\it Bottom: } For the central galaxy, the weak $H$, $K$, $G$ and $Na$ absorption lines allow us to infer a redshift $z=0.479$ with low confidence. For the satellite galaxy, the velocity field visible both with the $[OII]$ and $H\alpha$ emission lines, as well as the weak $[OIII]$ doublet allow us to infer a redshift $z=0.47319$. }
\end{figure}
\clearpage
\begin{figure}
\begin{tabular}{c}
\begin{tabular}{cc}
\includegraphics{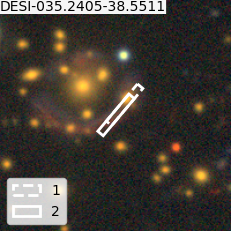} &
\includegraphics{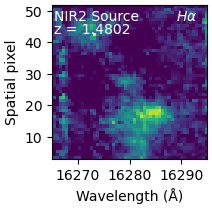} 
\end{tabular} \\
\includegraphics[width=\textwidth]{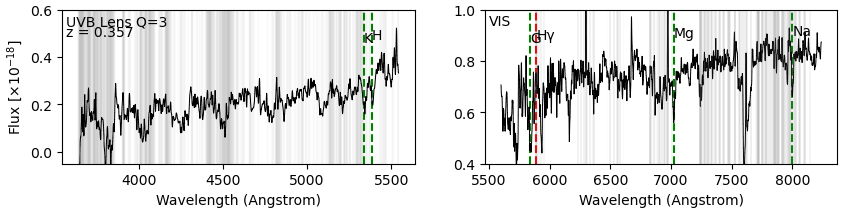} \\
\includegraphics[width=\textwidth]{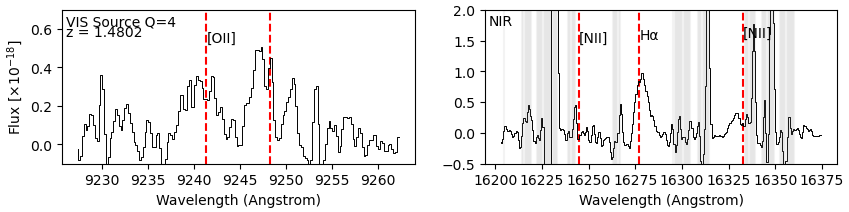} \\
\end{tabular}
\caption{ {\it Top: } RGB image of gravitational lens system DESI-035.2405-38.5511 taken from DESI Legacy Imaging Surveys. This galaxy cluster lens system is also known as the Molten Ring \citep{diaz2021}. The slit position 2 covers the right side of the giant arc and a satellite galaxy. The slit position 1 only covers the satellite galaxy. {\it Bottom: } For the lens, the $H$, $K$ in UVB and $G$, $Mg$ and $Na$ absorption lines in VIS allow us to infer a redshift $z0.357$, consistent with the Gemini data $z=0.35732$ \citep{diaz2021}.  For the source, the $[OII]$, $H\alpha$ and $[NeIII]$ emissions allow us to infer a redshift $z=1.4802$. This redshift mostly driven by the $H\alpha$ emission in the NIR is in perfect agreement with the CO(5-4) line reported in \cite{diaz2021}, but  $\sim 70$ km/s larger than the redshift $z=1.47935$ obtained from their VLT/FORS2 observations of the $[OII]$ emission. The asymmetric shape of the $H\alpha$ line is due to the velocity field of the arc. We estimate a rotation velocity of $\sim 160$ km/s, in agreement with the $145\pm20$ km/s reported in \cite{diaz2021}. }
\end{figure}
\clearpage
\begin{figure}
\begin{tabular}{c}
\includegraphics{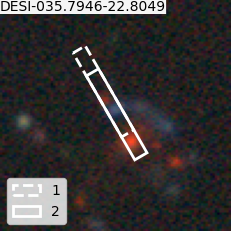} \\
\includegraphics[width=\textwidth]{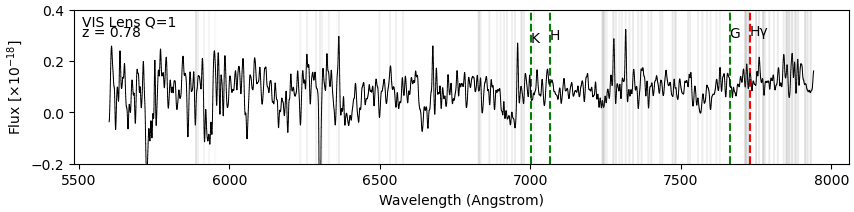} \\
\includegraphics[width=\textwidth]{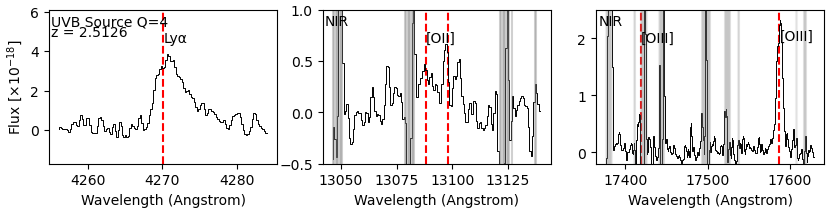} \\
\end{tabular}
\caption{ {\it Top: } RGB image of gravitational lens system DESI-035.7946-22.8049 taken from DESI Legacy Imaging Surveys. The slit position 1 covers the arc and a small satellite galaxy northward, for which we could not measure a redshift. The slit position 2 covers the central galaxy, the arc and the small galaxy.  {\it Bottom: } For the lens, we infer a redshift $z=0.78$ with low confidence based on the $G$ and $H\gamma$ lines. The $H$ and $K$ absorption lines are very weak. For the source, the $Ly\alpha$ in the UVB, the $[OII]$ and $[OIII]$ lines in the NIR allow us to infer a redshift $z=2.5126$. We also detect a radial image of the $[OIII]$ line on top of the lens. }
\end{figure}
\clearpage
\begin{figure}
\begin{tabular}{c}
\includegraphics{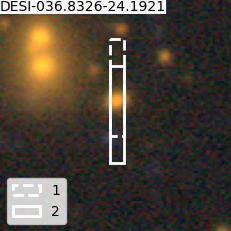} \\
\includegraphics[width=\textwidth]{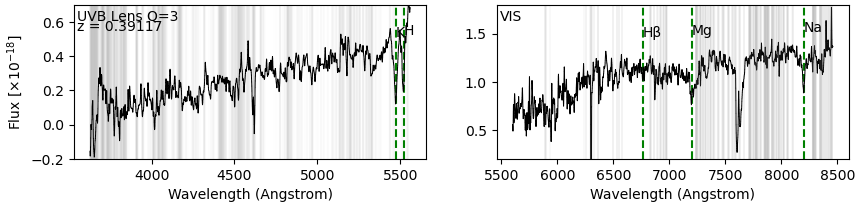} \\
\includegraphics[width=\textwidth]{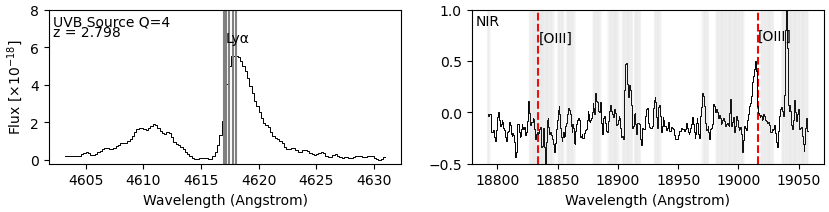} \\
\end{tabular}
\caption{ {\it Top: } RGB image of gravitational lens system DESI-036.8326-24.1921 taken from DESI Legacy Imaging Surveys. This galaxy cluster is also known as ACT-CL J0227.3-2411 \citep{hilton2021}. The slit position 1 covers a satellite galaxy. The slit position 2 covers the satellite galaxy as well as a barely visible blue arc in its southern corner.  {\it Bottom: } For the lens, deep $H$, $K$ in the UV and $Mg$, $Na$ absorption lines in the VIS allow us to infer a redshift $z=0.39117$; For the source, a strong $Ly\alpha$ line in the UV and $[OIII]$ emission lines in the NIR allow us to infer a redshift $z=2.798$. }
\end{figure}
\clearpage
\begin{figure}
\begin{tabular}{c}
\includegraphics{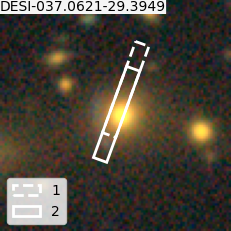} \\
\includegraphics[width=\textwidth]{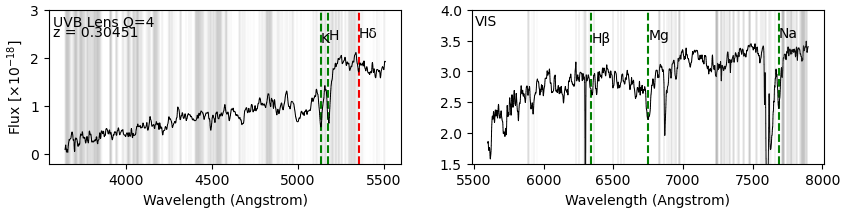} \\
\end{tabular}
\caption{ {\it Top: } RGB image of gravitational lens system DESI-037.0621-29.3949 taken from DESI Legacy Imaging Surveys. The 2 slit positions cover the central galaxy and the arc southward. A small $MgII$ absorption in the UV, and $H\alpha$ in absorption in the NIR were not sufficient to infer a redshift for the arc. {\it Bottom: } Strong $H$ and $K$ lines in the UV, and $Mg$ and $Na$ lines in the VIS allow us to infer a redshift $z=0.30451$. The dip at $\lambda \sim 7600\AA$ is a data reduction artifact at the overlap location of two successive dispersion orders of the XShooter echelle spectrograph. }
\end{figure}
\clearpage
\begin{figure}
\begin{tabular}{c}
\includegraphics{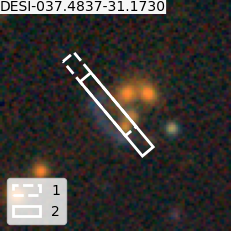} \\
\includegraphics[width=\textwidth]{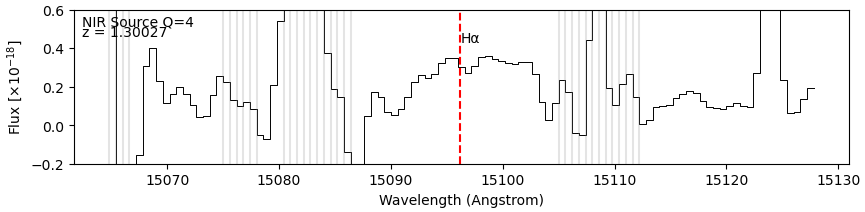} \\
\end{tabular}
\caption{ {\it Top: } RGB image of gravitational lens system DESI-037.4837-31.1730 taken from DESI Legacy Imaging Surveys. The 2 slit positions cover a group satellite galaxy and the northern part of a $\sim 6\arcsec$ long arc. We detect a continuum for the lens, but the lack of emission line does not allow us to infer a redshift. {\it Bottom: } For the source, a clear $H\alpha$ emission in the NIR allow us to infer a redshift $z=1.30027$. Other Balmer lines in the VIS fall under sky lines, and $[OII]$ line is not present. }
\end{figure}
\clearpage
\begin{figure}
\begin{tabular}{c}
\includegraphics{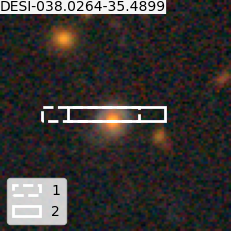} \\
\includegraphics[width=\textwidth]{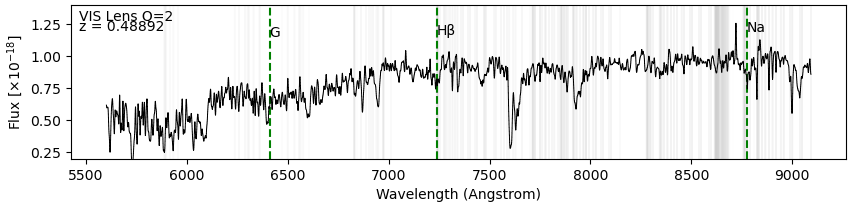} \\
\includegraphics[width=\textwidth]{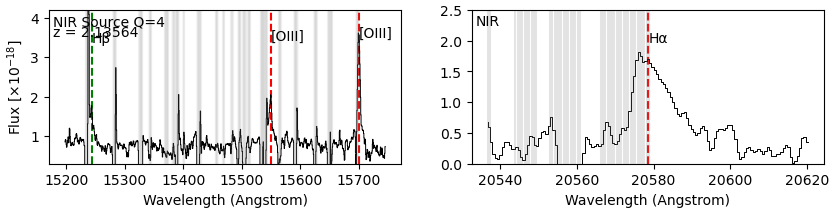} \\
\end{tabular}
\caption{ {\it Top: } RGB image of gravitational lens system DESI-038.0264-35.4899 taken from DESI Legacy Imaging Surveys. The 2 slit positions cover the lens galaxy and the Einstein ring northward. {\it Bottom: } For the lens, deep $G$, $H\beta$ and $Na$ absorption lines allow us to infer a redshift $z=0.48892$. For the source, strong $H\alpha$, $[OIII]$ and $H\beta$ lines in the NIR allow us to infer a redshift $z=2.13564$. }
\end{figure}
\clearpage
\begin{figure}
\begin{tabular}{c}
\includegraphics{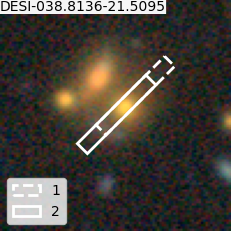} \\
\includegraphics[width=\textwidth]{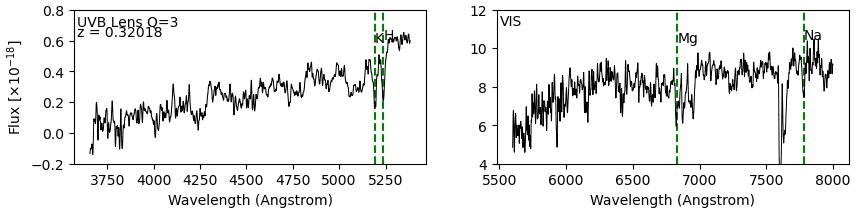} \\
\includegraphics[width=\textwidth]{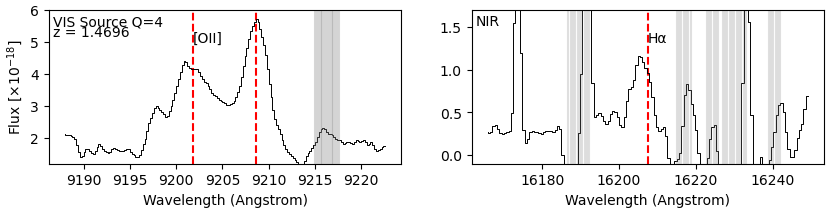} \\
\end{tabular}
\caption{ {\it Top: } RGB image of gravitational lens system DESI-038.8136-21.5095 taken from DESI Legacy Imaging Surveys. The 2 slit positions cover the central galaxy of the group and the arc South East. {\it Bottom: } For the lens, deep $H$ and $K$ line in the UV and $Mg$, $Na$ absorption lines in the VIS allow us to infer a redshift $z=0.32018$. For the source, strong $[OII]$ and $H\alpha$ emission in the VIS and NIR respectively allow us to infer a redshift $z=1.4696$. }
\end{figure}
\clearpage
\begin{figure}
\begin{tabular}{c}
\begin{tabular}{cc}
\includegraphics{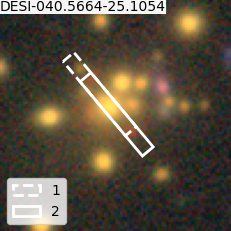} &
\includegraphics{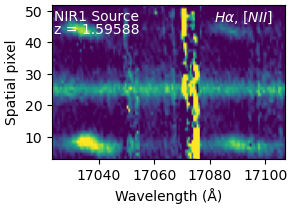} 
\end{tabular} \\
\includegraphics[width=\textwidth]{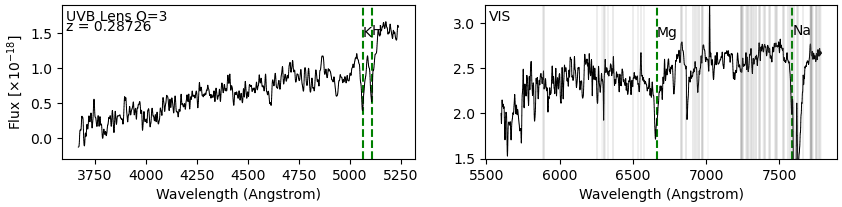} \\
\includegraphics[width=\textwidth]{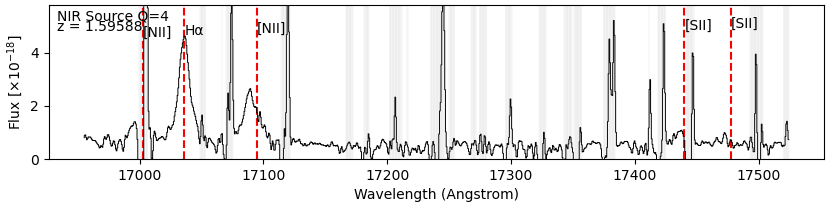} \\
\end{tabular}
\caption{ {\it Top: } RGB image of gravitational lens system DESI-040.5664-25.1054 taken from DESI Legacy Imaging Surveys. Slit position 1 covers the cluster central galaxy, the northern and southern red images of the source and a satellite galaxy, for which we could not measure a redshift. Slit position 2 covers the central galaxy and the southern image of the source. {\it Bottom: } For the lens, deep $H$ and $K$ lines in the UV and $Mg$ and $Na$ lines in the VIS allow us to infer a redshift $z=0.28726$. For the source, strong $H\alpha$ and $[NII]$ lines, and weak $[SII]$ lines allow us to infer a redshift $z=1.59588$. We observe the source velocity field of both multiple images in the $H\alpha$ and $[NII]$ lines.}
\end{figure}
\clearpage
\begin{figure}
\begin{tabular}{c}
\includegraphics{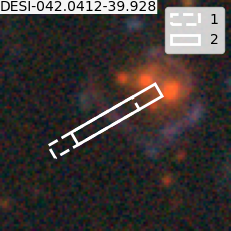} \\
\includegraphics[width=\textwidth]{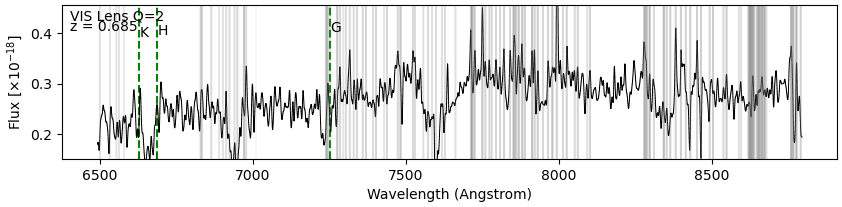} \\
\includegraphics[width=\textwidth]{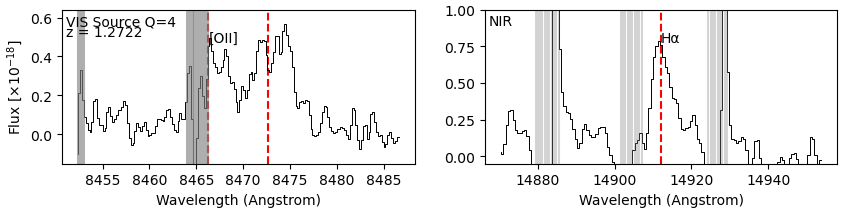} \\
\end{tabular}
\caption{ {\it Top: } RGB image of gravitational lens system DESI-042.0412-39.928 taken from DESI Legacy Imaging Surveys. This compact group lens system is part of the \cite{jacobs2019} DES lens sample.  is modeled in \cite{urcelay2025}. The 2 slit positions cover a fraction of the central galaxy and cross one of the multiple image of the system. {\it Bottom: } Spectra of the lens and source.}
\end{figure}
\clearpage
\begin{figure}
\begin{tabular}{c}
\includegraphics{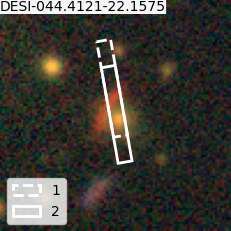} \\
\includegraphics[width=\textwidth]{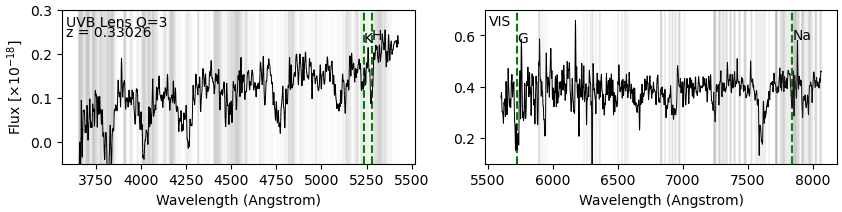} \\
\includegraphics[width=\textwidth]{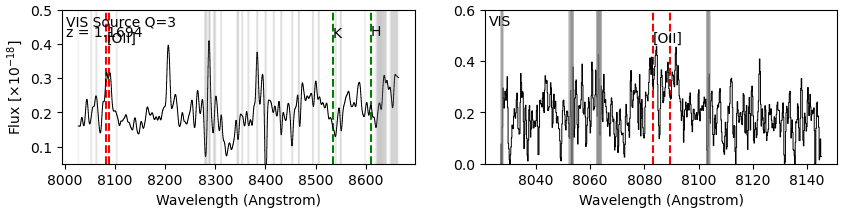} \\
\end{tabular}
\caption{ {\it Top: } RGB image of gravitational lens system DESI-044.4121-22.1575 taken from DESI Legacy Imaging Surveys. This lens system is located on the West side of the galaxy cluster MACS0257 at redshift $z=0.33026$ \citep{richard2021}. {\it Bottom: } For the lens, the $H $ and $K$ lines in the UV, and the $G$ and $Na$ absorption lines in the VIS allow us to infer a redshift $z=0.33026$ in agreement with the cluster redshift. For the source, $[OII]$ in emission and $H$ and $K$ in absorption allow us to infer a redshift $z=1.1694$ in agreement with \citep{richard2021}. }
\end{figure}
\clearpage
\begin{figure}
\begin{tabular}{c}
\includegraphics{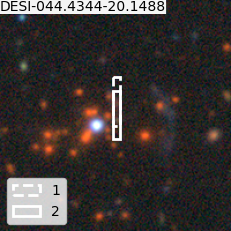} \\
\includegraphics[width=\textwidth]{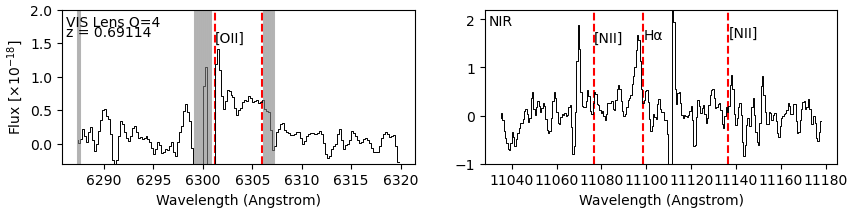} \\
\includegraphics[width=\textwidth]{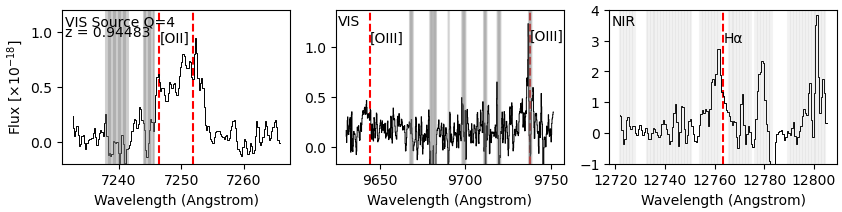} \\
\end{tabular}
\caption{ {\it Top: } RGB image of gravitational lens system DESI-044.4344-20.1488 taken from DESI Legacy Imaging Surveys. This galaxy cluster is also known as ACT-CL J0257.7-2009 and is estimated with a photometric redshift of $z=0.7$ \citep{hilton2021}. The 2 slit positions cover a $\sim 3\arcsec$ long red-blue arc in their northern side, and a blue satellite galaxy in their southern side. {\it Bottom: } For the lens, $H\alpha$ and $[OII]$ emission in the NIR and VIS respectively allow us to infer a redshift $z=0.69114$. For the source, $[OII]$, $[OIII]$ and $H\alpha$ allow us to infer a redshift $z=0.94483$.}
\end{figure}
\clearpage
\begin{figure}
\begin{tabular}{c}
\includegraphics{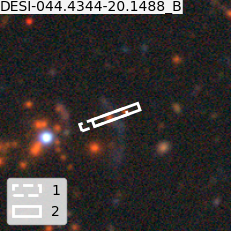} \\
\includegraphics[width=\textwidth]{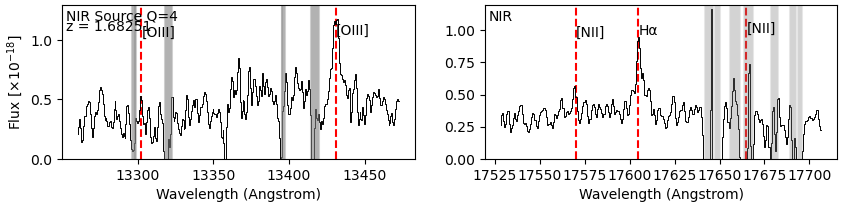} \\
\end{tabular}
\caption{ {\it Top: } RGB image of gravitational lens system DESI-044.4344-20.1488\_B taken from DESI Legacy Imaging Surveys. This galaxy cluster is also known as ACT-CL J0257.7-2009 and is estimated with a photometric redshift of $z=0.7$ \citep{hilton2021}. The 2 slit positions cross the $\sim 10\arcsec$ long blue tangential arc, and cover two red satellite galaxies for which we detect a continuum but no emission line to estimate a redshift.  {\it Bottom: } For the source, $[OIII]$ and $H\alpha$ lines allow us to infer a redshift $z=1.68251$. }
\end{figure}
\clearpage
\begin{figure}
\begin{tabular}{c}
\includegraphics{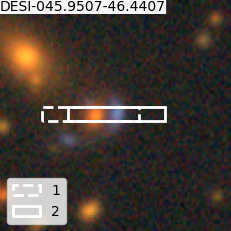} \\
\includegraphics[width=\textwidth]{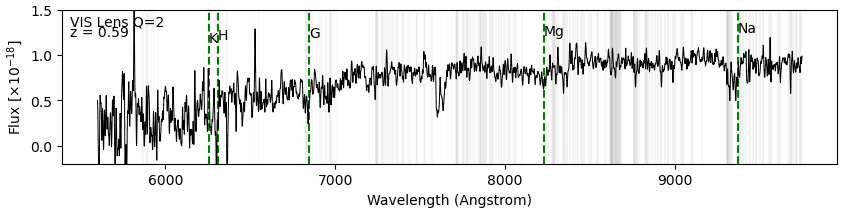} \\
\includegraphics[width=\textwidth]{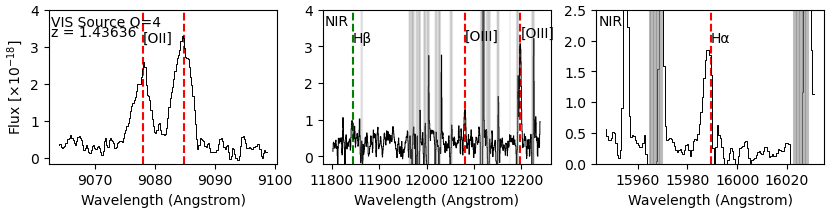} \\
\end{tabular}
\caption{ {\it Top: } RGB image of gravitational lens system DESI-045.9507-46.4407 taken from DESI Legacy Imaging Surveys. This system is part of the \cite{jacobs2019} DES lens sample. The 2 slit positions cover a satellite galaxy and one blue arc. {\it Bottom: } For the lens, the shape of the continuum, the $G$ and $Na$ absorption lines allow us to infer a redshift $z=0.59$. For the source, strong $[OII]$, $[OIII]$ and $H\alpha$ emissions allow us to infer a redshift $z=1.43636$. }
\end{figure}
\clearpage
\begin{figure}
\begin{tabular}{c}
\includegraphics{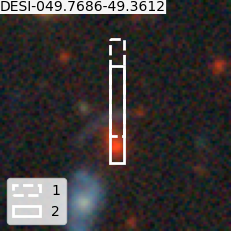} \\
\includegraphics[width=\textwidth]{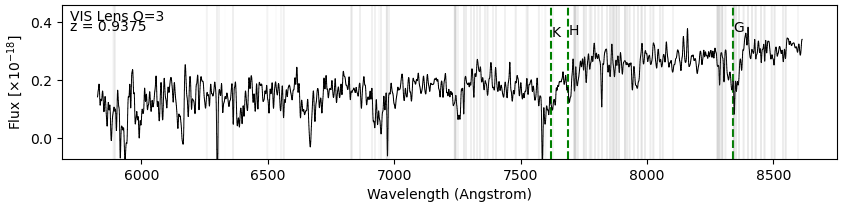} \\
\includegraphics[width=\textwidth]{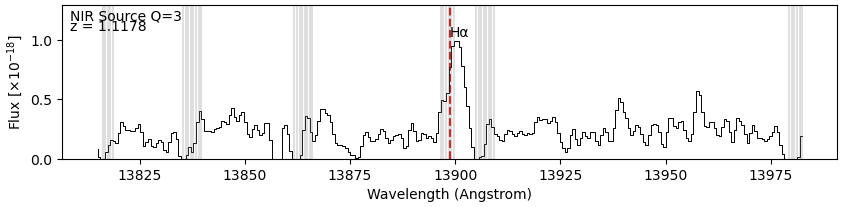} \\
\end{tabular}
\caption{ {\it Top: } RGB image of gravitational lens system DESI-049.7686-49.3612 taken from DESI Legacy Imaging Surveys. The slit position 1 crosses the blue tangential arc and covers a small fraction of the lens. The slit position 2 covers both the lens and source. {\it Bottom: } For the lens, $H$, $K$ and $G$ absorption lines in the VIS spectrum allow us to infer a redshift $z=0.9375$. For the source, a strong $H\alpha$ line in the NIR, confirmed by a weak $MgII$ absorption in VIS allow us to infer a redshift $z=1.1178$. }
\end{figure}
\clearpage
\begin{figure}
\begin{tabular}{c}
\includegraphics{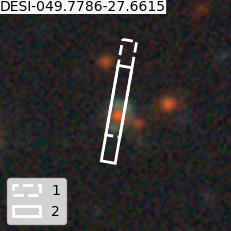} \\
\includegraphics[width=\textwidth]{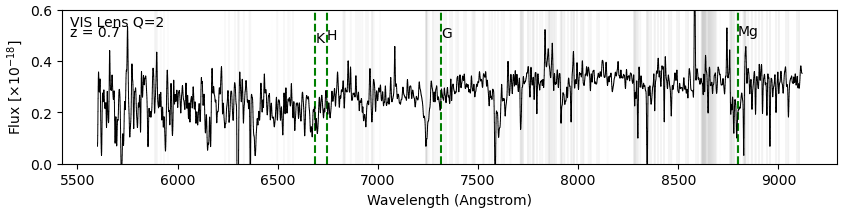} \\
\includegraphics[width=\textwidth]{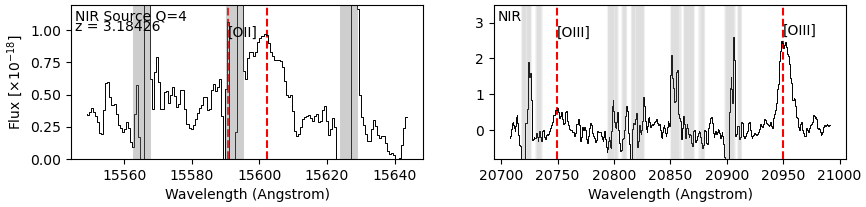} \\
\end{tabular}
\caption{ {\it Top: } RGB image of gravitational lens system DESI-049.7786-27.6615 taken from DESI Legacy Imaging Surveys. The 2 slit positions cover the lens and the Einstein ring on both sides of the lens. {\it Bottom: } For the lens, the shape of the continuum, and the $G$ absorption line allow us to infer a redshift $z=0.7$. For the source, $[OII]$ and $[OIII]$ emissions allow to infer a redshift $z=3.18426$. We note the slightly skewed shape of the $[OIII]$ lines which suggests an outflow. }
\end{figure}
\clearpage
\begin{figure}
\begin{tabular}{c}
\includegraphics{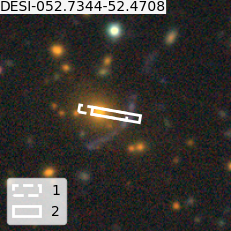} \\
\includegraphics[width=\textwidth]{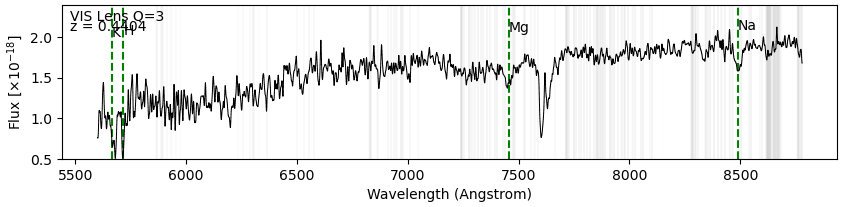} \\
\includegraphics[width=\textwidth]{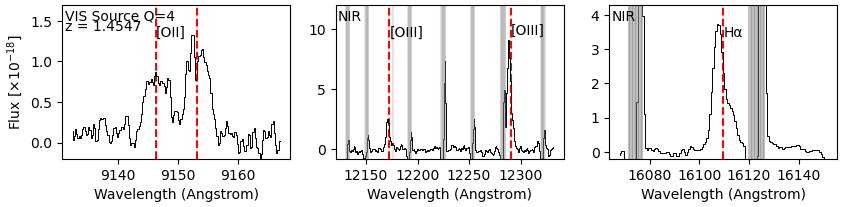} \\
\end{tabular}
\caption{ {\it Top: } RGB image of gravitational lens system DESI-052.7344-52.4708 taken from DESI Legacy Imaging Surveys. This system, also known as ACT-CL J0330.9-5228,  is part of the XMM Cluster and X-CLASS surveys and is estimated at redshift $z=0.44$  \citep{mehrtens2012, koulouridis2021}. The giant arc is also part of the DES Bright Arcs Survey \citep{diehl2017}. The 2 slit positions cover the $\sim 10\arcsec$ long arc and the central galaxy. {\it Bottom: } For the lens, $H$, $K$, $Mg$ and $Na$ absorption lines allow us to infer a redshift $z=0.4404$. For the source, $[OII]$ in the VIS, $[OIII]$ and $H\alpha$ in the NIR allow us to infer a redshift $z=1.4547$. We note the strongly skewed shape of the $H\alpha$ line, which suggests an outflow. }
\end{figure}
\clearpage
\begin{figure}
\begin{tabular}{c}
\includegraphics{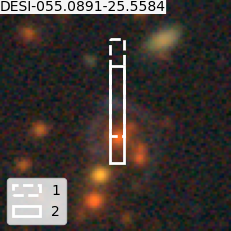} \\
\includegraphics[width=\textwidth]{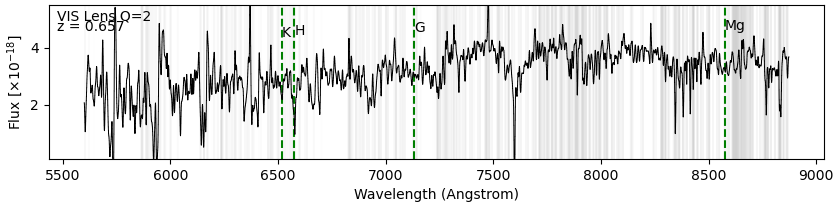} \\
\includegraphics[width=\textwidth]{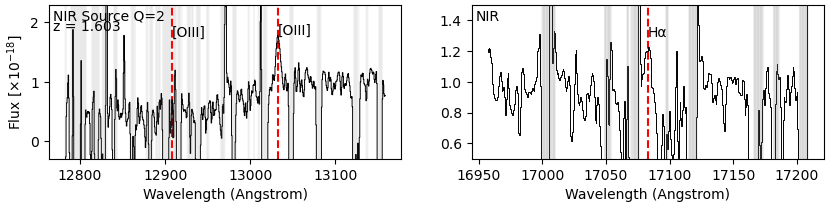} \\
\end{tabular}
\caption{ {\it Top: } RGB image of gravitational lens system DESI-055.0891-25.5584 taken from DESI Legacy Imaging Surveys. This system is part of the \cite{jacobs2019} DES lens sample. The 2 slit positions cover the central galaxy of the group and cross the tangential arc northward. {\it Bottom: } For the lens, the $H$, $K$, $G$ and $Mg$ absorption lines allow us to infer a redshift $z=0.657$. For the source, $H\alpha$ and $[OIII]$ lines in the NIR allow us to infer a redshift $z=1.603$. }
\end{figure}
\clearpage
\begin{figure}
\begin{tabular}{c}
\includegraphics{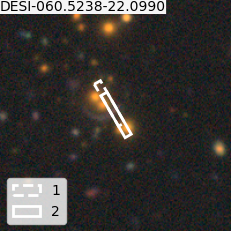} \\
\includegraphics[width=\textwidth]{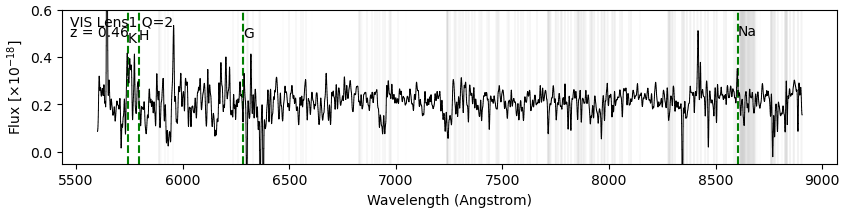} \\
\includegraphics[width=\textwidth]{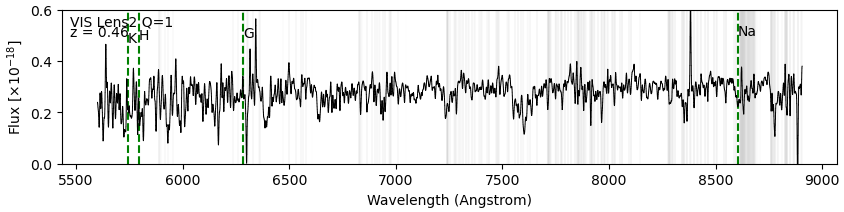} \\
\includegraphics[width=\textwidth]{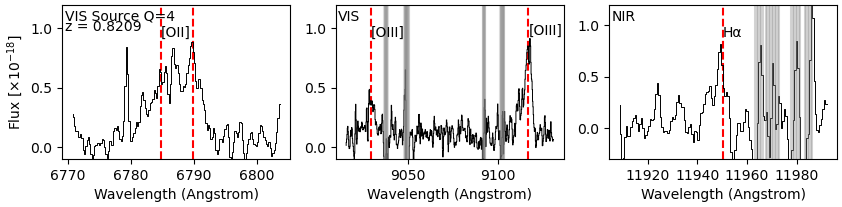} \\
\end{tabular}
\caption{ {\it Top: } RGB image of gravitational lens system DESI-060.5238-22.0990 taken from DESI Legacy Imaging Surveys. This system is part of the \cite{jacobs2019} DES lens sample. The slit position 1 covers the tangential arc in blue, and a fraction of one central galaxy of the group (Lens 2). The slit position 2 covers the tangential arc, the central galaxy (Lens 2) and a satellite galaxy southward (Lens 1). {\it Bottom: } For the two galaxies Lens 1 and 2, $H$, $K$, $G$ and $Na$ absorption lines in VIS allow us to infer a redshift $z=0.46$. For the source, $[OII]$ and $[OIII]$ in VIS, and $H\alpha$ in NIR allow us to infer a redshift $z=0.8209$. We also find an $H\alpha$ emission line at the location of the central galaxy, which reveals the existence of a radial image for this lens system. }
\end{figure}
\clearpage
\begin{figure}
\begin{tabular}{c}
\includegraphics{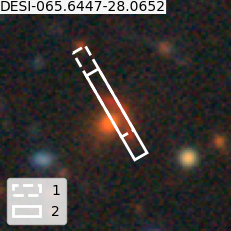} \\
\includegraphics[width=\textwidth]{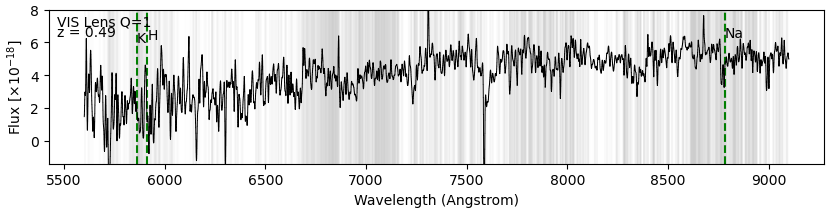} \\
\includegraphics[width=\textwidth]{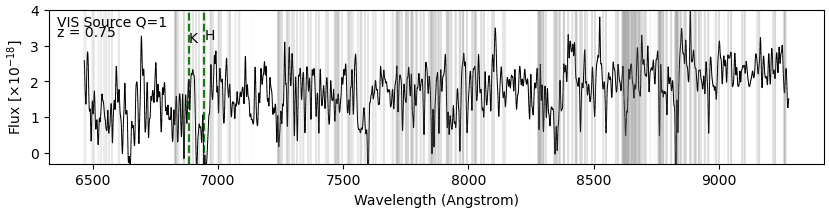} \\
\end{tabular}
\caption{ {\it Top: } RGB image of gravitational lens system DESI-065.6447-28.0652 taken from DESI Legacy Imaging Surveys. The 2 slit positions cover the red $\sim 4\arcsec$ long arc and the lens. {\it Bottom: } For the lens, $H$, $K$ and $Na$ absorption lines allow us to infer a redshift $z=0.49$. For the source, the shape of the continuum, $H$ and $K$ absorption lines allow us to infer a redshift $z=0.75$ with little confidence $Q=1$.}
\end{figure}
\clearpage
\begin{figure}
\begin{tabular}{c}
\includegraphics{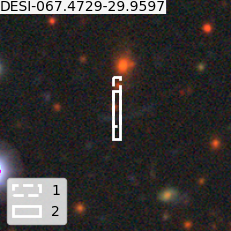} \\
\includegraphics[width=\textwidth]{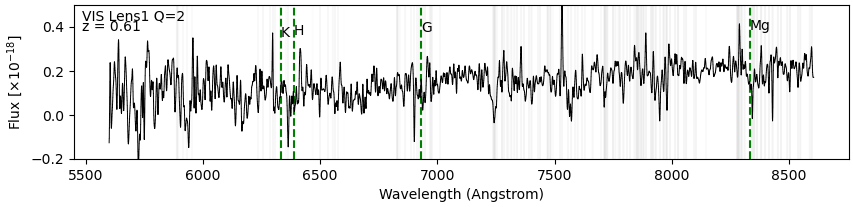} \\
\includegraphics[width=\textwidth]{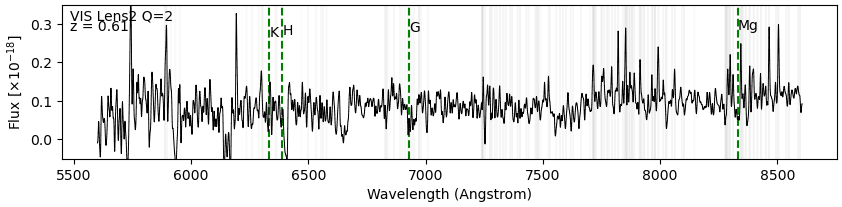} \\
\end{tabular}
\caption{ {\it Top: } RGB image of gravitational lens system DESI-067.4729-29.9597 taken from DESI Legacy Imaging Surveys. 
Slit position 1 covers a satellite galaxy (Lens 2) close to the central galaxy of the group, another satellite further South (Lens 1) and the tangential arc below. Slit position 2 covers Lens1 and the source.We detect no emission line at the location of the source to estimate a redshift. {\it Bottom: } For Lens 1 and 2, $H$, $K$, $G$ and $Mg$ lines allow us to infer a redshift $z=0.61$ for both with little confidence. }
\end{figure}
\clearpage
\begin{figure}
\begin{tabular}{c}
\includegraphics{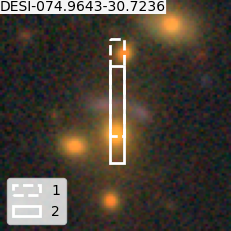} \\
\includegraphics[width=\textwidth]{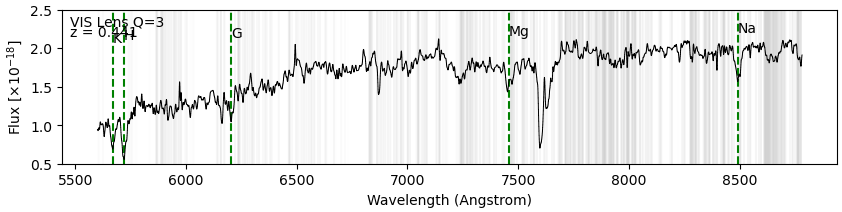} \\
\includegraphics[width=\textwidth]{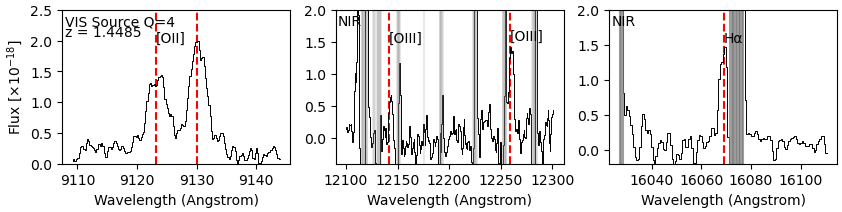} \\
\end{tabular}
\caption{ {\it Top: } RGB image of gravitational lens system DESI-074.9643-30.7236 taken from DESI Legacy Imaging Surveys. The 2 slit positions cover the central galaxy and the blue arc northward. The satellite galaxy in the northern part of slit position 1 is not observed, due to a slight rotation of the slit. {\it Bottom: } For the lens, $H$, $K$, $Mg$ and $Na$ absorption lines allow us to infer a redshift $z=0.4410$. For the source, $[OII]$ in VIS, $[OIII]$ and $H\alpha$ in NIR allow us to infer a redshift $z=1.4485$. }
\end{figure}
\clearpage
\begin{figure}
\begin{tabular}{c}
\includegraphics{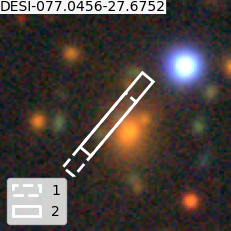} \\
\includegraphics[width=\textwidth]{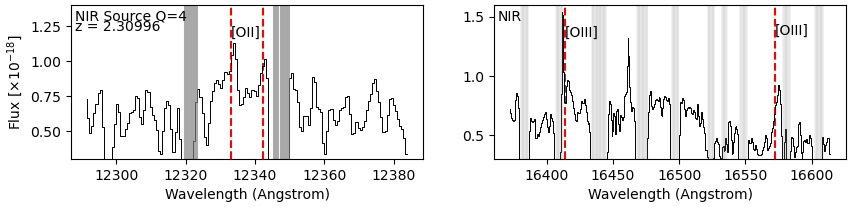} \\
\end{tabular}
\caption{ {\it Top: } RGB image of gravitational lens system DESI-077.0456-27.6752 taken from DESI Legacy Imaging Surveys. The 2 slit positions cover the faint blue arc on the East side of the lens galaxy. We do not have a spectrum for the lens. The photometric redshift given in the LegacySurvey DR9 is $z_\mathrm{phot} = 0.517$. {\it Bottom: } The $[OII]$ and $[OIII]$ emission lines allow us to infer a redshift $z=2.30996$ for the source.}
\end{figure}
\clearpage
\begin{figure}
\begin{tabular}{c}
\includegraphics{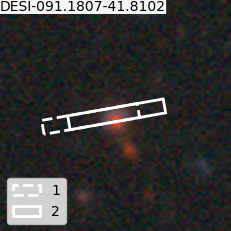} \\
\includegraphics[width=\textwidth]{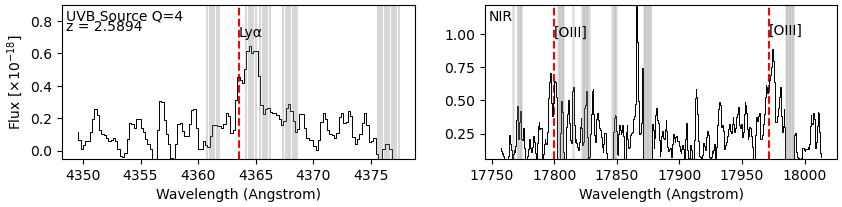} \\
\end{tabular}
\caption{ {\it Top: } RGB image of gravitational lens system DESI-091.1807-41.8102 taken from DESI Legacy Imaging Surveys. The 2 slit positions cover the red lens galaxy and the surrounding Einstein ring. The spectrum of the lens is too faint to derive a redshift. {\it Bottom: } For the source, a clear $Ly\alpha$ emission line in UV, confirmed by $[OIII]$ in the NIR allow us to infer a redshift $z=2.5894$.}
\end{figure}
\clearpage
\begin{figure}
\begin{tabular}{c}
\includegraphics{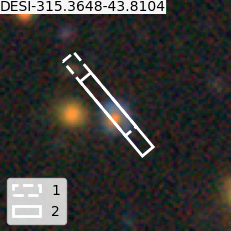} \\
\includegraphics[width=\textwidth]{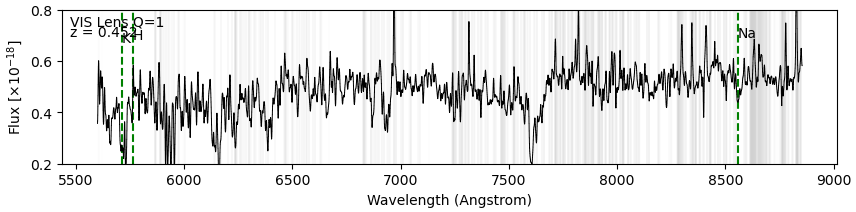} \\
\includegraphics[width=\textwidth]{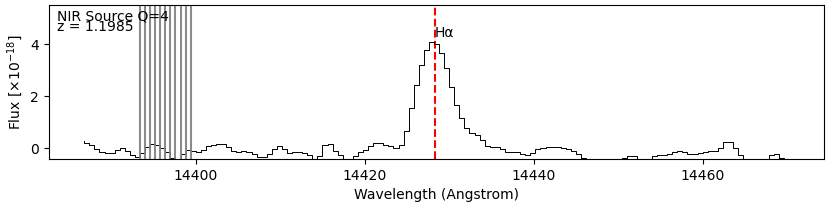} \\
\end{tabular}
\caption{ {\it Top: } RGB image of gravitational lens system DESI-315.3648-43.8104 taken from DESI Legacy Imaging Surveys. The 2 slit positions cover the lens and the surrounding Einstein ring. {\it Bottom: } For the lens, the shape of the continuum, the $Na$, $H$ and $K$ lines, as well as the LegacySurvey DR9 photometric redshift $z_\mathrm{phot} = 0.436$ of the same color nearby galaxy, allow us to infer a redshift $z=0.452$. For the source, a clear $Ly\alpha$ emission in UVB allow us to infer a redshift $z=1.1985$.  }
\end{figure}
\clearpage
\begin{figure}
\begin{tabular}{c}
\includegraphics{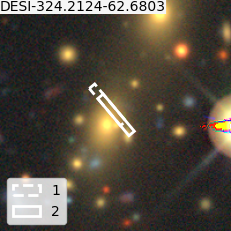} \\
\includegraphics[width=\textwidth]{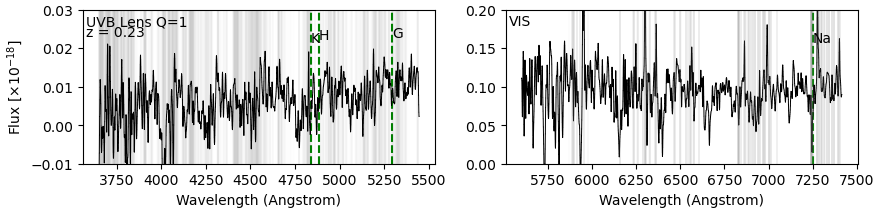} \\
\end{tabular}
\caption{ {\it Top: } RGB image of gravitational lens system DESI-324.2124-62.6803 taken from DESI Legacy Imaging Surveys. The 2 slit positions cover the northern part of the cluster central galaxy, and 2 thin radial arc in green. The arc spectra are buried in the central galaxy one, and we detect no emission line to derive a redshift.  {\it Bottom: } For the lens, the shape of the continuum, the $H$, $K$, $G$ in UVB and $Na$ in VIS allow us to infer a redshift $z=0.23$ in agreement with the LegacySurvey DR9 photometric redshift $z_\mathrm{phot} = 0.245$.}
\end{figure}
\clearpage
\begin{figure}
\begin{tabular}{c}
\includegraphics{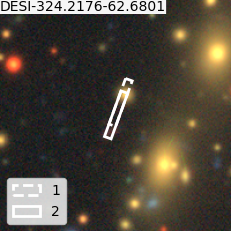} \\
\includegraphics[width=\textwidth]{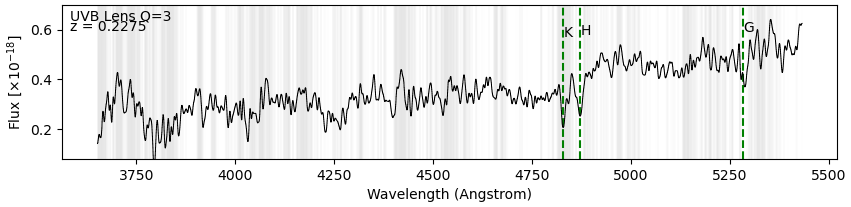} \\
\end{tabular}
\caption{ {\it Top: } RGB image of gravitational lens system DESI-324.2176-62.6801 taken from DESI Legacy Imaging Surveys. The 2 slit positions cover a cluster satellite galaxy and a faint arc in the southern side of the slits. The spectrum of the source is too faint to be extracted. {\it Bottom: } For the lens, $H$, $K$ and $G$ absorption lines in UVB allow us to infer a redshift $z=0.2275$.}
\end{figure}
\clearpage
\begin{figure}
\begin{tabular}{c}
\includegraphics{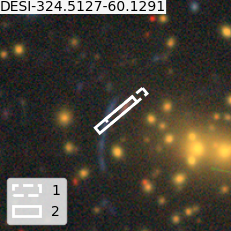} \\
\includegraphics[width=\textwidth]{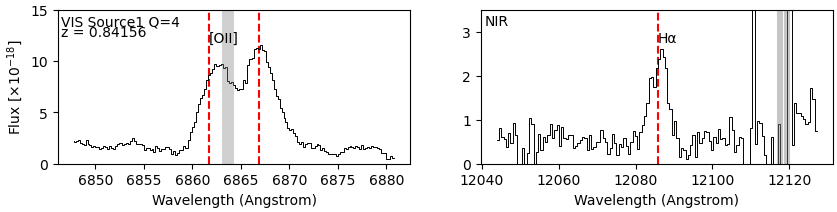} \\
\includegraphics[width=\textwidth]{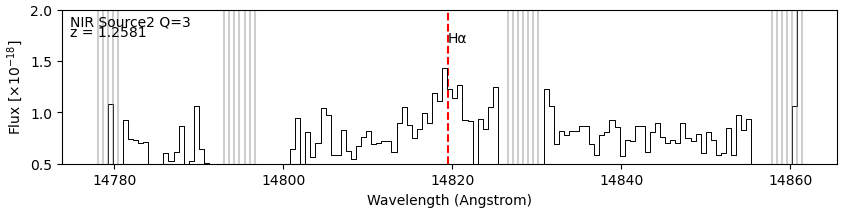} \\
\end{tabular}
\caption{ {\it Top: } RGB image of gravitational lens system DESI-324.5127-60.1291 taken from DESI Legacy Imaging Surveys. This galaxy cluster is also known as SPT-CL J2138-6007 at redshift $z=0.319$ \citep{ruel2014,bleem2015}. It is also part of the new RASS galaxy cluster catalog with low contamination \citep{klein2019}. The 2 slit positions cover 2 background sources: Source1 is red-blue color, and Source 2 is the giant blue arc. {\it Bottom: } For Source 1, $[OII]$ in VIS and $H\alpha$ in NIR allow us to infer a redshift $z=0.84156$. We also detect $[SII]$ and and $Pa\epsilon$. For Source 2, $H\alpha$ in NIR allow us to infer a redshift $z=1.2581$ with confidence $Q=3$ because the spectrum is on the edge of the slit. }
\end{figure}
\clearpage
\begin{figure}
\begin{tabular}{c}
\includegraphics{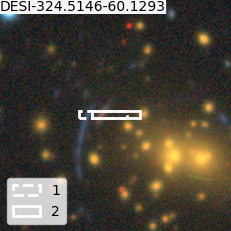} \\
\includegraphics[width=\textwidth]{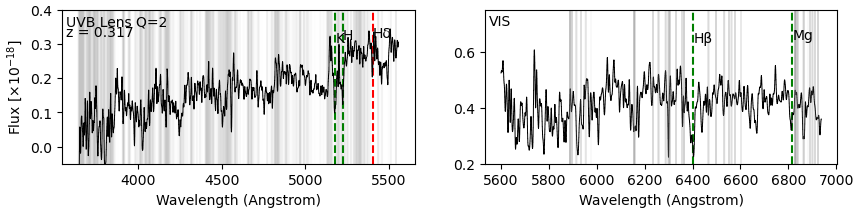} \\
\includegraphics[width=\textwidth]{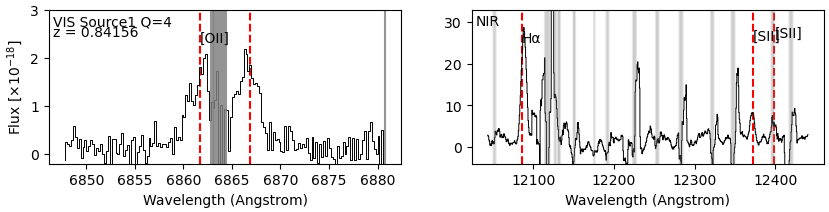} \\
\includegraphics[width=\textwidth]{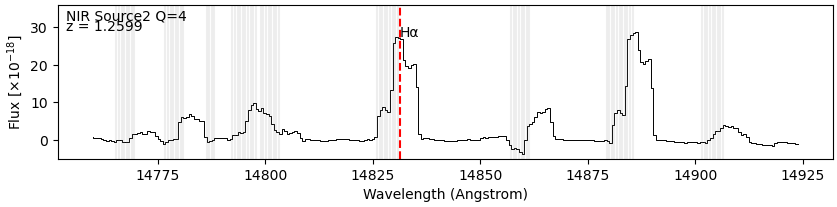} \\
\end{tabular}
\caption{ {\it Top: } RGB image of gravitational lens system DESI-324.5146-60.1293 taken from DESI Legacy Imaging Surveys. This galaxy cluster is also known as SPT-CL J2138-6007 at redshift $z=0.319$ \citep{ruel2014,bleem2015}. It is also part of the new RASS galaxy cluster catalog with low contamination \citep{klein2019}. The 2 slit positions cover 2 background sources: Source1 is red-blue color, and Source 2 is the giant blue arc. Slit position 2 also cover a cluster satellite galaxy. {\it Bottom: } For the lens, $H$, $K$ absorption lines in UVB and $Mg$ in VIS allow us to infer a redshift $z=0.317$. For source 1, $[OII]$ in VIS, $H\alpha$ and $[SII]$ in NIR allow us to infer a redshift $z=0.84156$. For source 2, $H\alpha$ in NIR allow us to infer a redshift $z=1.2599$. }
\end{figure}
\clearpage
\begin{figure}
\begin{tabular}{c}
\includegraphics{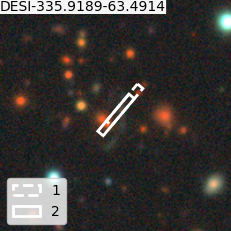} \\
\includegraphics[width=\textwidth]{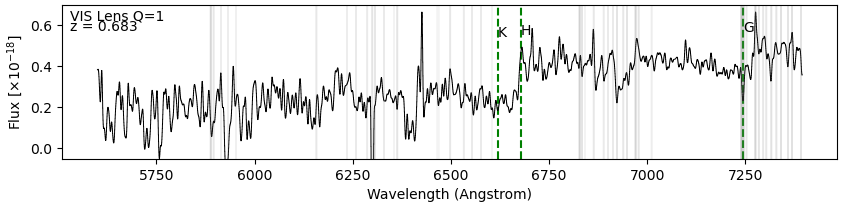} \\
\includegraphics[width=\textwidth]{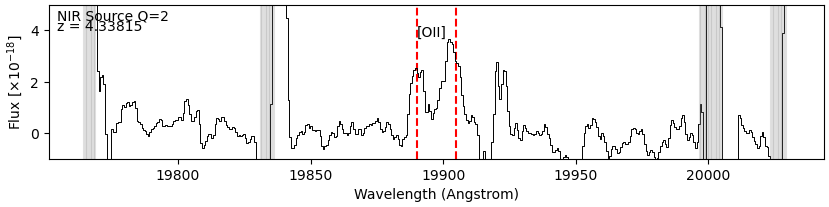} \\
\end{tabular}
\caption{ {\it Top: } RGB image of gravitational lens system DESI-335.9189-63.4914 taken from DESI Legacy Imaging Surveys. The 2 slit positions cross the $\sim 10\arcsec$ long arc and cover a group satellite galaxy southward. The galaxy northward was not observed due to some slight misalignment of the slit. {\it Bottom: } For the lens, the shape of the continuum, the $G$, $H$ and $K$ lines allow us to infer a redshift $z=0.683$. For the source, a single $[OII]$ emission in the NIR allow us to infer a redshift $z=4.33815$ with confidence $Q=2$ because the $\lambda = 19890\AA$ line is strongly affected by a sky line. The other option would be $H\alpha$ at redshift $z=2.0327$.  }
\end{figure}
\clearpage
\begin{figure}
\begin{tabular}{c}
\includegraphics{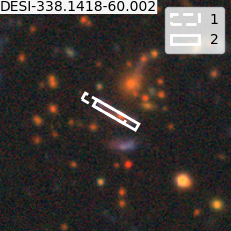} \\
\includegraphics[width=\textwidth]{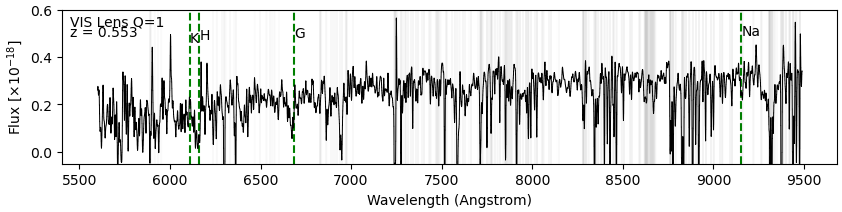} \\
\includegraphics[width=\textwidth]{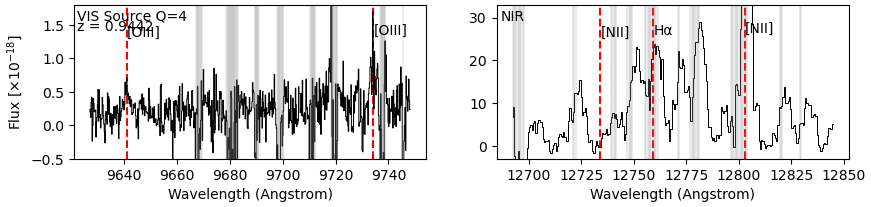} \\
\end{tabular}
\caption{ {\it Top: } RGB image of gravitational lens system DESI-338.1418-60.002 taken from DESI Legacy Imaging Surveys. The galaxy cluster is also known as ACT-CL J2232.5-5959 at redshift $z=0.594$ \citep{bleem2015}. It is also part of the ACT SZ cluster catalog \citep{hilton2021}. The 2 slit positions cover a satellite galaxy embedded in a $\sim 10\arcsec$ long arc. {\it Bottom: } For the lens, the shape of the continuum, $G$, $Na$, $H$ and $K$ absorption lines allow us to infer a redshift $z=0.553$. For the source, $[OIII]$ and $H\alpha$ emission lines in the NIR, allow us to infer a redshift $z=0.9442$. }
\end{figure}
\clearpage
\begin{figure}
\begin{tabular}{c}
\includegraphics{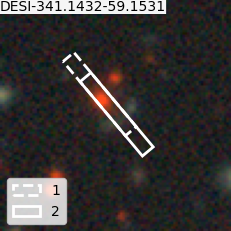} \\
\includegraphics[width=\textwidth]{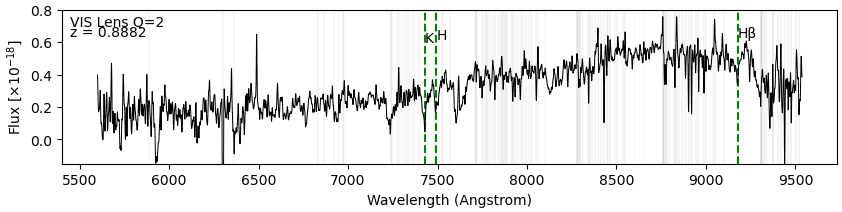} \\
\end{tabular}
\caption{ {\it Top: } RGB image of gravitational lens system DESI-341.1432-59.1531 taken from DESI Legacy Imaging Surveys. The 2 slit positions cover the group central galaxy and the arc southward. The spectrum of the source is too faint to be extracted. {\it Bottom: } For the lens, $H$, $K$ and $H\beta$ in absorption allow us to infer a redshift $z=0.8882$. }
\end{figure}
\clearpage
\begin{figure}
\begin{tabular}{c}
\includegraphics{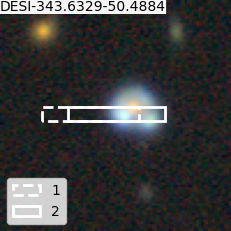} \\
\includegraphics[width=\textwidth]{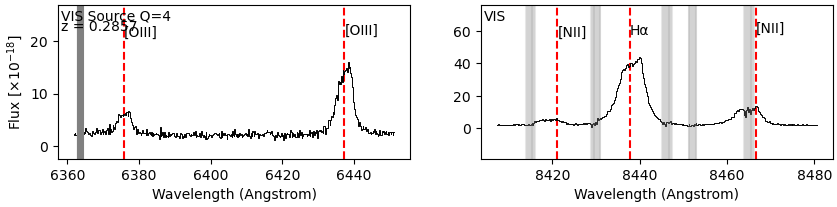} \\
\end{tabular}
\caption{ {\it Top: } RGB image of gravitational lens system DESI-343.6329-50.4884 taken from DESI Legacy Imaging Surveys. The 2 slit positions cover the lens galaxy and two bright spots of the surrounding Einstein ring. The flux of the galaxy is buried into the flux of the arc, and we could not extract a spectrum for it. We could not identify any obvious emission line for the lens. {\it Bottom: } For the source, strong $[OIII]$  and $H\alpha$ emission in VIS allow us to infer a redshift $z=0.2857$. $[OII]$ is also in the UV but strong affected by a sky line. We also observe $H\delta$ and $[NeIII]$ in the UV.}
\end{figure}
\clearpage
\begin{figure}
\begin{tabular}{c}
\includegraphics{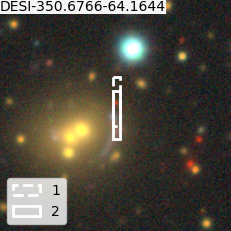} \\
\includegraphics[width=\textwidth]{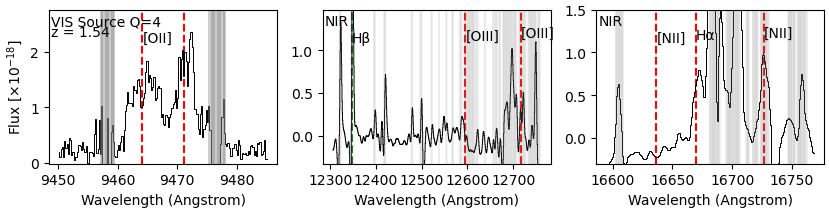} \\
\end{tabular}
\caption{ {\it Top: } RGB image of gravitational lens system DESI-343.6329-50.4884 taken from DESI Legacy Imaging Surveys. The 2 slit positions cover the northern part of a $\sim 10\arcsec$ long arc and a red galaxy. For this later, apart from a faint continuum, no line allow us to infer a redshift. {\it Bottom: } For the source, $[OII]$ in VIS, $[OIII]$, $H\beta$ and $H\alpha$ in NIR allow us to infer a redshift $z=1.54$.}
\end{figure}
\clearpage
\begin{figure}
\begin{tabular}{c}
\includegraphics{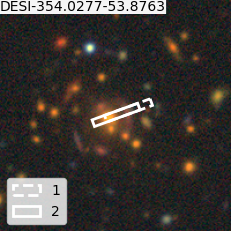} \\
\includegraphics[width=\textwidth]{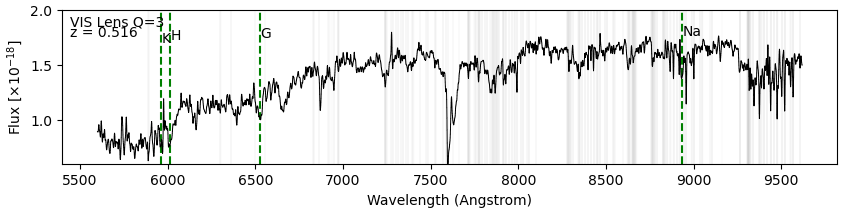} \\
\includegraphics[width=\textwidth]{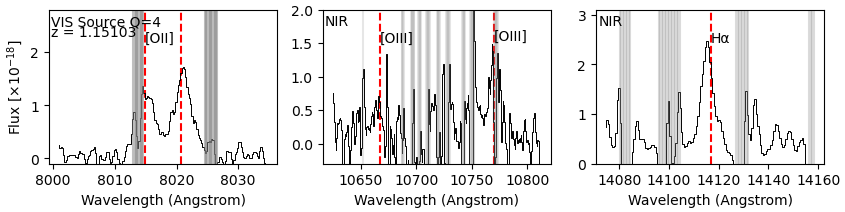} \\
\end{tabular}
\caption{ {\it Top: } RGB image of gravitational lens system DESI-354.0277-53.8763 taken from DESI Legacy Imaging Surveys. The galaxy cluster is also known as ACT-CL J2336.0-5352 and is estimated at photometric redshift $z=0.5206$ \citep{hilton2021}. The redshifts $z_A = 1.1528$ and $z_B = 0.8972$ of the two giant arcs A to the West and B to the East were first measured in the optical with Gemini-GMOS \citep{nord2016}. The 2 slit positions cover the central galaxy and cross a $\sim 7\arcsec$ long arc on its West side. {\it Bottom: } For the lens, $H$, $K$, $G$ and $Na$ absorption lines allow us to infer a redshift $z=0.516$. For the source, $[OII]$ in VIs, $[OIII]$ and $H\alpha$ in NIR allow us to infer a redshift $z=1.15103$. The  difference of $\sim 246$ km/s with \cite{nord2016} might be due to sky line or outflows affecting $[OII]$ and $H\alpha$ regions differently. }
\end{figure}
\clearpage
\begin{figure}
\begin{tabular}{c}
\begin{tabular}{cc}
\includegraphics{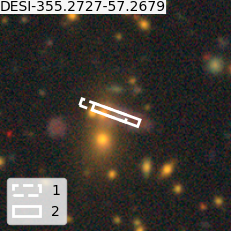} &
\includegraphics{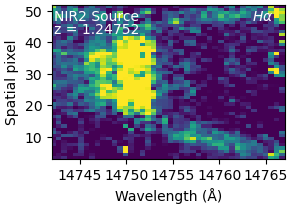} 
\end{tabular} \\
\includegraphics[width=\textwidth]{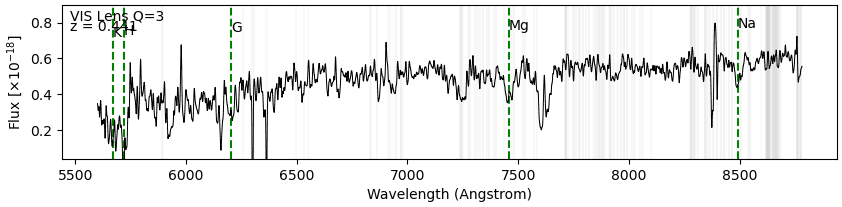} \\
\includegraphics[width=\textwidth]{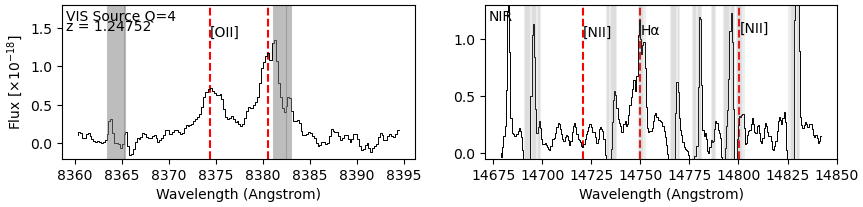} \\
\end{tabular}
\caption{ {\it Top: } RGB image of gravitational lens system DESI-355.2727-57.2679 taken from DESI Legacy Imaging Surveys. The galaxy cluster is also known as LCS-CL J234105-5716.0 and is estimated at photometric redshift $z=0.46$ \citep{bleem2015b}. The 2 slit positions cover a satellite galaxy northward of the BCG, and the East side of a giant arc next to it. {\it Bottom: } For the lens, $H$, $K$, $G$, $Mg$ and $Na$ absorption lines allow us to infer a redshift $z=0.4410$. For the source, $[OII]$ in VIS and $H\alpha$ and $[NII]$ in NIR allow us to infer a redshift $z=1.24752$. The asymmetric shape of the 
$H\alpha$ line is due to the rotating nature of the disk and the velocity field observed in the 2D spectrogram. }
\end{figure}
\clearpage
\begin{figure}
\begin{tabular}{c}
\begin{tabular}{cc}
\includegraphics{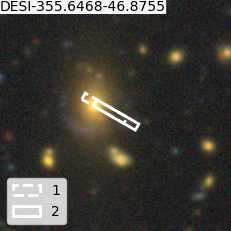} &
\includegraphics{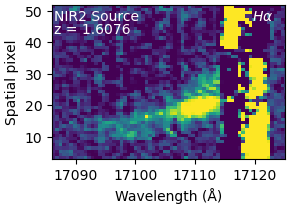}
\end{tabular} \\
\includegraphics[width=\textwidth]{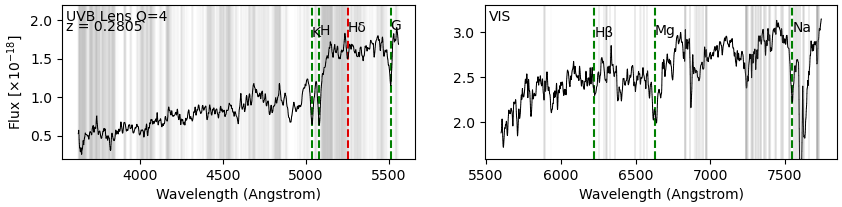} \\
\includegraphics[width=\textwidth]{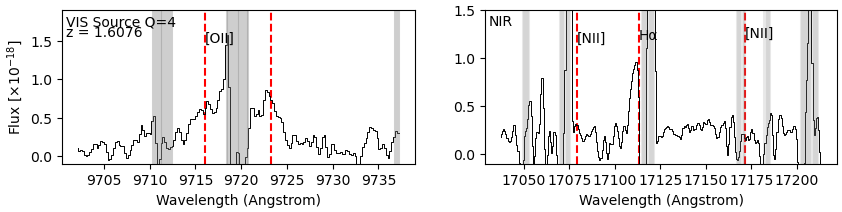} \\
\end{tabular} 
\caption{ {\it Top: } RGB image of gravitational lens system DESI-355.6468-46.8755 taken from DESI Legacy Imaging Surveys. The galaxy cluster is also known as ACT-CL J2342.5-46522 and is estimated at photometric redshift $z=0.2795$ \citep{hilton2021}. The 2 slit positions cover the cluster central galaxy and the West side of the giant arc southward. {\it Bottom: } For the lens, $H$, $K$, $G$  absorption lines in UV, $H\beta$, $Mg$ and $Na$ in VIS allow us to infer a redshift $z=0.2805$. For the source,  $[OII]$ in VIS, $H\alpha$ and $[NII]$ in NIR allow us to infer a redshift $z=1.6076$. We note a velocity field in the 2D spectrogram of the $H\alpha$ line. }
\end{figure}
\clearpage
\begin{figure}
\begin{tabular}{c}
\begin{tabular}{cc}
\includegraphics{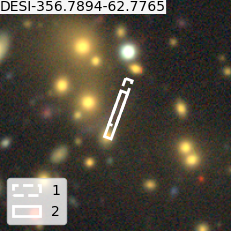} &
\includegraphics{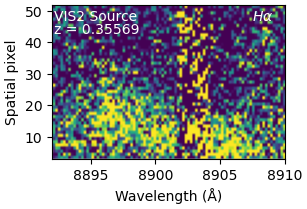} 
\end{tabular} \\
\includegraphics[width=\textwidth]{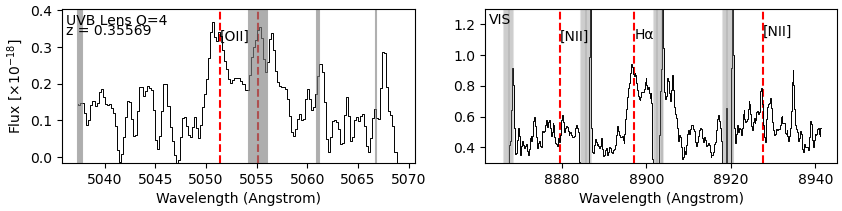} \\
\end{tabular}
\caption{ {\it Top: } RGB image of gravitational lens system DESI-356.7894-62.7765 taken from DESI Legacy Imaging Surveys. This galaxy cluster is also know as ACO 4036 \citep{abell1989}. The 2 slit positions cover a cluster satellite galaxy and the northern tail of a giant arc, East of the BCG. Unfortunately, this arc is too faint to measure a redshift. {\it Bottom: } For the lens, $[OII]$ and $H\alpha$ in VIS allow us to infer a redshift $z=0.35569$. The velocity field observed in the 2D spectrogram suggests that this galaxy is surrounded by a rotating disk.}
\end{figure}
\clearpage
\clearpage
\FloatBarrier
\twocolumn


\end{appendix}

\end{document}